\documentclass[sort&compress, 3p]{elsarticle}
\usepackage[utf8]{inputenc}
\usepackage{amsmath}
\usepackage{amssymb}
\usepackage{bm}
\usepackage{mathtools}
\usepackage{caption}
\usepackage{subcaption}
\usepackage{pdfpages}
\usepackage[capitalise]{cleveref}
\usepackage{siunitx}
\newtheorem{remark}{Remark}[section]
\newcommand{\deriv}[3][]{\ensuremath{\dfrac{\partial^{#1} {#2}}{\partial {#3}^{#1}}}}
\usepackage{algorithm}
\usepackage{algpseudocode}

\definecolor{mygreen}{RGB}{0,0,0}

\begin{document}
%%%%%%%% FRONT MATTER %%%%%%%%%%%
\begin{frontmatter}
\title{Learning constitutive models from microstructural simulations via a non-intrusive reduced basis method: Extension to geometrical parameterizations}
\author[1,3]{Theron Guo\corref{cor1}}\ead{t.guo@tue.nl}
\author[1]{Francesco A. B. Silva}\ead{f.a.b.silva@tue.nl}
\author[2,3]{Ond\v{r}ej Roko\v{s}}\ead{o.rokos@tue.nl}
\author[1,3]{Karen Veroy}\ead{k.p.veroy@tue.nl}
\cortext[cor1]{Corresponding author}
\address[1]{Centre for Analysis, Scientific Computing and Applications, Department of Mathematics and Computer Science, Eindhoven University of Technology, 5612 AZ Eindhoven, The Netherlands}
\address[2]{Mechanics of Materials, Department of Mechanical Engineering, Eindhoven University of Technology, 5612 AZ Eindhoven, The Netherlands}
\address[3]{Institute for Complex Molecular Systems, Eindhoven University of Technology, 5612 AZ Eindhoven, The Netherlands}
\begin{abstract}
Two-scale simulations are often employed to analyze the effect of the microstructure on a component's macroscopic properties. Understanding these structure-property relations is essential in the optimal design of materials for specific applications. However, these two-scale simulations are typically computationally expensive and infeasible in multi-query contexts such as optimization and material design. To make such analyses amenable, the microscopic simulations can be replaced by inexpensive-to-evaluate surrogate models. Such surrogate models must be able to handle microstructure parameters in order to be used for material design. A previous work focused on the construction of an accurate surrogate model for microstructures under varying loading and material parameters by combining proper orthogonal decomposition and Gaussian process regression. However, that method works only for a fixed geometry, greatly limiting the design space. This work hence focuses on extending the methodology to treat geometrical parameters. To this end, a method that transforms different geometries onto a parent domain is presented, that then permits existing methodologies to be applied. We propose to solve an auxiliary problem based on linear elasticity to obtain the geometrical transformations. The method has a good reducibility and can therefore be quickly solved for many different geometries. Using these transformations, combined with the nonlinear microscopic problem, we derive a fast-to-evaluate surrogate model with the following key features: (1) the predictions of the effective quantities are independent of the auxiliary problem, (2) the predicted stress fields automatically fulfill the microscopic balance laws and are periodic, (3) the method is non-intrusive, (4) the stress field for all geometries can be recovered, and (5) the sensitivities are available and can be readily used for optimization and material design. The proposed methodology is tested on several composite microstructures, where rotations and large variations in the shape of inclusions are considered. Finally, a two-scale example is shown, where the surrogate model achieves a high accuracy and significant speed up, thus demonstrating its potential in two-scale shape optimization and material design problems.
\end{abstract}
\end{frontmatter}

%%%%%%%% INTRODUCTION %%%%%%%%%%%
\section{Introduction}
\label{sec:introduction}

With recent advances in additive manufacturing and metamaterials, tailoring the microstructure of materials to obtain desired engineering properties has become possible and crucial. In general, the structure-property relations of such microstructures are not well understood and need to be numerically investigated by fully resolving all microstructural details. Computational homogenization (CH) is typically employed, where the macro- and microstructure are modeled simultaneously. This results in a two-scale formulation which is either solved by a nested Finite Element (FE) scheme, also known as FE$^2$ (see, e.g. \cite{Geers2010, Matous2017, Feyel1999, Miehe2002, Kouznetsova2002}), or by a combination of the FE method with a Fast Fourier Transform (FFT) solver, also known as FE-FFT (see, e.g. \cite{Moulinec1998,Mishra2016a}).

Due to high computational costs involved in solving two-scale simulations, computational homogenization cannot be readily used for material design or optimization problems, where numerous forward simulations must be run. For this reason, several methods have been proposed in the literature that attempt to replace the parametric microscopic model with a fast-to-evaluate while accurate surrogate model. Most of these methods require training data collected by running full microscopic simulations for different parameters. They generally differ in the following aspects:
\begin{enumerate}
    \item the amount of data needed for obtaining an accurate surrogate model,
    \item the ability to make physically sound predictions, and
    \item the way they handle parameters, which can be classified as loading parameters (the applied macroscopic strain) or microstructure descriptors (e.g., material properties of individual phases, volume ratios, etc.). If only loading parameters can be handled, the surrogate model has no design space and can only be used for forward computations.
\end{enumerate}

One popular framework, termed data-driven computational mechanics, was introduced in Kirchdoerfer et al. \cite{Kirchdoerfer2016} and extended in follow-up works; see, e.g., \cite{Kirchdoerfer2017, Eggersmann2019}. Given a collection of stress-strain pairs, this method uses a distance minimizing scheme to directly find the corresponding global stress and strain states of the macroscopic system that satisfy the balance laws as closely as possible, and hence bypasses the empirical material modeling step. This method typically requires very large datasets to be accurate and cannot handle any microstructure descriptors.

Another popular approach is to learn an effective constitutive model from the available data. Usually, neural networks are employed for this regression. In \cite{Wu2020a, Abueidda2021, Mozaffar2019}, the authors used recurrent neural networks, a special type of neural network, to learn a path-dependent elasto-plastic model for a composite RVE. Additionally, in Mozaffar et al. \cite{Mozaffar2019}, microstructure descriptors were included to predict the stresses for a class of composite RVEs. Similarly, in Le et al. \cite{Le2015}, the authors used neural networks to learn a constitutive model for a nonlinear elastic microstructure, considering also microstructural parameters. Recently, this methodology was also applied on mechanical metamaterials, where the authors considered topological microstructural parameters \cite{Kumar2020Inverse-designedMetamaterials, Fernandez2021a}. Despite the successes of this approach, it has two inherent drawbacks: requirements of large datasets for training and no possibility to guarantee that predictions fulfill physical laws.

To overcome these problems, several works recently proposed to embed the physics in neural networks by choosing special network architectures. In \cite{Linka2021}, by assuming a specific form of the strain energy density function, the authors incorporated theoretical knowledge from materials theory in order to predict physical results. A similar idea was implemented in \cite{Flaschel2020}, where a library of strain energy density functions, taken from literature, was used as the basis for the approximated constitutive model. Although these methods fulfill physics by construction and can treat some material parameters, they still require a large amount of data for the training phase, and the material parameters often have a rather difficult interpretation, i.e., different from microstructure descriptors such as fiber size or volume ratio.

The above-mentioned techniques essentially construct the surrogate model in a purely data-driven fashion that neglects the microscopic simulation entirely. On the other hand, also approaches exist that accelerate the microscopic simulation. One example is the Transformation Field Analysis (TFA) which was proposed in Dvorak \cite{dvorak1992} and later extended to the Nonuniform Transformation Field Analysis (NTFA) in Michel and Suquet \cite{Michel2003}. This method is specifically suited for models including internal variables such as plasticity. The internal variables are clustered together into groups to reduce the number of degrees of freedom and averaged evolution laws are developed for each group. A similar approach, termed self-consistent clustering analysis \cite{Liu2016}, was developed that finds these clusters from linear elastic precomputations. These methods require only little data and yield physical predictions. However, they are limited to a fixed microstructure and can therefore not be used for microstructural design.

Another popular technique for dimensionality reduction is the Reduced Basis (RB) method \cite{Quarteroni2015, Hesthaven2016}, which can be applied to general parametric partial differential equations (pPDEs). In this method, the solution to a PDE is sought on a reduced basis spanned by global parameter-independent basis functions. Given a collection of precomputed simulations, often referred to as snapshots, the proper orthogonal decomposition (POD) can be employed to find the reduced basis. The solution of the PDE for any parameter value can then be found by either solving the reduced problem (see, e.g., \cite{Hernandez2014, Soldner2017, Ryckelynck2009, Radermacher2013}) or by using a regression-based approach (see, e.g., \cite{Guo2018, Hesthaven2018, Swischuk2019, Guo2021b}). In the case where the PDE allows an affine decomposition, the method permits an offline--online decomposition and the reduced set of equations can be solved very efficiently online. However, in the case of a nonlinear problem, an affine decomposition usually does not exist and a further approximation called hyperreduction needs to be implemented. In the context of computational homogenization, Hernandez et al. used a discrete empirical interpolation method (DEIM) (see, e.g. \cite{Barrault2004, Chaturantabut2010}) on the stress field to recover the affine decomposition and accelerated an elasto-plastic RVE simulation \cite{Hernandez2014}. Unfortunately, the reduced stiffness matrix is not symmetric. Soldner et al. compared three different hyperreduction approaches for a hyperelastic microstructure and showed that the non-symmetric stiffness matrix leads to convergence problems in several scenarios \cite{Soldner2017}.

Instead of solving the reduced system, the regression-based approach directly predicts the parameter-dependent coefficients. With the already computed POD basis, the solution field can therefore be directly obtained. In our previous work \cite{Guo2021b}, we successfully utilized this approach on a hyperelastic composite microstructure with varying stiffness of the constituents, by combining POD with Gaussian Process Regression (GPR) \cite{Rasmussen2004}. However, this method is not able to treat geometrical parameters such as fiber radius or shape of inclusion, which is important for material design, where optimized shapes are sought.

Within the POD framework, geometrical parameters are typically addressed through transformations that map each snapshot onto a parent domain. The surrogate model can then be constructed and solved on the parent domain. There are typically two kinds of methods to describe such domain transformations: space deformations and surface-based deformations \cite{Botsch2010}. For the former, popular techniques comprise the Free Form Deformation (FFD) \cite{Sederberg1986Free-formModels} and Radial Basis Functions (RBF) \cite{buhmann2003radial}, where the transformation maps are governed by the movement of control points. These methods have been used in the context of RB in various papers, see, e.g., \cite{Rozza2008, Veroy2003, Rozza2020, Negri2015EfficientInterpolation, Manzoni2012ModelGeometries, Washabaugh2016OnGeometries, Demo2018ShapeDecomposition, Stabile2020EfficientMethods}. The biggest benefit is that the auxiliary problem that determines the transformation map is independent of the discretization mesh of the full simulation and depends only on the number of control points used. However, since the deformation can only be prescribed at the control points, many points might be needed to sufficiently describe complicated transformations. Moreover, the boundary nodes on opposite edges of the simulation mesh of a periodic domain, which is usually assumed for a RVE, should be transformed identically, which might not be straightforward to enforce.

For surface-based deformations, knowledge of a parent simulation mesh is needed and then an auxiliary linear elliptic PDE problem is solved to obtain the transformation. Such deformations have been used in shape optimization (see, e.g., \cite{Yao1989}) and for finite volume schemes (see, e.g., \cite{Jasak2006AutomaticMethod}). These auxiliary problems are typically more expensive to solve than the ones arising for the space deformation methods, since they scale with the full simulation mesh. On the other hand, they allow for full flexibility in terms of deformations, and periodicity can be easily enforced. Moreover, this auxiliary problem can be reduced with standard RB techniques \cite{Stabile2020EfficientMethods}, resulting in an efficient online stage. Hence, in this work, we propose an auxiliary problem based on linear elasticity to obtain the transformation maps and show how a fast-to-evaluate effective constitutive model can be constructed for a nonlinear microscopic problem. This surrogate model has the following features:
\begin{enumerate}
    \item the predictions of the effective quantities are independent of the auxiliary problem,
    \item the predicted stress fields automatically fulfill the microscopic balance laws and the periodic boundary conditions,
    \item it can handle both loading parameters and microstructure descriptors (e.g., stiffness and shapes of the constituents) and thus can be used for material design,
    \item it is non-intrusive and can be easily implemented into a macroscopic solver, and
    \item the microscopic stress field can be recovered and visualized.
\end{enumerate}

%%% Structure
The remainder of this paper is organized as follows. Section \ref{sec:problem} introduces the multi-scale problem based on first-order computational homogenization. In Section \ref{sec:surrogate}, the construction of the surrogate model for the microscopic simulation is presented in depth. Then, in Section \ref{sec:results}, the proposed method is validated on multiple composite microstructures, where variations in the shape of the inclusions are considered, and where a full two-scale example is presented. Section \ref{sec:conclusion} concludes this work with a summary on the findings and some final remarks.

%%% Notation
In this work, the following notational convention is adopted. To differentiate between macroscopic and microscopic variables, an overline is added for the macroscopic ones. Italic bold symbols are used for coordinates $\bm{X}$ and vectorial or tensorial fields, such as the displacement $\bm{u}$ or stress field $\bm{P}$. Upright bold symbols are used for algebraic vectors and matrices, such as the identity matrix $\mathbf{I}$ or the macroscopic deformation gradient at a fixed macroscopic point $\bar{\mathbf{F}}$. A field quantity $\bm{u}$ for given parameters $\bm{\mu}$ is denoted as $\bm{u}(\bm{X}; \bm{\mu})$. Given second-order tensors $\bm{A}$ and $\bm{B}$, fourth-order tensor $\bm{C}$, and vector $\bm{v}$, the following operations are used: $\bm{A}\bm{B} = A_{ij}B_{jk}$, $\bm{A}:\bm{B} = A_{ij}B_{ij}$, $\bm{A}\otimes\bm{B} = A_{ij}B_{kl}$, $\bm{A}\otimes\bm{v} = A_{ij}v_k$, $\bm{A}:\bm{C}:\bm{B} = A_{ij}C_{ijkl}B_{kl}$ and $\bm{A}\bm{v} = A_{ij}v_j$,
where the Einstein summation convention is used.

%%%%%%%% PROBLEM %%%%%%%%%%%
\section{Problem Statement}
\label{sec:problem}
When considering engineering systems with complex microstructures, the micro- and macrostructure are both modeled and solved simultaneously in a coupled manner. Governing equations at the level of individual scales are formulated as described below.

\subsection{Macroscopic Problem}
\begin{figure}[pth]
    \centering
    \includegraphics[width=0.5\textwidth]{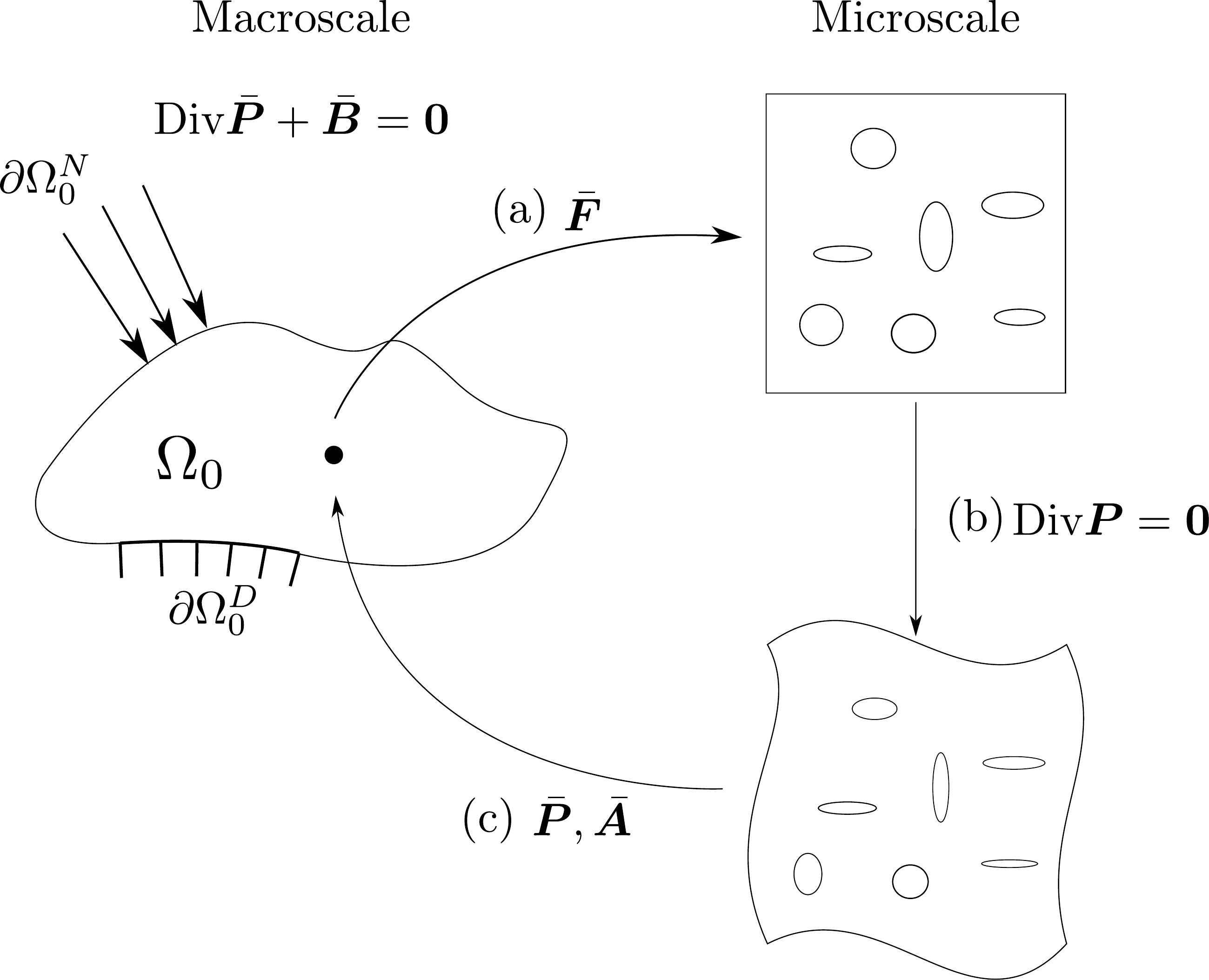}
    \caption{Coupling of two scales. (a) In every macroscopic point the macroscopic deformation gradient is used to specify the microscopic problem. (b) The microscopic problem is then solved to obtain a microscopic stress field. (c) The stress field is subsequently averaged to obtain the effective stress and stiffness which are transferred back to the macroscale. Note that only the effective stress and stiffness are needed for the solution of the macroscopic problem.}
    \label{fig:micromacro}
\end{figure}

Consider a solid body in the reference configuration $\bar{\Omega}_0$. Under prescribed boundary conditions, every point $\bar{\bm{X}} \in \bar{\Omega}_0$ is mapped by a deformation $\bar{\bm{\Phi}}$ onto $\bar{\bm{x}} = \bar{\bm{\Phi}}(\bar{\bm{X}}) \in \bar{\Omega}$, where $\bar{\Omega}$ is the current configuration. The displacement is then defined as $\bar{\bm{u}}(\bar{\bm{X}})\coloneqq\bar{\bm{x}}-\bar{\bm{X}}=\bar{\bm{\Phi}}(\bar{\bm{X}})-\bar{\bm{X}}$. The governing partial differential equation (PDE) describing the effective physics of the system is given by the quasi-static linear momentum balance,
\begin{alignat}{2}
    \begin{aligned}
    \text{Div} \bar{\bm{P}} + \bar{\bm{B}}_0 &= \bm{0}         &&\text{ on } \bar{\Omega}_0, \\
    \bar{\bm{P}}\bar{\bm{N}}_0      &= \bar{\bm{t}}_0 &&\text{ on } \partial\bar{\Omega}_0^N, \text{ and}\\
    \bar{\bm{u}}            &= \bar{\bm{u}}_0 &&\text{ on } \partial\bar{\Omega}_0^D,
    \end{aligned} \label{eq:macropde}
\end{alignat}
where $\text{Div}$ denotes the divergence operator with respect to the reference configuration, $\bar{\bm{P}}$ is the macroscopic first Piola-Kirchhoff (PK1) stress tensor, $\bar{\bm{B}}_0$ are the macroscopic body forces, $\bar{\bm{N}}_0$ is the outward normal on the surface of the body, $\bar{\bm{t}}_0$ and $\bar{\bm{u}}_0$ are the prescribed tractions and displacements, and $\partial\bar{\Omega}^N_0$, $\partial\bar{\Omega}^D_0$ denote the Neumann and Dirichlet boundaries with $\partial\bar{\Omega}_0 = \partial\bar{\Omega}^N_0 \cup \partial\bar{\Omega}^D_0$ and $\partial\bar{\Omega}^N_0 \cap\partial\bar{\Omega}^D_0 = \emptyset$. The stress tensor $\bar{\bm{P}}$ is a nonlinear operator that in general depends on parameters $\bm{\mu}$ and acts on the deformation gradient $\bar{\bm{F}}$ which is defined as
\begin{align}
    \bar{\bm{F}} \coloneqq \deriv{\bar{\bm{x}}}{\bar{\bm{X}}} = \mathbf{I}+\deriv{\bar{\bm{u}}}{\bar{\bm{X}}},
\end{align}
with $\mathbf{I}$ the identity matrix. The weak form of Eq.~\eqref{eq:macropde} is given as
\begin{align}
    \bar{G} \coloneqq \int_{\bar{\Omega}_0} \deriv{\delta\bar{\bm{u}}}{\bar{\bm{X}}}:\bar{\bm{P}} dV - \int_{\bar{\Omega}_0} \bar{\bm{B}}_0 \cdot \delta\bar{\bm{u}} dV - \int_{\partial\bar{\Omega}_0^N} \bar{\bm{t}}_0\cdot\delta\bar{\bm{u}} dA = 0, \qquad \forall \delta\bar{\bm{u}}\in H^1_0({\bar{\Omega}}_0),
\end{align}
where $H^1_0({\bar{\Omega}}_0)=\{\bm{v} \in H^1(\bar{\Omega}_0) \ | \ \bm{v}=0 \text{ on } \partial\bar{\Omega}_0^D\}$ is the test function space with $H^1(\bar{\Omega}_0)$ a Hilbert space, and a solution for the displacement, $\bar{\bm{u}}\in H^1({\bar{\Omega}}_0)$, is sought that fulfills $\bar{\bm{u}}=\bar{\bm{u}}_0$ on $\partial{{\bar{\Omega}}}_0^D$. For the Newton-Raphson method, the G\^{a}teaux derivative of $\bar{G}$ at the current deformation $\bar{\bm{u}}$ in the direction $\Delta\bar{\bm{u}}$ is needed,
\begin{align}
\begin{aligned}
    \left.D\bar{G}\right|_{\bar{\bm{u}}} \cdot \Delta \bar{\bm{u}}
    &= \int_{\bar{\Omega}_0} \deriv{(\delta\bar{\bm{u}})}{\bar{\bm{X}}}:\bar{\bm{A}}:\deriv{\Delta \bar{\bm{u}}}{\bar{\bm{X}}} dV,
\end{aligned}
\end{align}
where $\bar{\bm{A}} \coloneqq \deriv{\bar{\bm{P}}}{\bar{\bm{F}}}$ is the fourth-order stiffness tensor. To find a solution to this problem, a material model needs to be specified, also known as a constitutive law, which relates the stress tensor $\bar{\bm{P}}$ to the deformation gradient $\bar{\bm{F}}$ given a set of parameters $\bm{\mu}$ (e.g., material, geometry, or loading). This can be an empirical law, e.g., obtained by fitting experimental data; however, finding an empirical law with meaningful parameters is often rather difficult. Furthermore, phenomenological laws are often insufficiently rich to capture the full complexity of the microstructural behavior. Therefore, instead, homogenization techniques are used, where a microscopic problem, defined on a representative volume element (RVE), is solved for an effective stress $\bar{\bm{P}}$ and stiffness $\bar{\bm{A}}$, given the macroscopic deformation gradient $\bar{\bm{F}}$ and a set of parameters $\bm{\mu}$, resulting in a two-scale formulation. A visualization of the two-scale simulation is shown in Fig.~\ref{fig:micromacro}.

\begin{remark}
Utilizing the polar decomposition, the deformation gradient $\bar{\bm{F}}$ can always be multiplicatively decomposed into a rotation $\bar{\bm{R}}$ and stretch tensor $\bar{\bm{U}}$. Then, the stretch tensor $\bar{\bm{U}}$ is used to evaluate the effective stress and stiffness. To obtain the effective stress and stiffness with respect to the deformation gradient $\bar{\bm{F}}$, the quantities can be rotated accordingly with $\bar{\bm{R}}$, for details see \cite{itskov2007tensor}. The advantage of this decomposition is that the number of loading parameters is reduced, since the stretch tensor $\bar{\bm{U}}$ is symmetric, unlike $\bar{\bm{F}}$.
\end{remark}

\subsection{Microscopic Problem}
In first order homogenization, the microscopic displacement field $\bm{u}(\bar{\bm{X}}, \bm{X})$, with $\bar{\bm{X}}$ and $\bm{X}$ denoting the macroscopic and microscopic coordinates, is assumed to consist of a macroscopic mean field $(\bar{\bm{F}}(\bar{\bm{X}}) - \mathbf{I}) \bm{X}$ and a zero-mean microscopic fluctuation field $\bm{w}(\bar{\bm{X}}, \bm{X})$, i.e.
\begin{align}
    \bm{u}(\bar{\bm{X}}, \bm{X}) = (\bar{\bm{F}}(\bar{\bm{X}}) - \mathbf{I}) \bm{X} + \bm{w}(\bar{\bm{X}}, \bm{X}),
\end{align}
with the macroscopic deformation gradient $\bar{\bm{F}}(\bar{\bm{X}})$ depending only on the macroscopic point $\bar{\bm{X}}$. For conciseness, in the following the dependence on $\bar{\bm{X}}$ is omitted and the following equations are given for a fixed macroscopic point $\bar{\bm{X}}$.
The microscopic deformation gradient then reads
\begin{align}
    \bm{F}(\bm{X}) \coloneqq \mathbf{I} + \deriv{\bm{u}}{\bm{X}} = \bar{\mathbf{F}} + \deriv{\bm{w}}{\bm{X}} \label{eq:deformationcoupling},
\end{align}
with
\begin{align}
    \left<\bm{F}\right> = \bar{\mathbf{F}},
\end{align}
where $\left<(\bullet)\right>\coloneqq|\Omega|^{-1}\int_\Omega (\bullet) dV$ denotes the averaging operator with $|\Omega|$ the volume of the RVE $\Omega$.
The microscopic governing PDE has the same shape as \Cref{eq:macropde} and is defined on the RVE $\Omega$:
\begin{alignat}{2}
\begin{aligned}
    \text{Div} \bm{P}      &= \bm{0}         &&\text{ on } \Omega, \\
    \bm{u}^+ - \bm{u}^- &= (\bar{\mathbf{F}}-\mathbf{I})(\bm{X}^+ - \bm{X}^-) &&\text{ on } \partial\Omega,
\end{aligned} \label{eq:micropde}
\end{alignat}
where the body forces are neglected, $\bm{P}$ denotes the microscopic PK1 stress, and periodic boundary conditions are prescribed, with $(\bullet)^+$ and $(\bullet)^-$ denoting a quantity on opposite boundaries of the RVE. The weak form is then given as
\begin{align}
    G = \int_\Omega \deriv{\delta\bm{u}}{\bm{X}}:\bm{P} dV = 0,\qquad \forall \delta\bm{u}\in H^1_0(\Omega),
\end{align}
and the G\^{a}teaux derivative in the direction $\Delta\bm{u}$ around the current state $\bm{u}$ is given by
\begin{align}
\begin{aligned}
    \left.DG\right|_{\bm{u}}\cdot \Delta \bm{u} &= \int_\Omega \deriv{(\delta\bm{u})}{\bm{X}}:\bm{A}:\deriv{\Delta \bm{u}}{\bm{X}} dV.
\end{aligned}
\end{align}
The periodic boundary conditions can be enforced by using either Lagrange multipliers \cite{Miehe2002} or the condensation method \cite{Kouznetsova2001}.

In this work, we model each phase of the microstructure as a hyperelastic Neo-Hookean material with strain energy density function
\begin{align}
    W(\bm{F}, \bm{\lambda}) &= C_1(\text{Tr}(\bm{C})-3-2\ln{J})+D_1(J-1)^2, \label{eq:neohookeanenergy}
\end{align}
where $C_1$ and $D_1$ are the material parameters stored in $\bm{\lambda}=[C_1,D_1]^T$, $\text{Tr}(\bullet)$ denotes the trace operator, $\bm{C}=\bm{F}^T\bm{F}$ the right Cauchy-Green deformation tensor and $J=\text{det}(\bm{F})$ is the determinant of $\bm{F}$. The material parameters $C_1$ and $D_1$ can be rewritten into the corresponding Young's modulus $E$ and Poisson's ratio $\nu$ with
\begin{align}
    E = \frac{2C_1 (3D_1 + 2C_1)}{C_1+D_1}, \quad \nu=\frac{D_1}{2(C_1+D_1)}. \label{eq:lameconstants}
\end{align}
The stress and stiffness tensor are found by differentiating \cref{eq:neohookeanenergy}:
\begin{align}
    \bm{P} = \deriv{W}{\bm{F}}, \quad \bm{A} = \deriv{\bm{P}}{\bm{F}}.
\end{align}
After solving the microscopic problem, the effective stress $\bar{\mathbf{P}}$ can be obtained by averaging the microscopic stress, i.e.,
\begin{align}
    \bar{\mathbf{P}} = \left<\bm{P}\right>.
\end{align}
However, computing the effective stiffness $\bar{\mathbf{A}}$ is more complicated, since
\begin{align}
    \left<\bm{A}\right> \not = \bar{\mathbf{A}};
\end{align}
see, e.g., \cite{Saeb2016}. Different ways of computing a consistent effective stiffness have been derived in \cite{Miehe2002, Kouznetsova2001}. Another possible way is to numerically approximate it with a finite difference scheme \cite{Temizer2008}.

%%%%%%%% SURROGATE %%%%%%%%%%%
\section{Surrogate Modeling}
\label{sec:surrogate}
Since the microscopic problem needs to be solved at every quadrature point of the macroscopic problem in each Newton iteration, running even a single full two-scale simulation is expensive. In multi-query contexts such as optimization or material design, it is therefore needed to accelerate the simulation. Several methods have been proposed to replace the microscopic model with a cheap-to-evaluate surrogate model. One of the most powerful tools for dimensionality reduction is the so-called Reduced Basis (RB) method. The reduced basis is often obtained by employing a Proper Orthogonal Decomposition (POD). Even though powerful for loading and material parameters, geometrical parameters cannot be easily treated since the snapshots generally need to be transformed first onto a parent domain for accurate approximations. To address this problem, we present in this section a surface-based deformation method for finding such geometrical transformations by solving an auxiliary problem based on linear elasticity. We then subsequently show how to construct the surrogate model and replace the microscopic simulation.

\subsection{Proper Orthogonal Decomposition}
For convenience, we differentiate between macroscopic loading parameters $\bar{\mathbf{U}}$, material parameters $\bm{\lambda}$ and geometrical parameters $\bm{\mu}$. {\color{mygreen}Given a set of snapshots $\{\bm{P}^{i}\}^{N_s}_{i=1} \in \mathcal{V}_h$, where $\mathcal{V}_h$ is a discretized function space with $\dim{\mathcal{V}_h}=\mathcal{N}$, $N_s$ is the number of snapshots and $\bm{P}^i$ is the field variable corresponding to a given choice of parameters $(\bar{\mathbf{U}}^i, \bm{\lambda}^i, \bm{\mu}^i)$, the proper orthogonal decomposition utilizes the correlation between the snapshots to find an optimal subspace in $\mathcal{V}_h$.} Then, the field variable $\bm{P}$ can be approximated with
\begin{align}
    \bm{P}(\bm{X}; \bar{\mathbf{U}}, \bm{\lambda}, \bm{\mu}) \approx \sum_{n=1}^N \alpha_n(\bar{\mathbf{U}}, \bm{\lambda}, \bm{\mu}) \bm{B}_n(\bm{X}),
\end{align}
where the $\bm{B}_n(\bm{X})$ are global parameter-independent basis functions, $\alpha_n(\bar{\mathbf{U}}, \bm{\lambda}, \bm{\mu})$ are parameter-dependent coefficients, and $N$ is the number of basis functions, where, ideally, $N \ll 
\mathcal{N}$. More details on the computation of the basis functions $\bm{B}_n(\bm{X})$ can be found in \cite{Guo2021b}.

\subsection{Transformation of Snapshots}
\label{subsec:transformation}
For geometrical parameters, a reduced basis obtained with POD will in general perform poorly and might exhibit artificial oscillatory effects \cite{Sarna2021a}. Furthermore, changing the geometry will also affect the simulation mesh, leading to incompatible snapshots, since the field variable is only known at different discrete points and some type of extrapolation and interpolation needs to be adopted. In order to apply POD on such a set of snapshots, one would first need to interpolate each snapshot onto the same underlying grid of discrete points, leading to additional interpolation errors. To overcome these issues, the snapshots can be transformed onto a parent domain first. In most existing works, such transformations are obtained with space deformation techniques (see, e.g., \cite{Rozza2008, Veroy2003, Rozza2020, Negri2015EfficientInterpolation, Manzoni2012ModelGeometries, Washabaugh2016OnGeometries, Demo2018ShapeDecomposition}) relying only on a few control points. The biggest advantage of these methods is that obtaining the transformation map only scales with the number of control points and is independent of the original simulation mesh. However, several control points might be needed to describe complicated geometries.

In the case of surface-based deformation techniques, an auxiliary diffusion-type PDE problem is posed and solved on the simulation mesh of the original problem. Hence, the transformation map has exactly the same flexibility as the solution field in the original simulation. {\color{mygreen}Furthermore, one can ensure by construction that the obtained transformation is a bijective map}. To achieve these properties, we propose to find the transformations by solving an auxiliary problem based on linear elasticity, which is closely related to techniques presented in \cite{Yao1989, Jasak2006AutomaticMethod, Botsch2010}. Even though the auxiliary problem formulated this way is more expensive to solve than the one arising from space deformation techniques, the solution can be substantially accelerated with standard reduced basis (RB) methods, as shown in \cref{subsubsec:auxiliaryproblem} below.

The main idea is as follows: for any geometrical parameterization, the movement of certain parts of the geometry is known, while other parts are fixed. This movement can be prescribed onto a parent mesh and then an auxiliary problem is solved to obtain the transformation map for all geometrical parameters. The example below illustrates this approach.

\paragraph{2D RVE with Elliptical Inclusion}
\begin{figure}[pth]
    \centering
    \includegraphics[width=0.8\textwidth]{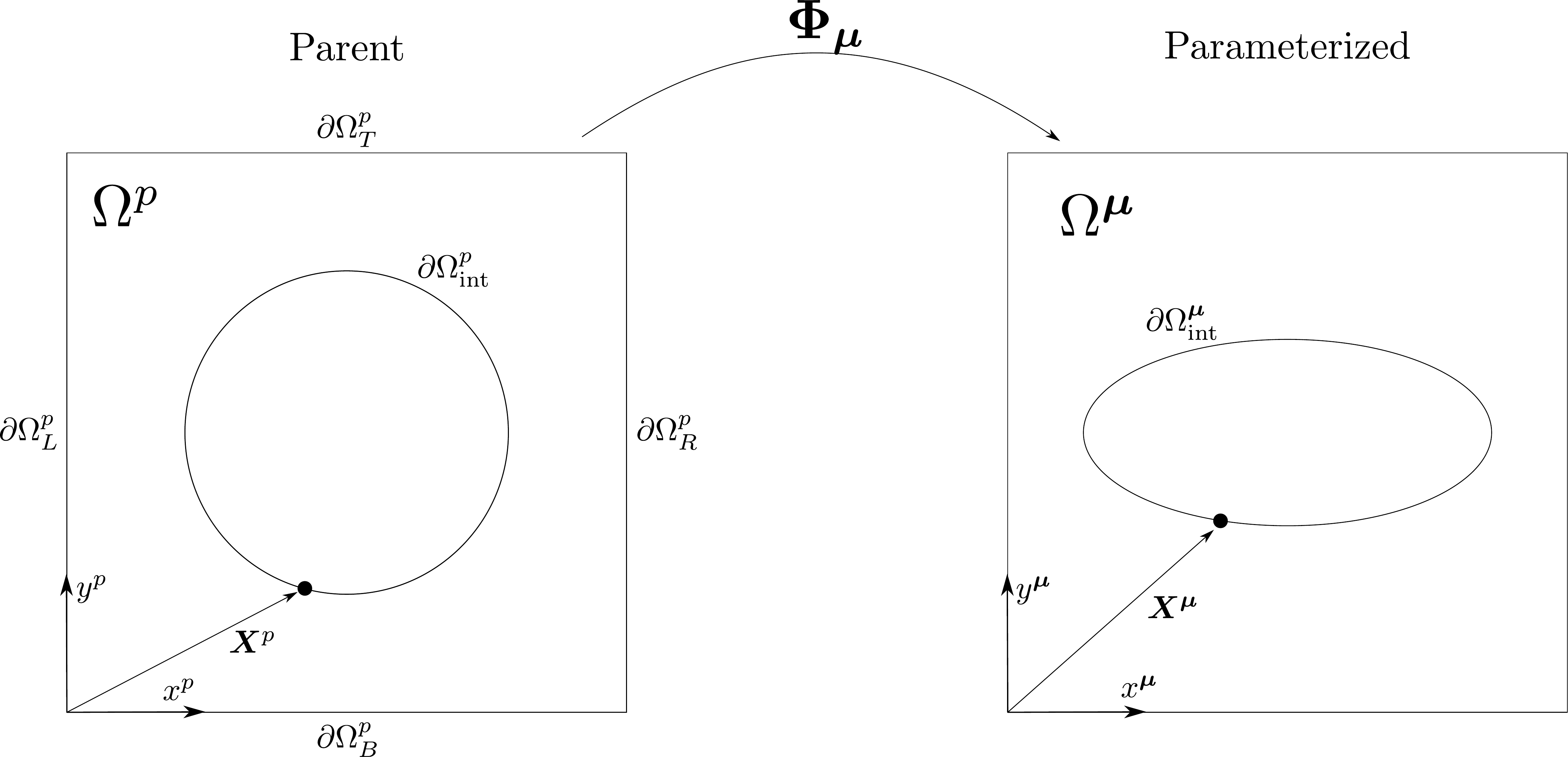}
    \caption{Definition of parent and parameterized domain. The transformation map $\bm{\Phi}_{\bm{\mu}}$ maps from the parent $\Omega^p$ to the parameterized $\Omega^{\bm{\mu}}$ domain. The transformation displacement $\bm{d}$ on the boundaries is fixed. The points on both the parent $\partial \Omega^p_{\text{int}}$ and parameterized interface $\partial \Omega^{\bm{\mu}}_{\text{int}}$ are known and used to prescribe the transformation along the interface. As an example, the indicated point $\bm{X}^p$ in the parent domain is displaced onto the indicated point $\bm{X}^{\bm{\mu}}$ in the parameterized domain.}
    \label{fig:2drve_displacement}
\end{figure}
Consider a RVE defined on $\Omega=[0, 1]^2$ consisting of an elliptical inclusion embedded in a homogeneous matrix. The shape of the inclusion is parameterized, i.e., the major and minor axis can be changed. Assuming a parent domain $\Omega^p$ with a given circular inclusion, it can be deformed into any of the parameterized domains $\Omega^{\bm{\mu}}$, by moving the points on the circular interface into the shape of the elliptical inclusion while keeping the outer boundaries constant, see \cref{fig:2drve_displacement}. By defining this transformation map as 
\begin{align}
\bm{\Phi}_{\bm{\mu}} : \Omega^p \rightarrow \Omega^{\bm{\mu}}, \bm{X}^p \mapsto \bm{X}^{\bm{\mu}} = \bm{\Phi}_{\bm{\mu}}(\bm{X}^p) = \bm{X}^p + \bm{d}(\bm{X}^p; \bm{\mu})
\label{eq:domaintrafo}
\end{align}
with $\bm{d}$ the transformation displacement, this can now be translated into the following linear-elastic auxiliary problem:
\begin{subequations}
 \label{eq:auxiliary_problem}
\begin{alignat}{2}
    \text{Div} \left(\mathbb{C} : \frac{1}{2}\left( \frac{\partial\bm{d}}{\partial\bm{X}^p} + \left(\frac{\partial\bm{d}}{\partial\bm{X}^p}\right)^T\right)\right) &= \bm{0}         &&\text{ in } \Omega^p, \label{eq:auxiliary_pde}\\
    \bm{d} &= \bm{0} &&\text{ on } \partial\Omega^p, \label{eq:auxiliary_bc}\\
    \bm{d} &= \bm{X}^{\bm{\mu}}(\bm{X}^p) - \bm{X}^p &&\text{ on } \partial\Omega^p_{\rm{int}}, \label{eq:auxiliary_prescribed_bc}
\end{alignat}
\end{subequations}
where $\partial\Omega^p=\partial\Omega^p_L \cup \partial\Omega^p_R \cup \partial\Omega^p_B\cup \partial\Omega^p_T$ denotes the union of the left, right, bottom and top RVE boundaries, and $\bm{X}^{\bm{\mu}}(\bm{X}^p)$ is known for all points on the parent interface $\partial\Omega^p_{\rm{int}}$. The elasticity tensor $\mathbb{C}$ is assumed to be constant throughout the whole domain, assumed in the form of Hooke's law, and specified by the Young's modulus and Poisson's ratio $\xi$. Since the problem is cast in a purely geometric manner, the Young's modulus has no influence on the transformation map, as it only changes the magnitude of the stresses that are of no interest here. The Poisson's ratio $\xi$ changes the compressibility of the material, hence affecting the transformation, and its influence on the approximation will be investigated in \cref{sec:results}. The boundary condition in \cref{eq:auxiliary_bc} is chosen such that the parameterized domain $\bm{\Phi}_{\bm{\mu}}(\Omega^p)=[0, 1]^2$ remains the same, i.e., covers the parent domain $\Omega^p$. This also means that the transformation preserves the volume, i.e. $|\Omega^p|=|\bm{\Phi}_{\bm{\mu}}(\Omega^p)|$. Moreover, this way a periodic quantity remains periodic after transformation. Finally, \cref{eq:auxiliary_prescribed_bc} prescribes the transformation displacements along the interface to deform the circle into an ellipse.
\begin{remark}
In principle, one could replace \cref{eq:auxiliary_bc} with periodic boundary conditions and fix the RVE at the corner points to allow more flexible transformations. However, the added complexity did not yield any significant improvement in accuracy in our test problems.
\end{remark}

\subsubsection{Discretization and Numerical Solution of Auxiliary Problem}
\label{subsubsec:auxiliaryproblem}
Discretizing the auxiliary problem in \cref{eq:auxiliary_problem} yields a linear system of equations
\begin{align}
    \mathbf{A}\mathbf{d} = \mathbf{b}(\bm{\mu}), \label{eq:auxiliary_problem_discretized}
\end{align}
where the size of $\mathbf{A}$, $\mathbf{d}$ and $\mathbf{b(\bm{\mu})}$ scale with the full mesh dimensionality $\mathcal{N}$ and \cref{eq:auxiliary_problem_discretized} needs to be solved for many right-hand sides, where each solution is typically computationally expensive. However, there are several ways to accelerate the solution: since the stiffness matrix $\mathbf{A}$ does not depend on the parameters $\bm{\mu}$ and is symmetric positive definite, for moderate values of $\mathcal{N}$, a Cholesky decomposition $\mathbf{A}=\mathbf{L}\mathbf{L}^T$ can be computed once and then a forward and backward substitution can be used to solve \cref{eq:auxiliary_problem_discretized} for any $\mathbf{b}(\bm{\mu})$. For even larger values of $\mathcal{N}$, POD can be used to find a reduced basis for the transformation displacement $\mathbf{d}=\mathbf{V}\hat{\mathbf{d}}$ with $\dim(\hat{\mathbf{d}})=N\ll \mathcal{N}$ and the reduced system is obtained via a Galerkin projection, i.e.,
\begin{align}
    \mathbf{V}^T\mathbf{AV}\hat{\mathbf{d}} &= \mathbf{V}^T \mathbf{b}(\bm{\mu}), \\
    \Rightarrow \hat{\mathbf{A}}\hat{\mathbf{d}} &= \mathbf{V}^T \mathbf{b}(\bm{\mu}),
\end{align}
where the reduced stiffness matrix $\hat{\mathbf{A}}\coloneqq \mathbf{V}^T\mathbf{AV}$ only needs to be computed once. In \cref{sec:results}, it will be shown that $N$ is equal to the number of geometrical parameters $\bm{\mu}$. However, as pointed out in \cite{Stabile2020EfficientMethods}, the term on the right hand side $\mathbf{V}^T \mathbf{b}(\bm{\mu})$ is in general not affinely decomposable, meaning that for each new value of $\bm{\mu}$, a matrix-vector product $\mathbf{V}^T \mathbf{b}(\bm{\mu})$ must be computed which depends linearly on the full problem size $\mathcal{O}(\mathcal{N})$. If this is prohibitive, then a discrete empirical interpolation method \cite{Barrault2004, Chaturantabut2010} could be used to approximate $\mathbf{b}(\bm{\mu})$ and then the complexity of the solution of the auxiliary problem becomes independent of $\mathcal{N}$.

After solving \cref{eq:auxiliary_problem_discretized}, the transformation displacements $\bm{d}(\bm{X}^p; \bm{\mu})$ are specified everywhere and hence the map $\bm{\Phi}_{\bm{\mu}}$ is obtained. To obtain the snapshots on the parent domain $\bm{P}^p(\bm{X}^p)$, the original snapshots $\bm{P}(\bm{X}^{\bm{\mu}})$, computed on the parameterized domain $\Omega^{\bm{\mu}}$, have to be evaluated at the transformed parent coordinates, i.e.,
\begin{align}
    \bm{P}(\bm{X}^{\bm{\mu}}) = \bm{P}(\bm{\Phi}_{\bm{\mu}}(\bm{X}^p)) \eqqcolon \bm{P}^p(\bm{X}^p). \label{eq:transformed_snapshot}
\end{align}

\begin{remark}
\label{remark:mesh_generation}
If the original snapshot $\bm{P}(\bm{X}^{\bm{\mu}})$ is obtained on an independent mesh, one needs to interpolate it onto the transformed parent coordinates $\bm{\Phi}_{\bm{\mu}}(\bm{X}^p)$. However, this interpolation introduces non-physical effects, i.e., periodicity and/or linear momentum balance might not be fulfilled anymore. To ensure a physical transformation, one can use the auxiliary problem in \cref{eq:auxiliary_problem} to generate a simulation mesh for the physical simulation, because that way the snapshots are directly found on the transformed parent coordinates and no interpolation step is needed anymore.
\end{remark}

\subsection{Surrogate Model for Microsimulation}
\label{subsec:surrogate}
In order to replace the microscopic simulation, a surrogate model for the effective stress and stiffness is required, which represents an effective constitutive law. In our previous work \cite{Guo2021b}, we constructed the surrogate model by finding a reduced representation of the stress field via POD,
\begin{align}
    \bm{P}(\bm{X};\bar{\mathbf{U}},\bm{\lambda}) \approx \sum_{n=1}^N \alpha_n(\bar{\mathbf{U}},\bm{\lambda}) \bm{B}_n(\bm{X}), \label{eq:P_approx_old}
\end{align}
where geometrical parameters were not considered. The effective stress was then expressed as
\begin{align}
    \bar{\mathbf{P}}(\bar{\mathbf{U}},\bm{\lambda}) \approx \sum_{n=1}^N \alpha_n(\bar{\mathbf{U}},\bm{\lambda}) \bar{\mathbf{B}}_n, \label{eq:P_eff_old}
\end{align}
where $\bar{\mathbf{B}}_n \coloneqq |\Omega|^{-1}\int_\Omega \bm{B}_n d\bm{X}$. Regression maps $(\bar{\mathbf{U}},\bm{\lambda}) \mapsto \alpha_n$ for the coefficients $\alpha_n(\bar{\mathbf{U}},\bm{\lambda})$ were learned with Gaussian Process Regression (GPR) \cite{Rasmussen2004, Guo2018}. Since each stress field snapshot used for the POD is periodic and fulfills the linear momentum balance, the basis functions $\bm{B}_n(\bm{X})$ fulfill these conditions as well. This is because a sum of periodic functions is still periodic, while the linear momentum balance follows from the linearity of the approximation and the fact that each $\bm{B}_n$ satisfies the linear momentum balance.

For geometrically parameterized domains, however, stress snapshots cannot be used directly. In order to satisfy balance of linear momentum in the parameterized domain, stress snapshots need to be mapped and transformed from the parent to parameterized domain through a corresponding mapping $\bm{\Phi}_{\bm{\mu}}$ and its deformation gradient $\bm{F}_{\bm{\mu}} \coloneqq \dfrac{\partial \bm{\Phi}_{\bm{\mu}}}{\partial \bm{X}^p}$, derived as follows. The weak form of the linear momentum balance on the parameterized domain $\Omega^{\bm{\mu}}$ reads
\begin{align}
    \int_{\Omega^{\bm{\mu}}} \frac{\partial(\delta\bm{u}(\bm{X}^{\bm{\mu}}))}{\partial \bm{X}^{\bm{\mu}}} : \bm{P}(\bm{X}^{\bm{\mu}};\bar{\mathbf{U}},\bm{\lambda},\bm{\mu}) d\bm{X}^{\bm{\mu}} = 0, \label{eq:internalForce_parameterized}
\end{align}
where $\delta\bm{u} \in H^1_0(\Omega^{\bm{\mu}})$ is a test function and $\bm{X}^{\bm{\mu}}$ denotes the coordinates on the parameterized domain. By introducing the transformation from \cref{eq:domaintrafo} and $d\bm{X}^{\bm{\mu}}=\left|\det{\bm{F}_{\bm{\mu}}}\right| d\bm{X}^p$, the left hand side of \cref{eq:internalForce_parameterized} becomes
\begin{align}
\begin{aligned}
    \int_{\Omega^{\bm{\mu}}} \frac{\partial(\delta\bm{u}(\bm{X}^{\bm{\mu}}))}{\partial \bm{X}^{\bm{\mu}}} : \bm{P}(\bm{X}^{\bm{\mu}};\bar{\mathbf{U}},\bm{\lambda},\bm{\mu}) d\bm{X}^{\bm{\mu}}
    &= \int_{\Omega^p} \frac{\partial(\delta\bm{u}^p(\bm{X}^p))}{\partial \bm{X}^p} \bm{F}_{\bm{\mu}}^{-1} : \bm{P}(\bm{\Phi}_{\bm{\mu}}(\bm{X}^p);\bar{\mathbf{U}},\bm{\lambda},\bm{\mu}) \left|\det{\bm{F}_{\bm{\mu}}}\right| d\bm{X}^p \\
    &= \int_{\Omega^p} \frac{\partial(\delta\bm{u}^p(\bm{X}^p))}{\partial \bm{X}^p} : \bm{P}(\bm{\Phi}_{\bm{\mu}}(\bm{X}^p);\bar{\mathbf{U}},\bm{\lambda},\bm{\mu}) \bm{F}_{\bm{\mu}}^{-T} \left|\det{\bm{F}_{\bm{\mu}}}\right| d\bm{X}^p,
\end{aligned}\label{eq:internalForce_parent}
\end{align}
where $\delta\bm{u}^p(\bm{X}^p) = \delta\bm{u}(\bm{\Phi}_{\bm{\mu}}(\bm{X}^p))$ is a test function in the parent configuration. From \cref{eq:internalForce_parent} we can see that if the stress field $\bm{P}(\bm{\Phi}_{\bm{\mu}}(\bm{X}^p);\bar{\mathbf{U}},\bm{\lambda},\bm{\mu})$ were directly approximated using POD, then the linear momentum balance \cref{eq:internalForce_parameterized} on the parameterized domain would not necessarily hold due to the effects of $\bm{F}_{\bm{\mu}}$. However, if these effects are approximated together with the stress as a weighted stress, i.e.,
\begin{align}
    \bm{P}(\bm{\Phi}_{\bm{\mu}}(\bm{X}^p);\bar{\mathbf{U}},\bm{\lambda},\bm{\mu}) \bm{F}_{\bm{\mu}}^{-T} \left|\det{\bm{F}_{\bm{\mu}}}\right| \approx \sum_{n=1}^N \alpha_n(\bar{\mathbf{U}},\bm{\lambda},\bm{\mu}) \bm{B}_n(\bm{X}^p), \label{eq:P_approx_new}
\end{align}
then each basis function $\bm{B}_n$ automatically fulfills the linear momentum balance on any of the parameterized domains, since the basis functions are computed from linear combinations of snapshots that fulfill the balance equation on different domains. This can be shown by inserting \cref{eq:P_approx_new} into \cref{eq:internalForce_parent}:
\begin{align}
    &\int_{\Omega^p} \frac{\partial(\delta\bm{u}^p(\bm{X}^p))}{\partial \bm{X}^p} : \bm{P}(\bm{\Phi}_{\bm{\mu}}(\bm{X}^p);\bar{\mathbf{U}},\bm{\lambda},\bm{\mu}) \bm{F}_{\bm{\mu}}^{-T} \left|\det{\bm{F}_{\bm{\mu}}}\right| d\bm{X}^p \\
    \approx & \sum_{n=1}^N \alpha_n(\bar{\mathbf{U}},\bm{\lambda},\bm{\mu}) \int_{\Omega^p} \frac{\partial(\delta\bm{u}^p(\bm{X}^p))}{\partial \bm{X}^p} : \bm{B}_n(\bm{X}^p) d\bm{X}^p. \label{eq:P_approx_linearcomb}
\end{align}
Since the basis functions are linear combinations of $N_s$ training snapshots, i.e.,
\begin{align}
    \bm{B}_n(\bm{X}^p) = \sum_{l=1}^{N_s} a_{nl} \bm{P}^{(l)} \bm{F}_{\bm{\mu}^{(l)}}^{-T} \left|\det{\bm{F}_{\bm{\mu}^{(l)}}}\right|, \label{eq:basisfunc_linearcomb}
\end{align}
where $a_{nl}$ are the corresponding coefficients determined by POD and the superscript $(l)$ denotes the $l$-th weighted stress snapshot, obtained for parameters $(\bar{\mathbf{U}}^{(l)},\bm{\lambda}^{(l)},\bm{\mu}^{(l)})$, inserting \cref{eq:basisfunc_linearcomb} into \cref{eq:P_approx_linearcomb} yields
\begin{align}
    &\sum_{n=1}^N \alpha_n(\bar{\mathbf{U}},\bm{\lambda},\bm{\mu}) \int_{\Omega^p} \frac{\partial(\delta\bm{u}^p(\bm{X}^p))}{\partial \bm{X}^p} : \bm{B}_n(\bm{X}^p) d\bm{X}^p \\
    = &\sum_{n=1}^N \alpha_n(\bar{\mathbf{U}},\bm{\lambda},\bm{\mu}) \sum_{l=1}^{N_s} a_{nl} \underbrace{\int_{\Omega^p} \frac{\partial(\delta\bm{u}^p(\bm{X}^p))}{\partial \bm{X}^p} : \bm{P}^{(l)} \bm{F}_{\bm{\mu}^{(l)}}^{-T} \left|\det{\bm{F}_{\bm{\mu}^{(l)}}}\right| d\bm{X}^p}_{=0},
\end{align}
where each of the integrals equals $0$, as every training snapshot fulfills \cref{eq:internalForce_parameterized}. Therefore, any predicted weighted stress will always fulfill \cref{eq:internalForce_parameterized} for any parameterized domains $\Omega^{\bm{\mu}}$. To obtain the stress field from the predicted weighted stress, one therefore has to multiply it with $\bm{F}_{\bm{\mu}}^T \left|\det{\bm{F}_{\bm{\mu}}}\right|^{-1}$. Furthermore, periodicity follows from the periodicity of the stress field and the transformation map.

\subsubsection{Effective Stress} 
The effective (average) stress on the parameterized domain $\Omega^{\bm{\mu}}$ is computed as
\begin{align}
    \bar{\mathbf{P}}(\bar{\mathbf{U}},\bm{\lambda},\bm{\mu}) &= |\Omega^{\bm{\mu}}|^{-1}\int_{\Omega^{\bm{\mu}}} \bm{P}(\bm{X}^{\bm{\mu}};\bar{\mathbf{U}},\bm{\lambda},\bm{\mu}) d\bm{X}^{\bm{\mu}}, \\ 
    \intertext{where $|\Omega^{\bm{\mu}}|$ denotes the volume of the RVE. Note that $|\Omega^{\bm{\mu}}|=|\Omega^p|$ for all $\bm{\mu}$ due to the volume-preserving transformation map. Pulling the integral back into the parent domain gives}
    \bar{\mathbf{P}}(\bar{\mathbf{U}},\bm{\lambda},\bm{\mu}) &= |\Omega^p|^{-1}\int_{\Omega^p} \bm{P}(\bm{\Phi}_{\bm{\mu}}(\bm{X}^p);\bar{\mathbf{U}},\bm{\lambda},\bm{\mu})\left| \det{\bm{F}_{\bm{\mu}}} \right| d\bm{X}^p, \\ 
    \intertext{and using the approximation for $\bm{P}(\bm{\Phi}_{\bm{\mu}}(\bm{X}^p);\bar{\mathbf{U}},\bm{\lambda},\bm{\mu})$ from \cref{eq:P_approx_new} then yields}
    \bar{\mathbf{P}}(\bar{\mathbf{U}},\bm{\lambda},\bm{\mu}) &\approx |\Omega^p|^{-1}\int_{\Omega^p} \left(\sum_{n=1}^N \alpha_n(\bar{\mathbf{U}},\bm{\lambda},\bm{\mu}) \bm{B}_n(\bm{X}^p)\right) \bm{F}_{\bm{\mu}}^T \frac{1}{|\det{\bm{F}_{\bm{\mu}}}|} |\det{\bm{F}_{\bm{\mu}}}| d\bm{X}^p \\
    &= |\Omega^p|^{-1}\int_{\Omega^p} \left(\sum_{n=1}^N \alpha_n(\bar{\mathbf{U}},\bm{\lambda},\bm{\mu}) \bm{B}_n(\bm{X}^p)\right) \bm{F}_{\bm{\mu}}^T d\bm{X}^p. \\
    \intertext{Since $\alpha_n$ does not depend on $\bm{X}^p$, it can be taken out of the integral, yielding}
    \bar{\mathbf{P}}(\bar{\mathbf{U}},\bm{\lambda},\bm{\mu}) &= |\Omega^p|^{-1}\sum_{n=1}^N \alpha_n(\bar{\mathbf{U}},\bm{\lambda},\bm{\mu}) \int_{\Omega^p} \bm{B}_n(\bm{X}^p) \bm{F}_{\bm{\mu}}^T d\bm{X}^p.
    \label{eq:effectivestress_1}
\end{align}
In order to have a rapid online phase, the integrals must be precomputed in the offline stage. Due to the specific form of the basis functions $\bm{B}_n$, it can be shown that the integral in \cref{eq:effectivestress_1} is invariant with respect to $\bm{F}_{\bm{\mu}}^T$, i.e.,
\begin{align}
    \int_{\Omega^p} \bm{B}_n(\bm{X}^p) \bm{F}_{\bm{\mu}}^T d\bm{X}^p = \int_{\Omega^p} \bm{B}_n(\bm{X}^p) d\bm{X}^p, \label{eq:invariance}
\end{align}
which can be precomputed since $\bm{B}_n$ is known; the proof of the identity in \cref{eq:invariance} is provided in \ref{sec:appendix_A}. With
\begin{align}
    \bar{\mathbf{B}}_n &\coloneqq |\Omega^p|^{-1} \int_{\Omega^p} \bm{B}_n(\bm{X}^p) d\bm{X}^p, \label{eq:constant_1}
\end{align}
the expression for the effective stress in \cref{eq:effectivestress_1} becomes
\begin{align}
    \bar{\mathbf{P}}(\bar{\mathbf{U}},\bm{\lambda},\bm{\mu}) = \sum_{n=1}^N \alpha_n(\bar{\mathbf{U}},\bm{\lambda},\bm{\mu}) \bar{\mathbf{B}}_n,
\label{eq:effective_stress_2}
\end{align}
which remarkably has exactly the same form as \cref{eq:P_eff_old}, with the additional geometrical parameter dependence in the coefficients $\alpha_n$. Hence, the prediction of the effective stresses is completely independent of the auxiliary problem, which only needs to be computed to recover the stress field. 

\subsubsection{Effective Stiffness and Sensitivities}
The effective constitutive stiffness and sensitivities with respect to material and geometrical parameters are given as
\begin{align}
    \bar{\bm{A}} = \frac{\partial \bar{\mathbf{P}}}{\partial \bar{\mathbf{U}}} &= \sum_{n=1}^N \bar{\mathbf{B}}_n \otimes \frac{\partial\alpha_n}{\partial \bar{\mathbf{U}}}, \label{eq:effective_sensitivities_1} \\
    \frac{\partial \bar{\mathbf{P}}}{\partial\bm{\lambda}} &= \sum_{n=1}^N \bar{\mathbf{B}}_n \otimes \frac{\partial\alpha_n}{\partial \bm{\lambda}},\label{eq:effective_sensitivities_2} \\
    \frac{\partial \bar{\mathbf{P}}}{\partial \bm{\mu}} &= \sum_{n=1}^N \bar{\mathbf{B}}_n \otimes \frac{\partial\alpha_n}{\partial \bm{\mu}}.\label{eq:effective_sensitivities_3}
\end{align}

Given coefficients $\alpha_n$ and their derivatives, the effective stress, stiffness and sensitivities can therefore be directly obtained. {\color{mygreen}These coefficients can be approximated with different regression models, such as radial basis functions \cite{Guenot2013}, GPR \cite{Guo2018, Kast2020} or neural networks \cite{Hesthaven2018}. A comparison in this context was carried out in \cite{Berzins2020}, revealing that GPRs can be as accurate as NNs, while being easier to train. Moreover, a GP model also returns an uncertainty measure for every prediction, which can be utilized to construct active learning schemes, see, e.g., \cite{Guo2018, Kast2020, Yang2020}. Due to these advantages, in this work, we learn regression models for $\alpha_n(\bar{\mathbf{U}},\bm{\lambda},\bm{\mu})$ with GPRs. Note that in this work, we do not use the uncertainty measure to adaptively enrich the training set, as in \cite{Guo2018, Kast2020, Yang2020}; this will be explored in future work.} A broad overview and theory on GPRs can be found in \cite{Rasmussen2004}, and is omitted here for brevity. For all GPR models, the Python library GPy \cite{gpy2014} with automatic relevance determination squared exponential kernels has been used. {\color{mygreen}The optimal hyperparameters of the kernels are determined by a maximum likelihood estimation, as presented in \cite{Rasmussen2004, Guo2018}, with the L-BFGS-B algorithm. In the numerical tests, we did not encounter any problems during hyperparameter tuning and the convergence of the L-BFGS-B algorithm was smooth.}

\subsection{Offline--Online Decomposition}
For convenience, the full offline--online decomposition is summarized in \cref{alg:podgpr}.

\begin{algorithm}
\caption{Offline--online decomposition of the proposed PODGPR framework with microstructures parameterized with external loading $\bar{\mathbf{U}}$, microstructural material parameters $\bm{\lambda}$ and geometrical features $\bm{\mu}$.}\label{alg:podgpr}
\begin{algorithmic}[1]
\Require
\State Define a parent domain $\Omega^p$ and its finite element discretization.
\State Generate parameter samples $\{\bar{\mathbf{U}}^i,\bm{\lambda}^i,\bm{\mu}^i\}_{i=1}^{N_s}$ from a random distribution.
\State For each different set of geometrical parameters $\bm{\mu}^i$, solve the auxiliary problem in \cref{eq:auxiliary_problem} to obtain the transformation map $\bm{\Phi}_{\bm{\mu}^i}$.
\State Use the transformation map $\bm{\Phi}_{\bm{\mu}^i}$ to generate simulation meshes for each parameter sample $\bm{\mu}^i$ and then run full simulations to obtain stress snapshots $\bm{P}^i(\bm{\Phi}_{\bm{\mu}^i}(\bm{X}^p);\bar{\mathbf{U}}^i,\bm{\lambda}^i,\bm{\mu}^i)$.
\State Compute POD of the weighted stress snapshots $\bm{P}^i(\bm{\Phi}_{\bm{\mu}^i}(\bm{X}^p);\bar{\mathbf{U}}^i,\bm{\lambda}^i,\bm{\mu}^i) \bm{F}_{\bm{\mu}^i}^{-T}|\det \bm{F}_{\bm{\mu}^i}|$ on the parent domain, cf. \cref{eq:P_approx_new}.
\State Project weighted stress snapshots onto POD basis and learn GPRs for the POD coefficients.
\State Compute $\bar{\mathbf{B}}_n$ using \cref{eq:constant_1}.
\Ensure
\setcounter{ALG@line}{0}
\State Given a new parameter set $(\bar{\mathbf{U}}^*,\bm{\lambda}^*,\bm{\mu}^*)$, evaluate $\alpha_n$ using GPRs.
\State Compute effective stress using \cref{eq:effective_stress_2} and effective sensitivities with \cref{eq:effective_sensitivities_1,eq:effective_sensitivities_2,eq:effective_sensitivities_3}
\end{algorithmic}
\end{algorithm}

%%%%%%%% EXAMPLES %%%%%%%%%%%
\section{Example Problems}
\label{sec:results}
In this section, the proposed method, in the following referred to as PODGPR, is first applied onto two geometrically parameterized microstructures to showcase the generality of the approach and its accuracy. Then, a two-scale Cook's membrane problem is shown to illustrate the speed-up and its potential applications. All examples are defined in 2D under plane strain conditions, although the proposed methodology can easily be extended to 3D microstructures.

For convenience, all quantities in all examples are normalized, and dimensionless quantities are considered. At the same time, all RVEs are assumed to be of size $[0,1]^2$.

The nonlinear physical simulations are solved within the Finite Element framework MOOSE \cite{permann2020moose} and the linear auxiliary problems are solved with an in-house code\footnote{The implementation can be found on https://github.com/theronguo/auxiliary-problem.} written in Python. As mentioned in Remark~\ref{remark:mesh_generation}, to obtain physically consistent transformations without the need for interpolation, we use the auxiliary problem to generate simulation meshes. To quantify the quality of the approximation, the following two error measures are introduced:
\begin{enumerate}
    \item Relative error of stress field
    \begin{align}
        \mathcal{E}_{\bm{P}} = \frac{||\bm{P}^{\rm{truth}} - \bm{P}^{\rm{surrogate}}||_{L^2(\Omega^{\bm{\mu}})}}{||\bm{P}^{\rm{truth}}||_{L^2(\Omega^{\bm{\mu}})}},
    \end{align}
    \item Relative error of effective stress
    \begin{align}
        \mathcal{E}_{\bar{\mathbf{P}}} = \frac{||\bar{\mathbf{P}}^{\rm{truth}} - \bar{\mathbf{P}}^{\rm{surrogate}}||_F}{||\bar{\mathbf{P}}^{\rm{truth}}||_F},
    \end{align}
\end{enumerate}
where $(\bullet)^{\text{truth}}$ and $(\bullet)^{\text{surrogate}}$ indicate the full and approximate solution, and $||(\bullet)||_{L^2(\Omega^{\bm{\mu}})}$ and $||(\bullet)||_F$ denote the $L^2(\Omega^{\bm{\mu}})$ and the Frobenius norm, with $\Omega^{\bm{\mu}}$ the parameterized domain. The average errors for given testing datasets are defined as:
\begin{align}
    \bar{\mathcal{E}}_{\bm{P}} = \frac{1}{N_{\rm{test}}} \sum_{n=1}^{N_{\rm{test}}} \mathcal{E}_{\bm{P}}^n, \qquad \bar{\mathcal{E}}_{\bar{\mathbf{P}}} = \frac{1}{N_{\rm{test}}} \sum_{n=1}^{N_{\rm{test}}} \mathcal{E}_{\bar{\mathbf{P}}}^n,
\end{align}
where $N_{\rm{test}}$ is the number of testing snapshots and $\mathcal{E}_{\bm{P}}^n$ and $\mathcal{E}_{\bar{\mathbf{P}}}^n$ correspond to the relative errors of the $n$-th snapshot.

For comparison, we also trained several deep feed-forward neural networks for the effective stress using the same data for each example, similar to \cite{Guo2021b}. All considered neural networks have as many inputs as the number of parameters, 4 outputs for each stress component and 2 hidden layers each with $N_n$ neurons. We trained four architectures with $N_n\in\{50, 100, 200, 300\}$. These four architectures will be referred to as NN1, NN2, NN3 and NN4. ELU activation functions are applied on each layer apart from the last layer. For the optimization, the mean squared error loss function is chosen and optimized with the \textit{Adam} optimizer \cite{Kingma2014} with a learning rate of $1\times10^{-4}$ and a batch size of 32 for 10000 epochs. The training is performed with the Python package PyTorch \cite{NEURIPS2019_9015}.

\subsection{Composite Microstructure With an Elliptical Fiber}
\label{subsec:problem_1}
\begin{figure}[ht]
     \centering
     \begin{subfigure}[c]{0.32\textwidth}
         \centering
         \includegraphics[width=\textwidth]{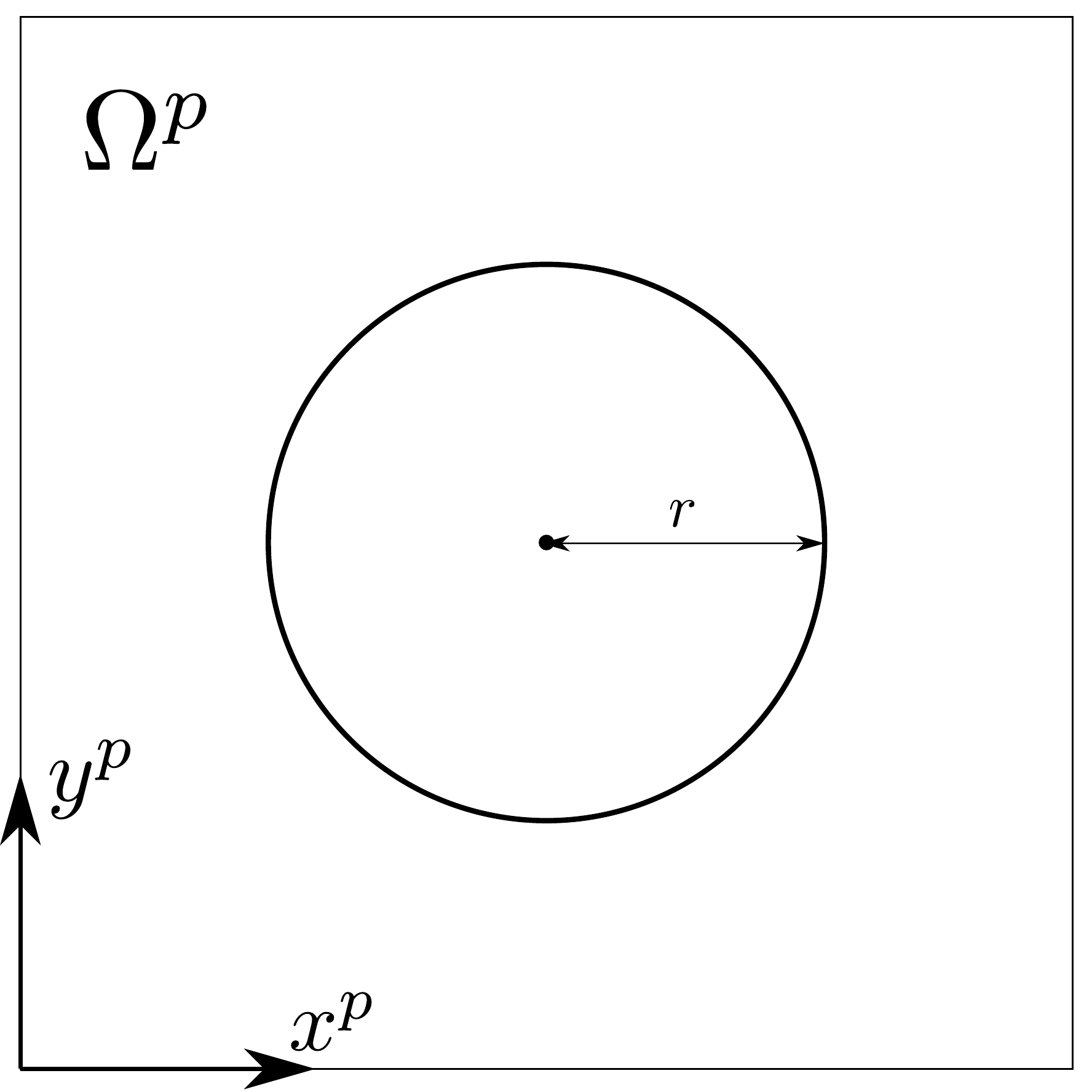}
         \caption{}
         \label{fig:problem_1_ref1}
     \end{subfigure}
     \hfill
     \begin{subfigure}[c]{0.31\textwidth}
         \centering
         \includegraphics[width=\textwidth]{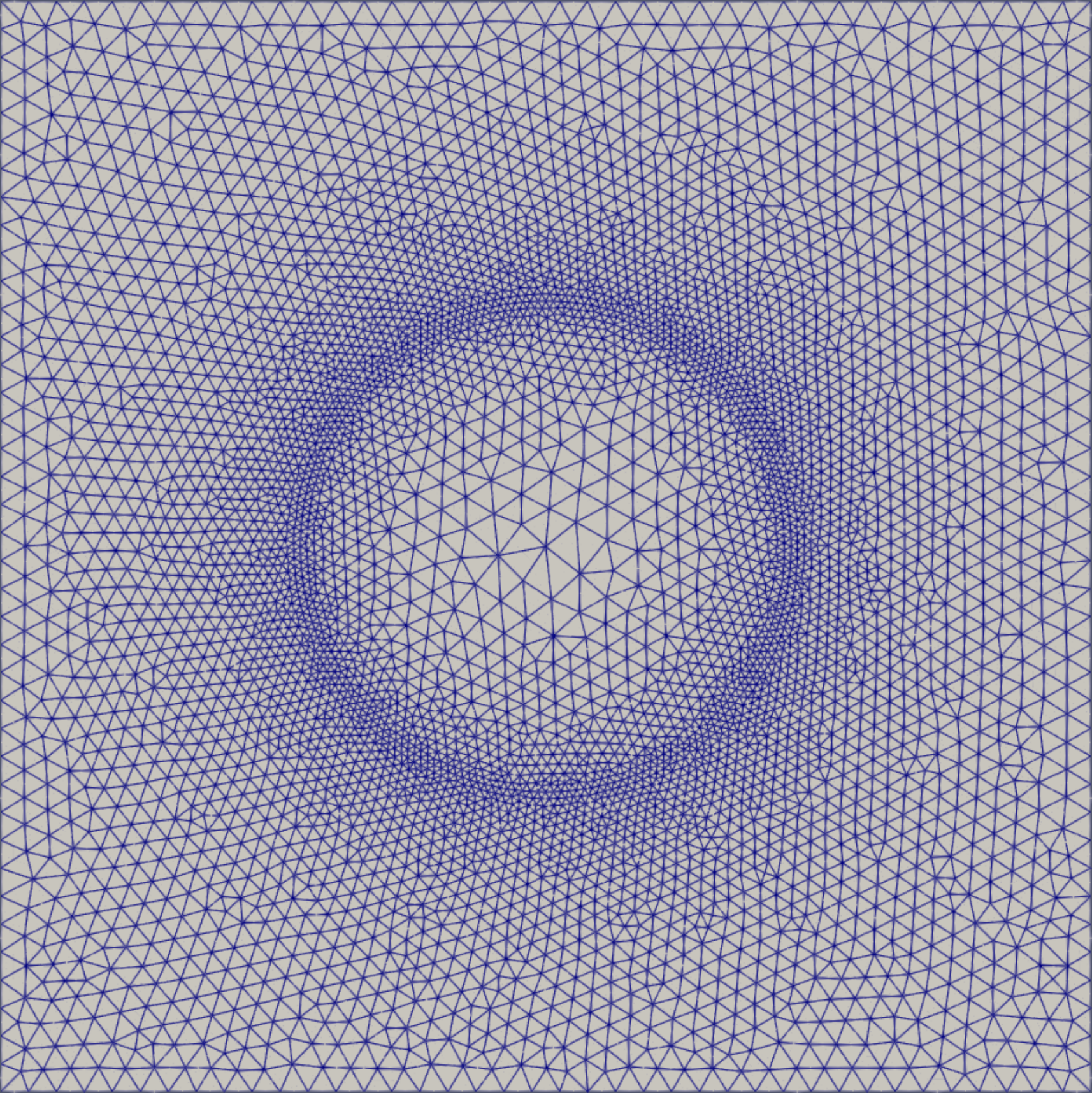}
         \caption{}
         \label{fig:problem_1_ref2}
     \end{subfigure}
     \hfill
     \begin{subfigure}[c]{0.31\textwidth}
         \centering
         \includegraphics[width=\textwidth]{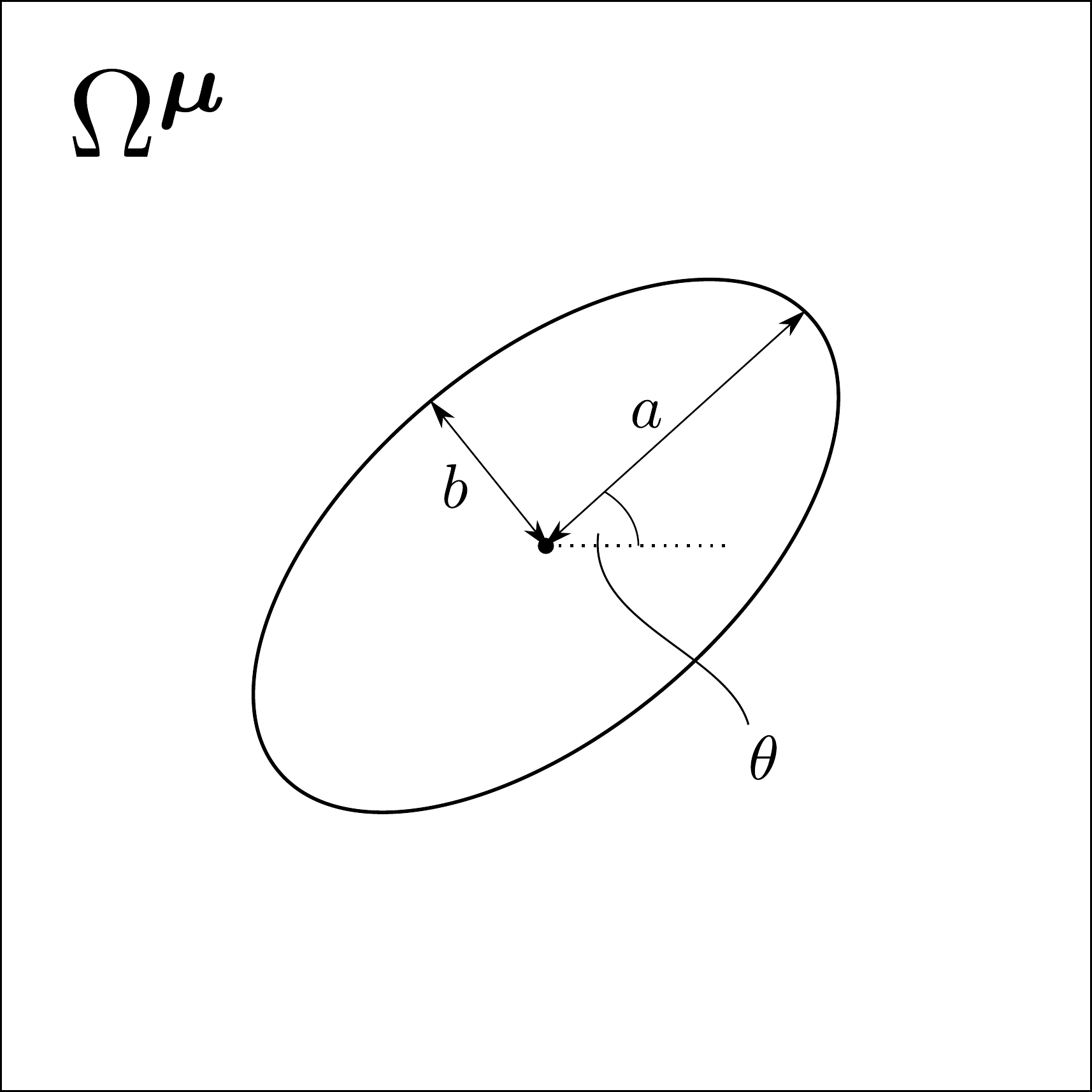}
         \caption{}
         \label{fig:problem_1_sketch}
     \end{subfigure}
        \caption{Parent and parameterized domain. (a) The chosen parent domain with a circular interface with fixed radius $a=b=r=0.225$ with (b) its corresponding mesh, consisting of a total of 20769 nodes and 10665 6-node triangular elements. (c) The parameterized domain, characterized by semi-major axis $a$, semi-minor axis $b$ and angle $\theta$.}
        \label{fig:three graphs}
\end{figure}

In this example, a composite structure, consisting of a soft matrix and an elliptical stiff fiber around the center of the domain $\mathbf{X}_c=[0.5, 0.5]^T$, is considered. Three geometrical parameters that parameterize the fiber shape, the semi-major axis $a$, semi-minor axis $b$ and a rotation angle $\theta$, are considered, see \cref{fig:problem_1_sketch}. Together with the three loading directions $\bar{U}_{xx}$, $\bar{U}_{yy}$ and $\bar{U}_{xy}$, this problem has 6 parameters. For the matrix material a Neo-Hookean material model with $C_1=1$ and $D_1=1$ is chosen, while for the fiber material a Neo-Hookean material model with $C_1=100$ and $D_1=100$ is assumed, corresponding to a stiffness that is 100 times higher than the matrix material. Both materials have Poisson's ratio $0.25$. The considered parameter ranges are given in \cref{table:problem_1_params}. {\color{mygreen}Lower and upper bounds for $\bar{\mathbf{U}}$ are chosen such that the solution of the microstructural problem converges for all parameter configurations. For some geometrical parameters, a few elements of the parent mesh might become highly distorted after applying the geometrical transformation. Together with the high contrast of material stiffness of both materials, larger magnitudes than 0.15 of the components $\bar{\mathbf{U}} - \mathbf{I}$ lead to convergence issues. If larger magnitudes have to be considered, a different parent domain could be employed or a smaller parameter space of the geometrical parameters could be chosen.}

\begin{table}[ht]
\centering
\caption{Example 1 - 6 parameters with corresponding ranges. The parameters $a$, $b$ and $\theta$ are the geometrical parameters describing the interface while $\bar{U}_{xx}$, $\bar{U}_{yy}$ and $\bar{U}_{xy}$ are external loading parameters.}
\begin{tabular}{c|c|c|c|c|c}
 $a$ &  $b$ &  $\theta$ & $\bar{U}_{xx}$ &  $\bar{U}_{yy}$ &  $\bar{U}_{xy}$ \\
 \hline
 $[0.1, 0.35]$ & $[0.1, 0.35]$ & $[-\pi/2,\pi/2]$ & $[0.85, 1.15]$ & $[0.85, 1.15]$ & $[-0.15, 0.15]$
\end{tabular}
\label{table:problem_1_params}
\end{table}

\subsubsection{Setup of the Auxiliary Problem}
For the parent domain, a domain with a circular inclusion with radius $r=0.225$ is chosen, see \cref{fig:problem_1_ref1}. The corresponding mesh, consisting of 20769 nodes and 10665 6-node triangular elements, is shown in \cref{fig:problem_1_ref2}. The transformation of the circular interface into the elliptical interface can be given as:
\begin{align}
    \mathbf{X}^{\bm{\mu}}(\mathbf{X}^p) = \begin{bmatrix} \cos{\theta} & -\sin{\theta} \\ \sin{\theta} & \cos{\theta} \end{bmatrix} \begin{bmatrix} a/r & 0 \\ 0 & b/r \end{bmatrix} \begin{pmatrix} \tilde{r}(\mathbf{X}^p - \mathbf{X}_c) \cos(\tilde{\theta}(\mathbf{X}^p - \mathbf{X}_c) - \theta) \\ \tilde{r}(\mathbf{X}^p - \mathbf{X}_c) \sin(\tilde{\theta}(\mathbf{X}^p - \mathbf{X}_c) - \theta) \end{pmatrix} + \mathbf{X}_c,
\end{align}
where
\begin{align}
    \tilde{r}(\mathbf{X}^p) = \sqrt{\mathbf{X}^{p^T} \mathbf{X}^p} \\ 
    \tilde{\theta}(\mathbf{X}^p) = \text{arctan2}(x^p, y^p),
\end{align}
with $\mathbf{X}^p = [x^p, y^p]^T$ a column vector of each of the nodal positions located at the interface. The arctan2-function is an extension of the arctan-function and is defined as,
\begin{align}
    (x,y) \mapsto 
    \begin{cases} 
        \arctan(y/x) & x>0, \\
        \arctan(y/x)+\pi & x<0, y\geq0, \\
        \arctan(y/x)-\pi & x<0, y<0, \\
        \pi/2 & x=0, y>0, \\
        -\pi/2 & x=0, y<0, \\
        \text{undefined} & x=0, y=0.
    \end{cases}
\end{align}
The Poisson's ratio $\xi$ for the auxiliary problem is first chosen to be 0.3. In \cref{subsubsec:influence_poisson} below, its influence on the accuracy of obtained results is discussed.

\subsubsection{Data Generation}
In total 1000 training snapshots are generated, of which $2^6=64$ snapshots are at the corners of the 6-dimensional parameter space and the rest are sampled from a Sobol sequence \cite{sobol1967distribution}. For testing, another 500 snapshots are generated from a random uniform distribution.

\subsubsection{Results}
\begin{figure}[ht]
     \centering
     \begin{subfigure}[b]{0.35\textwidth}
         \centering
         \includegraphics[width=\textwidth]{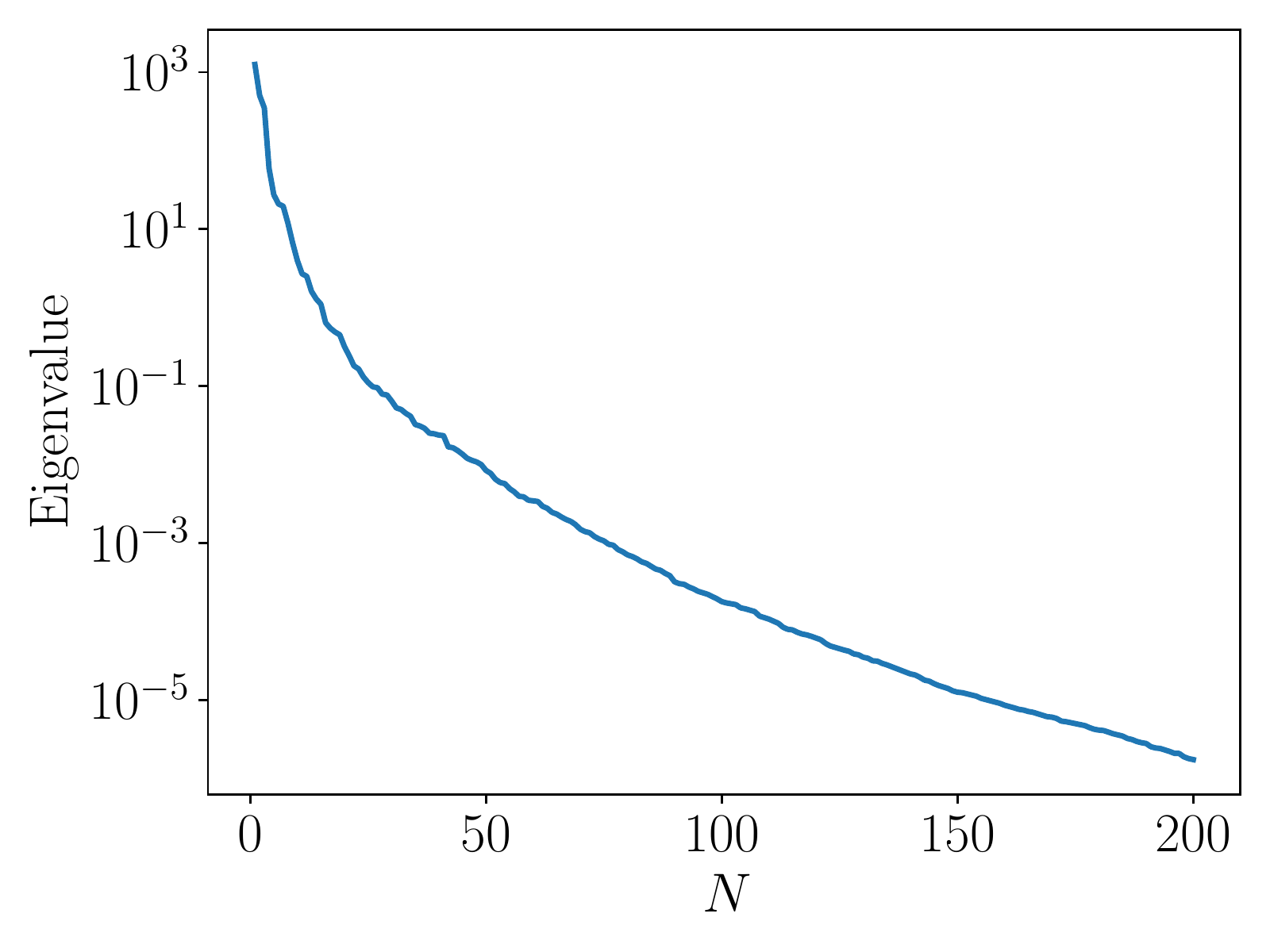}
         \caption{Weighted stress}
         \label{fig:problem_1_eig_1}
     \end{subfigure}
     \qquad
     \begin{subfigure}[b]{0.35\textwidth}
         \centering
         \includegraphics[width=\textwidth]{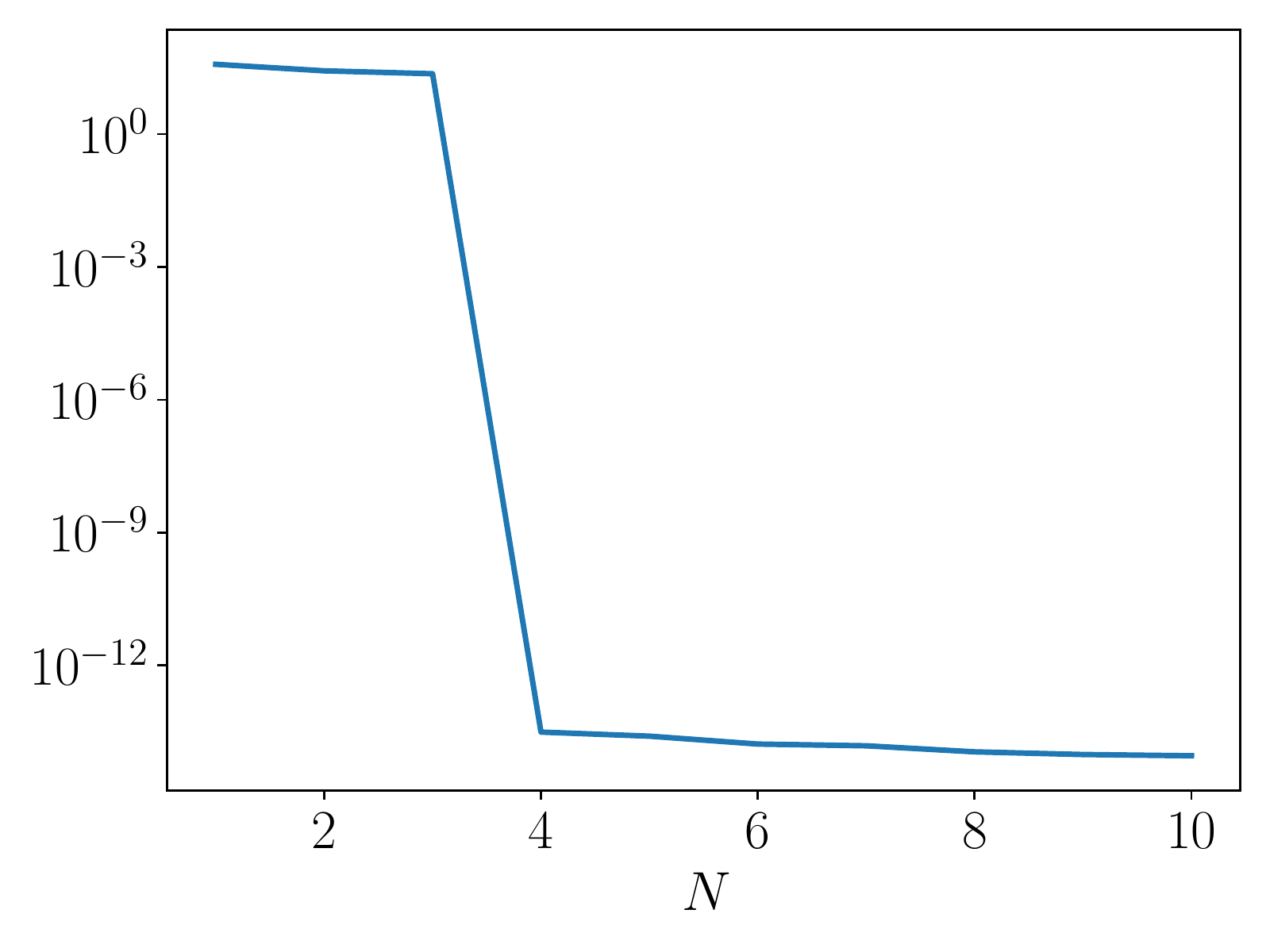}
         \caption{Transformation displacement}
         \label{fig:problem_1_eig_2}
     \end{subfigure}
    \caption{Eigenvalues of the correlation matrix for (a) weighted stress and (b) transformation displacement. For the weighted stress the eigenvalues decay exponentially, while for the transformation displacement only the first three modes are nonzero. This means that the transformation displacement can be represented with just three basis functions.}
    \label{fig:problem_1_eigs}
\end{figure}

\paragraph{POD of Weighted Stress and Transformation Displacement}
The eigenvalues of the correlation matrix for the weighted stress are given in \cref{fig:problem_1_eig_1}. It can be observed that the eigenvalues decay exponentially, indicating a good reducibility. To show that the auxiliary problem in \cref{eq:auxiliary_problem_discretized} can be reduced drastically, the eigenvalues of the correlation matrix for the transformation displacement are also shown in \cref{fig:problem_1_eig_2}. All but three eigenvalues are essentially zero, showing that the auxiliary problem can be solved with three basis functions.

\begin{figure}[ht]
     \centering
     \begin{subfigure}[b]{0.35\textwidth}
         \centering
         \includegraphics[width=\textwidth]{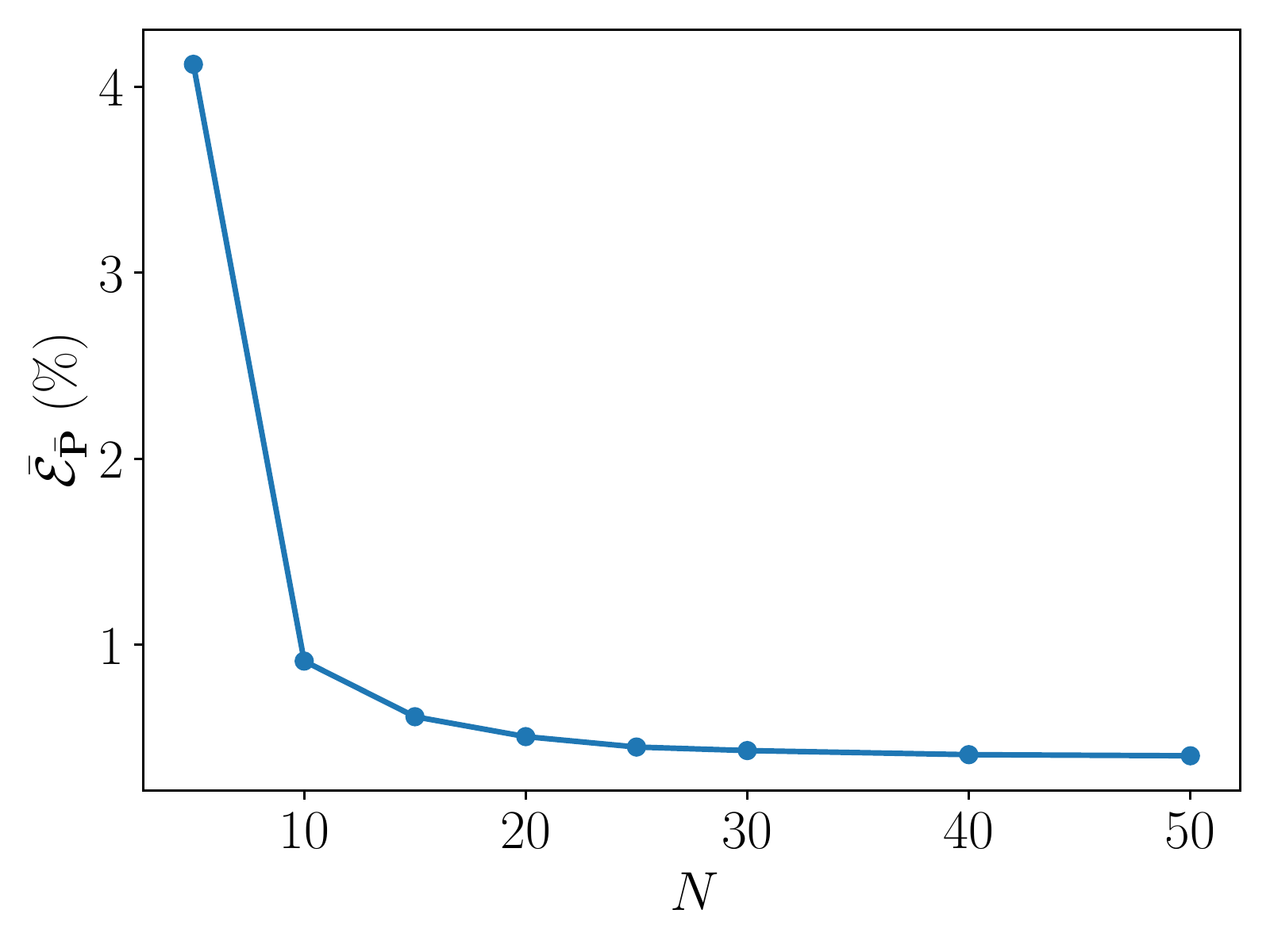}
         \caption{}
         \label{fig:problem_1_errN}
     \end{subfigure}
     \qquad
     \begin{subfigure}[b]{0.35\textwidth}
         \centering
         \includegraphics[width=\textwidth]{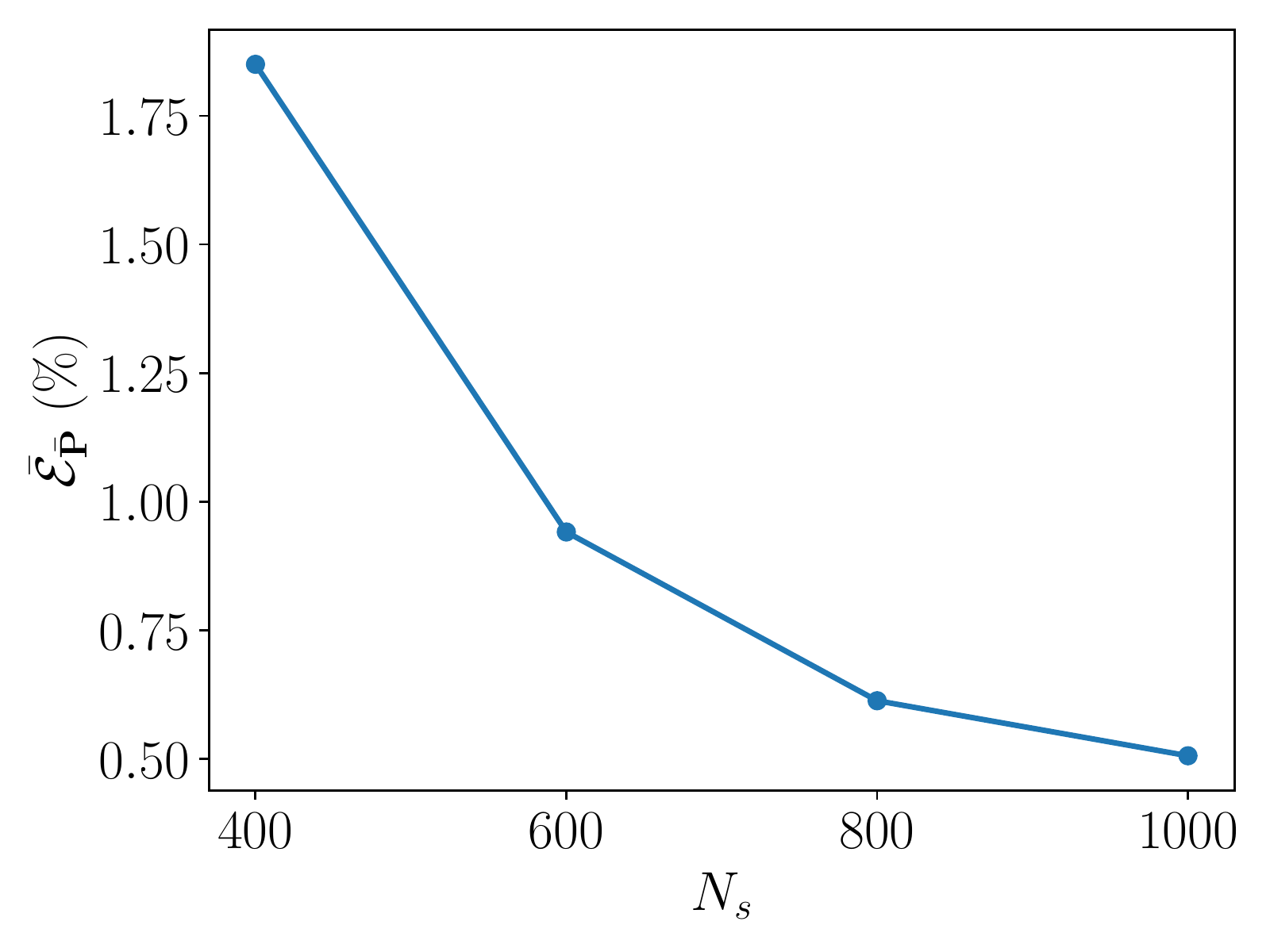}
         \caption{}
         \label{fig:problem_1_errNs}
     \end{subfigure}
    \caption{(a) Average error of the effective stress over number of basis functions with $N_s=1000$ training snapshots. The error curve decays rapidly for the first few basis functions and then flattens out. (b) Average error of the effective stress over number of training snapshots used for $N=20$ basis functions. The error increases drastically when fewer training snapshots are used.}
    \label{fig:problem_1_errs}
\end{figure}

\paragraph{Approximation Errors}
The average approximation error of the effective stress on the 500 testing snapshots of PODGPR for different numbers of basis functions $N$ and training snapshots $N_s$ is shown in \cref{fig:problem_1_errs}. All $N_s$ training snapshots are used for both the POD and GPR. In \cref{fig:problem_1_errN}, the error decays rapidly for the first few basis functions. For $N=20$ an error of roughly 0.5\% is reached. However, taking into account more basis functions barely improves the performance since the coefficients get increasingly more oscillatory with increasing number and hence more difficult to approximate with a GPR model, see \cite{Guo2021b}. From \cref{fig:problem_1_errNs}, we see that a higher number of snapshots is crucial for the accuracy of PODGPR. Data shown corresponds to $N=20$ basis functions. For small datasets the error increases exponentially, indicating a poor approximation of the first 20 POD coefficients.

In \cref{table:problem_1_errors}, the approximation quality of PODGPR with $N=20$ basis functions is compared with the four neural networks. The best approximation error (defined as the error of projecting the truth solution onto the reduced basis) with $N=20$ basis functions is also given. PODGPR approximates the effective stress better than all the NN models at least by a factor of 2. Moreover, it is nearly able to reach the error of the best approximation in both error measures, indicating that with $N_s=1000$ training snapshots the first 20 POD coefficients can be well captured. The error in the stress field is less than 0.1\%, while for the effective stress the error is 0.5\%. Furthermore, by comparing the results with \cref{fig:problem_1_errNs}, it can be seen that the best results obtained by the neural networks are reached by PODGPR with only $N_s=600$ snapshots, showing that PODGPR is more data efficient than the neural networks in this case.

\begin{table}[pth]
\centering
\caption{Approximation errors for different methods. The trained surrogate models are tested on 500 testing snapshots. PODGPR outperforms all NN models by a factor 2.}
\begin{tabular}{c|c|c|c|c|c|c}
       & Best approximation & PODGPR & NN1 & NN2 & NN3 & NN4\\ \hline
$\bar{\mathcal{E}}_{\bm{P}}$ & $8.17\times 10^{-4}$ & $9.49\times 10^{-4}$ &   n.a.  & n.a. & n.a. & n.a. \\
$\bar{\mathcal{E}}_{\bar{\mathbf{P}}}$ & $2.74\times 10^{-3}$ &  $5.06\times 10^{-3}$ &  $2.66\times 10^{-2}$   &   $1.53\times 10^{-2}$  & $1.02\times 10^{-2}$ & $9.8\times 10^{-3}$
\end{tabular}
\label{table:problem_1_errors}
\end{table}

\subsubsection{Influence of Poisson's Ratio on the Auxiliary Problem}
\label{subsubsec:influence_poisson}

\begin{figure}[ht]
     \centering
     \begin{subfigure}[c]{0.23\textwidth}
         \centering
         \includegraphics[width=\textwidth]{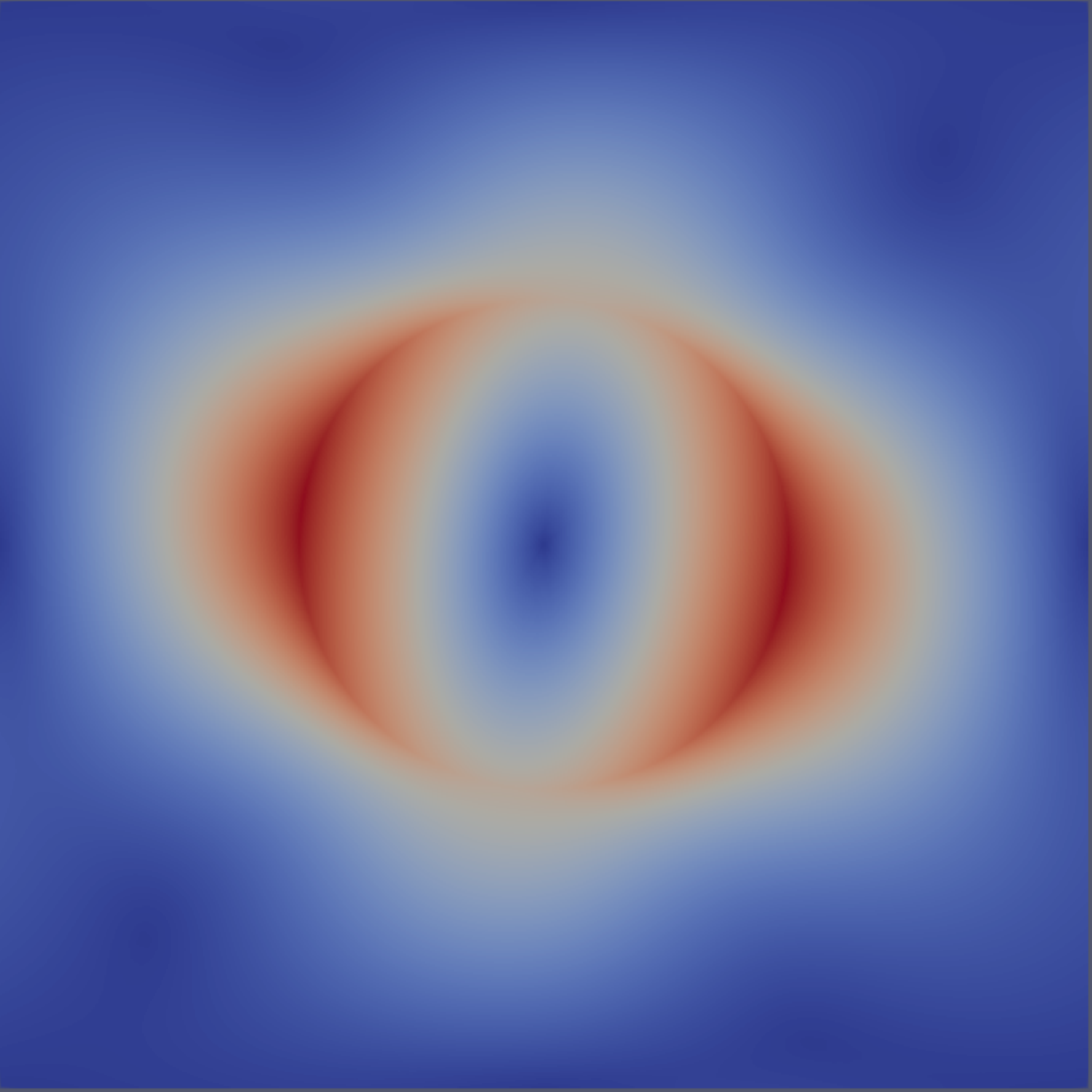}
         \caption{$\xi=0.0$}
         \label{fig:problem_1_poisson_1}
     \end{subfigure}
     \begin{subfigure}[c]{0.23\textwidth}
         \centering
         \includegraphics[width=\textwidth]{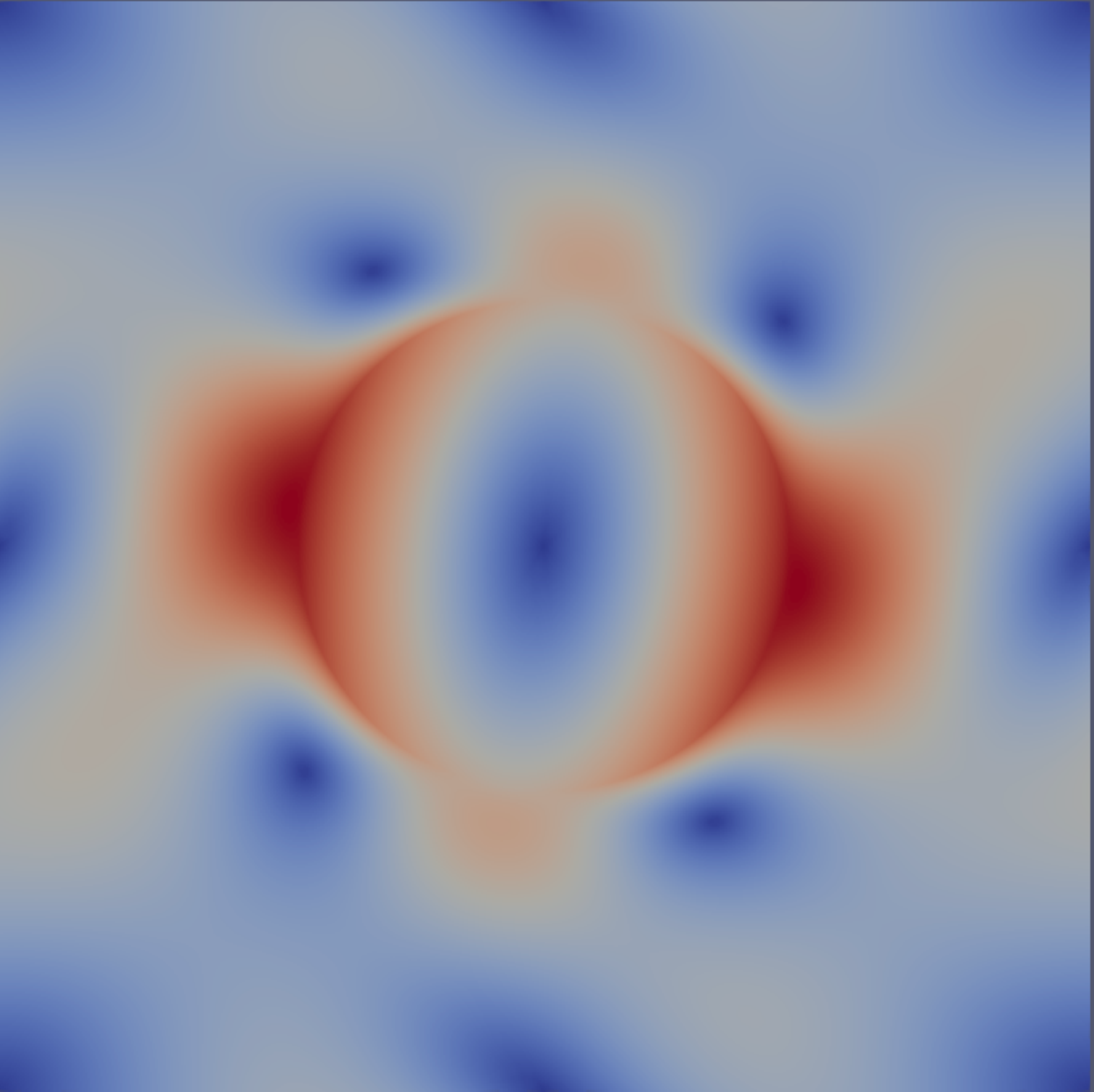}
         \caption{$\xi=0.49$}
         \label{fig:problem_1_poisson_2}
     \end{subfigure}
     \begin{subfigure}[c]{0.1\textwidth}
         \centering
         \includegraphics[width=\textwidth]{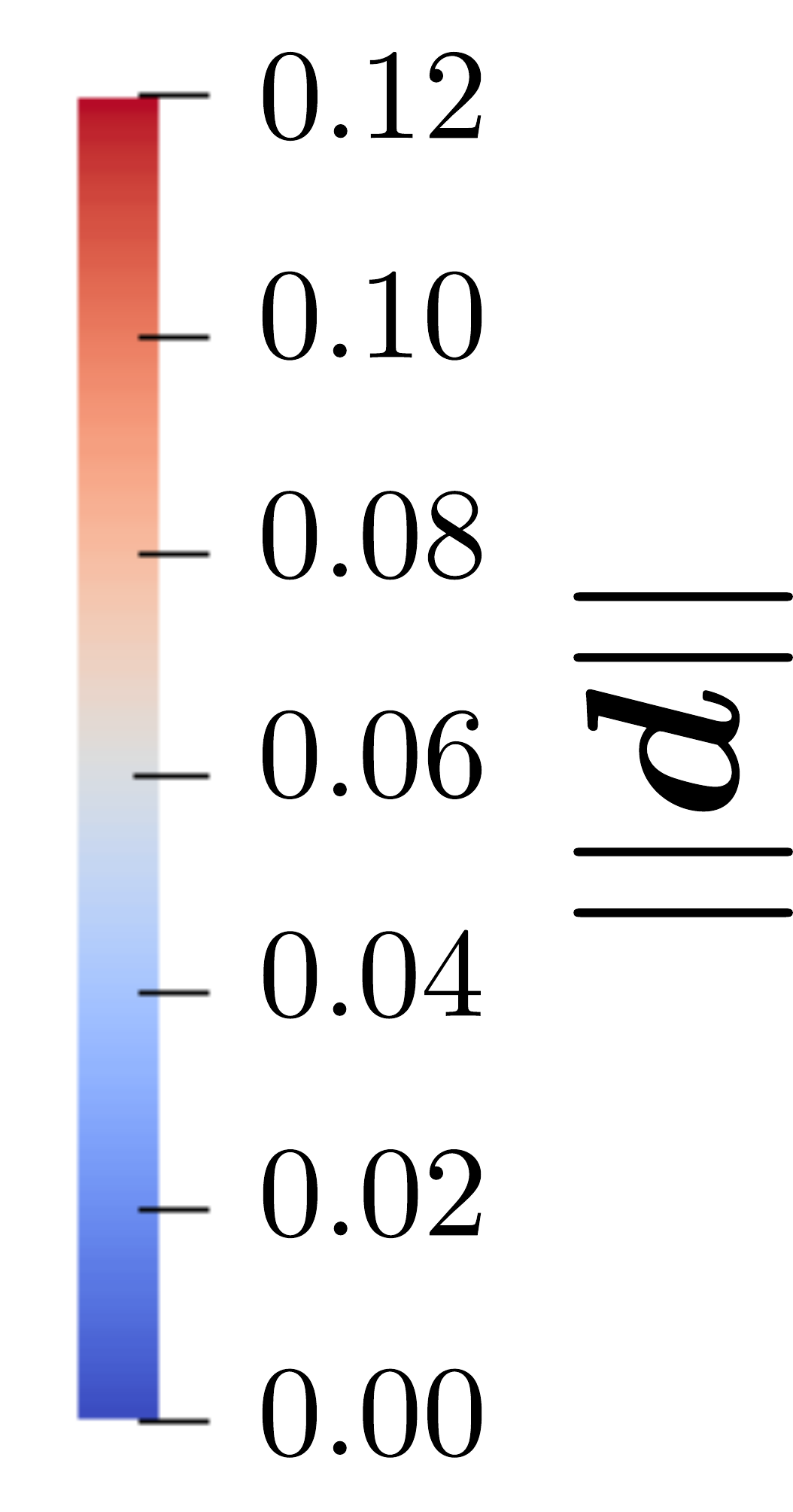}
         \label{fig:problem_1_poisson_3}
     \end{subfigure}
     \hfill
     \begin{subfigure}[c]{0.4\textwidth}
         \centering
         \includegraphics[width=\textwidth]{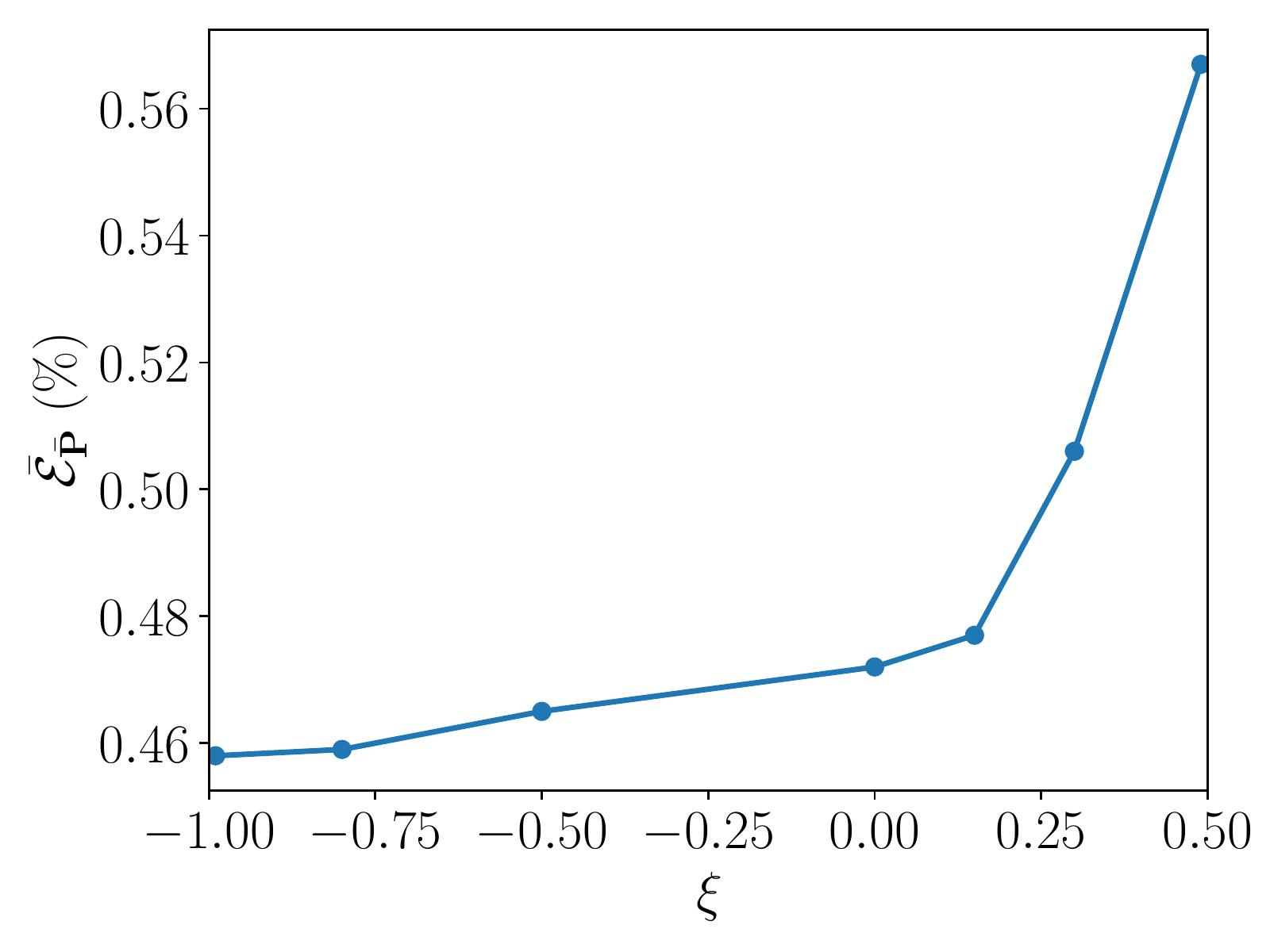}
         \caption{Error for different Poisson's ratios}
         \label{fig:problem_1_poisson_4}
     \end{subfigure}
    \caption{Norm of transformation displacement with $a=0.104$, $b=0.291$, $\theta=-8.44^\circ$ for (a) $\xi=0.0$ and (b) $\xi=0.49$. While the former leads to localized deformations, the latter affects the entire domain. (c) The mean error plotted over the Poisson's ratio $\xi$. The errors decrease with decreasing $\xi$, but remain on the same order of magnitude $\mathcal{O}(10^{-3})$.}
    \label{fig:problem_1_poisson}
\end{figure}

In this section, the influence of the Poisson's ratio on the approximation of the effective stress is investigated, considering $\xi\in\{-0.99, -0.8, -0.5, 0.0, 0.15, 0.3, 0.49\}$. Example transformation maps for $\xi\in\{0.0,0.49\}$ are shown in \cref{fig:problem_1_poisson_1,fig:problem_1_poisson_2} and the obtained errors are plotted over the Poisson's ratio in \cref{fig:problem_1_poisson_4}. For this example, the lower the Poisson's ratio the better the approximation. Nevertheless, all error values are close to each other (ranging from $0.46\%$ to $0.58\%$), although the transformation displacement field differs significantly, see \cref{fig:problem_1_poisson_1,fig:problem_1_poisson_2}. From this empirical result, it seems that the choice of the Poisson's ratio is insignificant and, since there is no practical way of finding the best value, $\xi=0.3$ is adopted hereafter.

\subsection{Composite Microstructure with a B-Spline Controlled Inclusion Shape}
\begin{figure}[ht]
     \centering
     \begin{subfigure}[c]{0.31\textwidth}
         \centering
         \includegraphics[width=\textwidth]{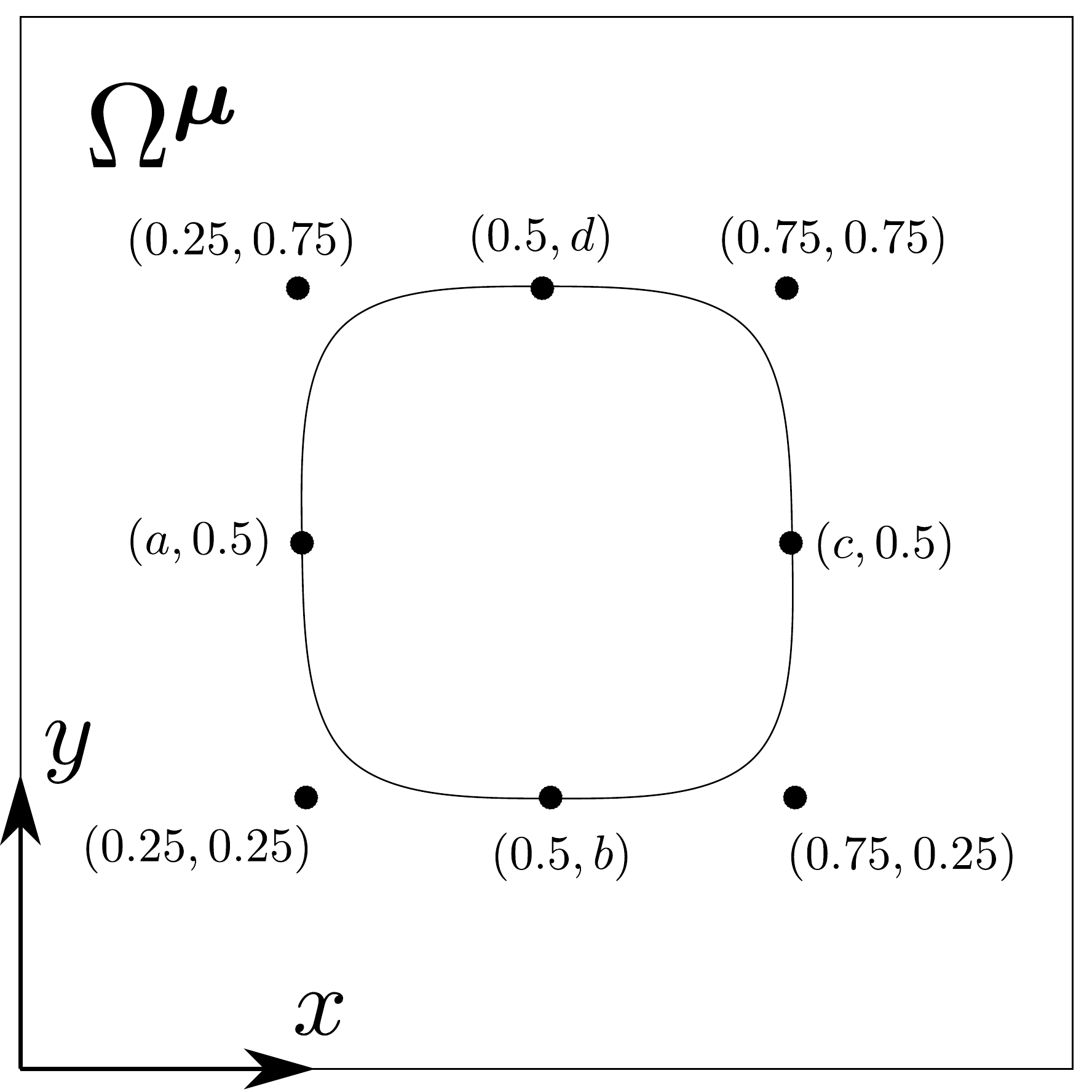}
         \caption{}
         \label{fig:problem_2_sketch_1}
     \end{subfigure}
     \qquad
     \begin{subfigure}[c]{0.3\textwidth}
         \centering
         \includegraphics[width=\textwidth]{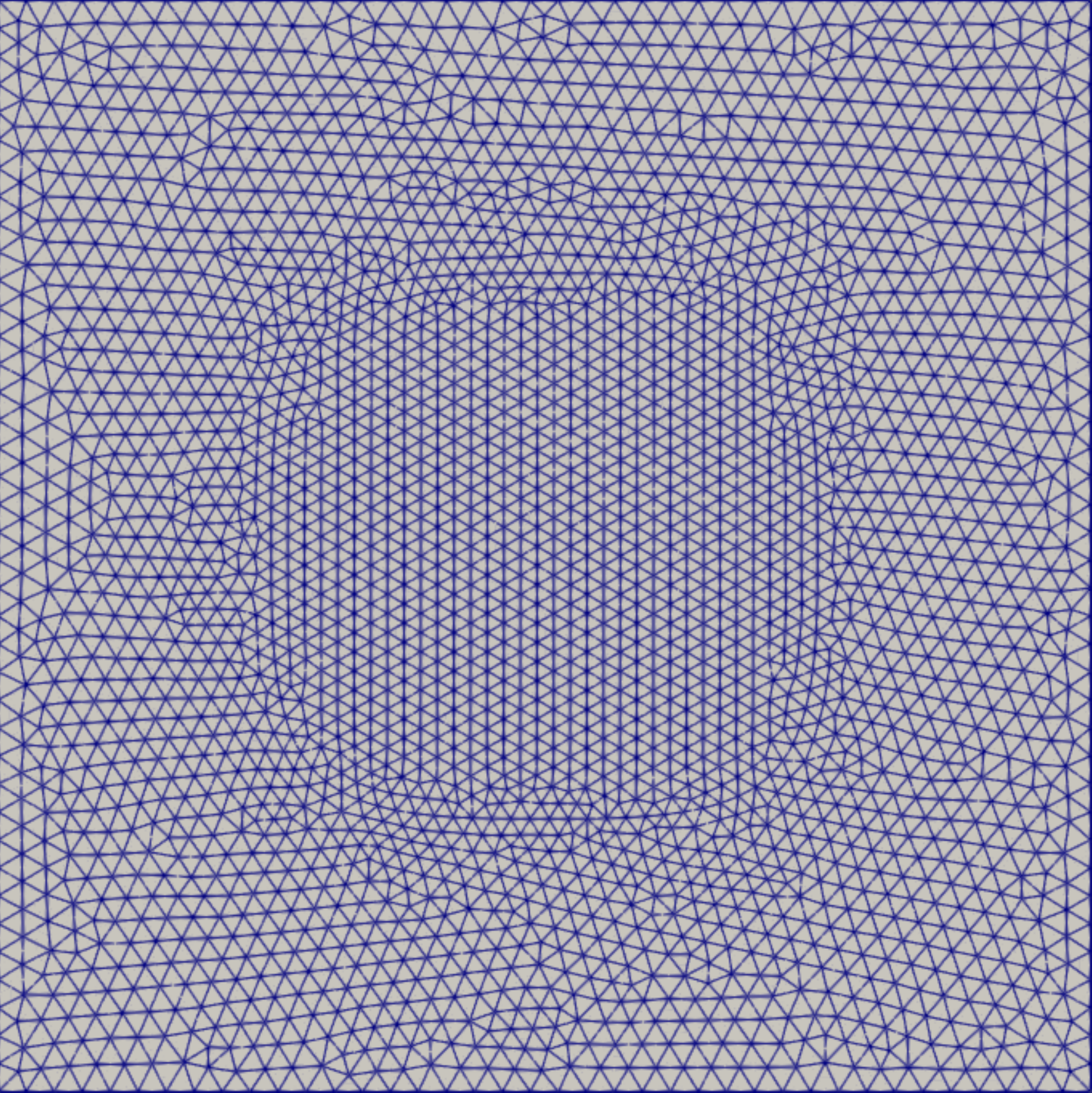}
         \caption{}
         \label{fig:problem_2_sketch_2}
     \end{subfigure}
    \caption{Parameterized domain. (a) The interface is spanned by eight control points. Out of those, four control points can move in one direction, which are controlled by the geometrical parameters $a$, $b$, $c$ and $d$. (b) The parent geometry with $a=b=0.25$ and $c=d=0.75$ is chosen and the mesh consists of 11296 nodes and 5833 6-node triangular elements.}
    \label{fig:problem_2_sketch}
\end{figure}

\begin{figure}[ht]
    \centering
    \begin{subfigure}[b]{0.32\textwidth}
        \centering
        \includegraphics[width=\textwidth]{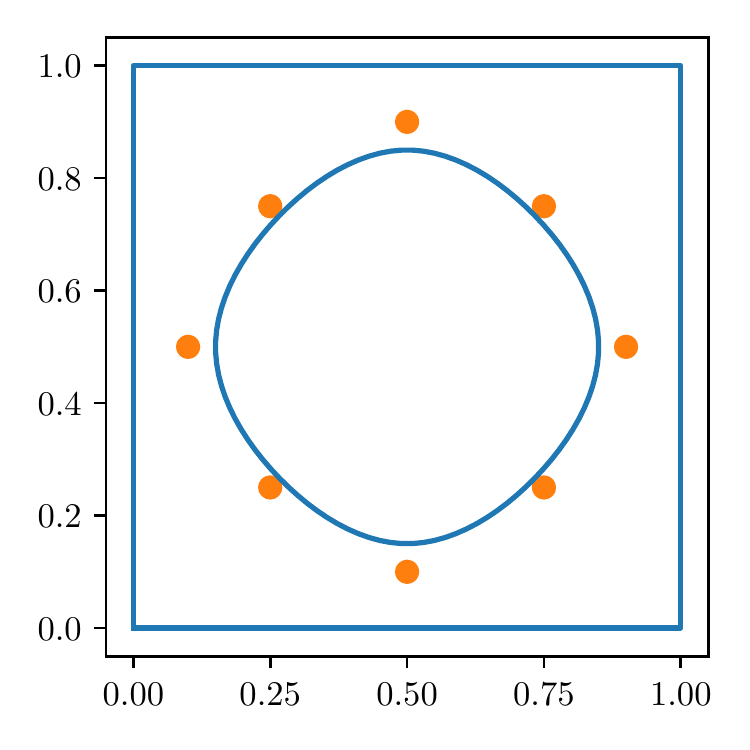}
        \caption{$a=b=0.1$, $c=d=0.9$}
        \label{fig:problem_2_examples1}
    \end{subfigure}
    \hfill
    \begin{subfigure}[b]{0.32\textwidth}
        \centering
        \includegraphics[width=\textwidth]{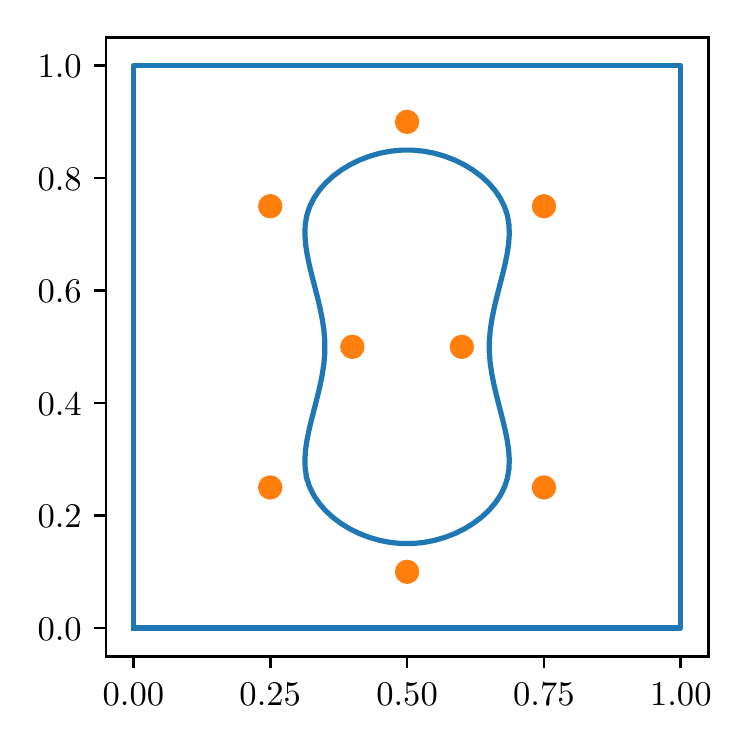}
        \caption{$a=0.4$, $b=0.1$, $c=0.6$, $d=0.9$}
        \label{fig:problem_2_examples2}
    \end{subfigure}
    \hfill
    \begin{subfigure}[b]{0.32\textwidth}
        \centering
        \includegraphics[width=\textwidth]{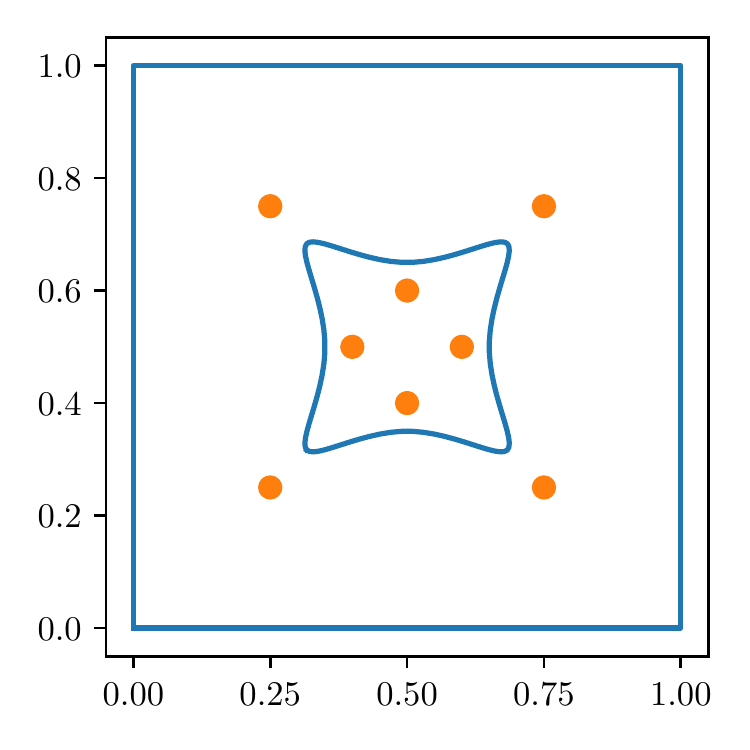}
        \caption{$a=b=0.4$, $c=d=0.6$}
        \label{fig:problem_2_examples3}
     \end{subfigure}
     \caption{Example geometries. The control points are shown in orange color and the resulting interface in blue color.}
    \label{fig:problem_2_examples}
\end{figure}

In the second example, an inclusion with a shape that is described by a B-spline with eight control points is considered, see \cref{fig:problem_2_sketch_1}. The $x$-coordinate of the left and right control point and the $y$-coordinate of the top and bottom control point are parameterized, resulting in four geometrical parameters $a$, $b$, $c$ and $d$. The curve is then interpolated with cubic polynomials using the Python library NURBS-Python \cite{Bingol2019NURBS-Python:Python}. The same material parameters are chosen as in the first example, with parameter ranges given in \cref{table:problem_2_params}. A few example geometries are shown in \cref{fig:problem_2_examples} to show the variety of shapes covered by this parameterization. {\color{mygreen}The lower and upper bounds for $\bar{\mathbf{U}}$ are chosen in the same way as in Example \ref{subsec:problem_1}.}

\begin{table}[ht]
\centering
\caption{Example 2 - 7 parameters with corresponding ranges. The parameters $a$, $b$, $c$ and $d$ are geometrical parameters describing the interface, see \cref{fig:problem_2_sketch} for the explanation, while $\bar{U}_{xx}$, $\bar{U}_{yy}$ and $\bar{U}_{xy}$ are external loading parameters.}
\begin{tabular}{c|c|c|c|c|c|c}
 $a$ & $b$ & $c$ & $d$ & $\bar{U}_{xx}$ & $\bar{U}_{yy}$ & $\bar{U}_{xy}$\\
 \hline
 $[0.1, 0.4]$ & $[0.1, 0.4]$ & $[0.6, 0.9]$ & $[0.6, 0.9]$ & $[0.85, 1.15]$ & $[0.85, 1.15]$ & $[-0.15, 0.15]$
\end{tabular}
\label{table:problem_2_params}
\end{table}

\subsubsection{Setup of the Auxiliary Problem}
For the parent domain, the midpoint of the parameter domain is selected, i.e., $a=b=0.25$ and $c=d=0.75$. The corresponding geometry and mesh, consisting of 11296 nodes and 5833 6-node triangular elements, are shown in \cref{fig:problem_2_sketch}.

\subsubsection{Data Generation}
In total 1000 training snapshots are again generated from a Sobol sequence \cite{sobol1967distribution}, while another 500 snapshots are generated from a random uniform distribution for testing.

\subsubsection{Results}
\begin{figure}[ht]
     \centering
     \begin{subfigure}[b]{0.35\textwidth}
         \centering
         \includegraphics[width=\textwidth]{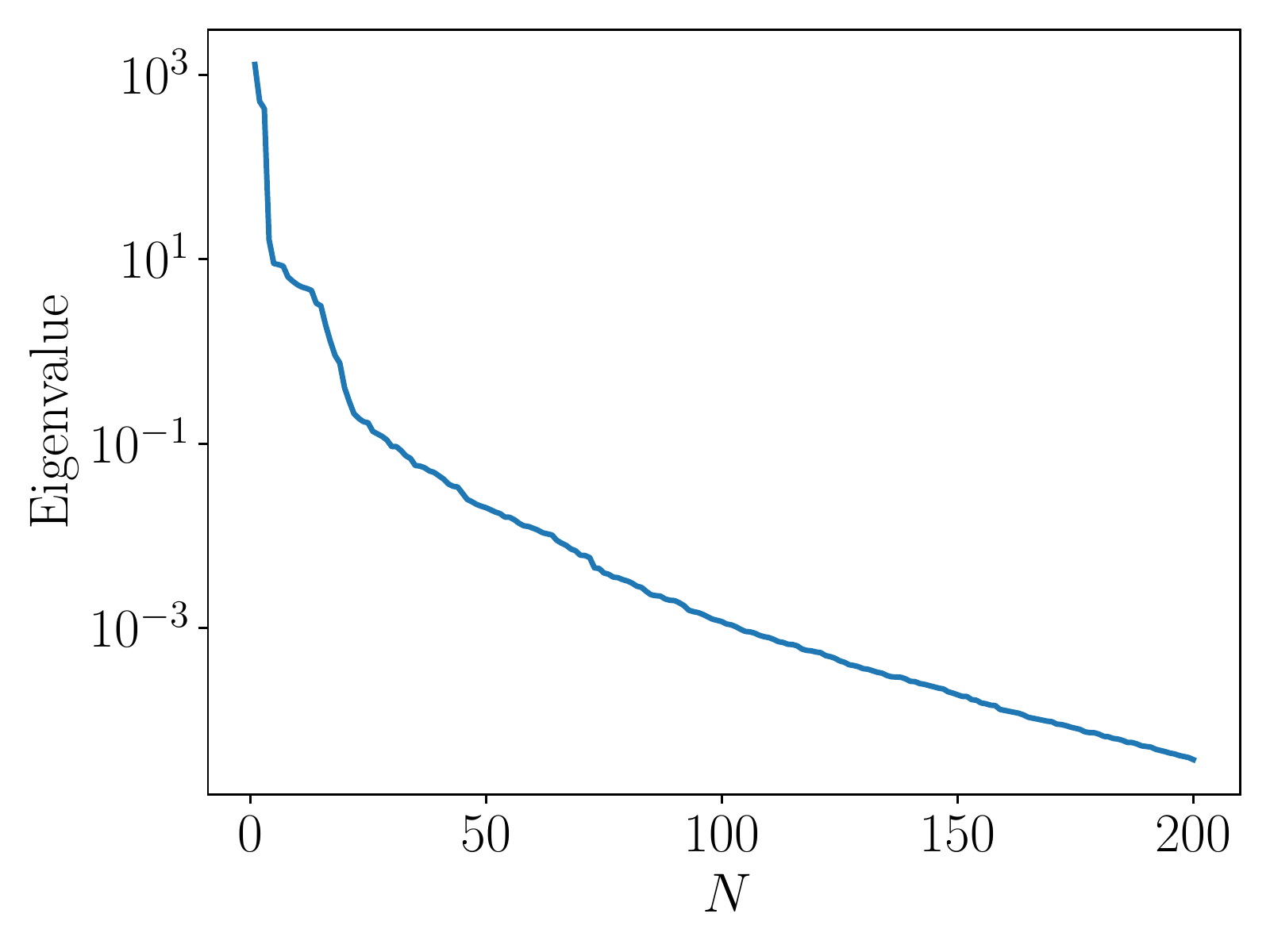}
         \caption{Weighted stress}
         \label{fig:problem_2_eig_1}
     \end{subfigure}
     \qquad
     \begin{subfigure}[b]{0.35\textwidth}
         \centering
         \includegraphics[width=\textwidth]{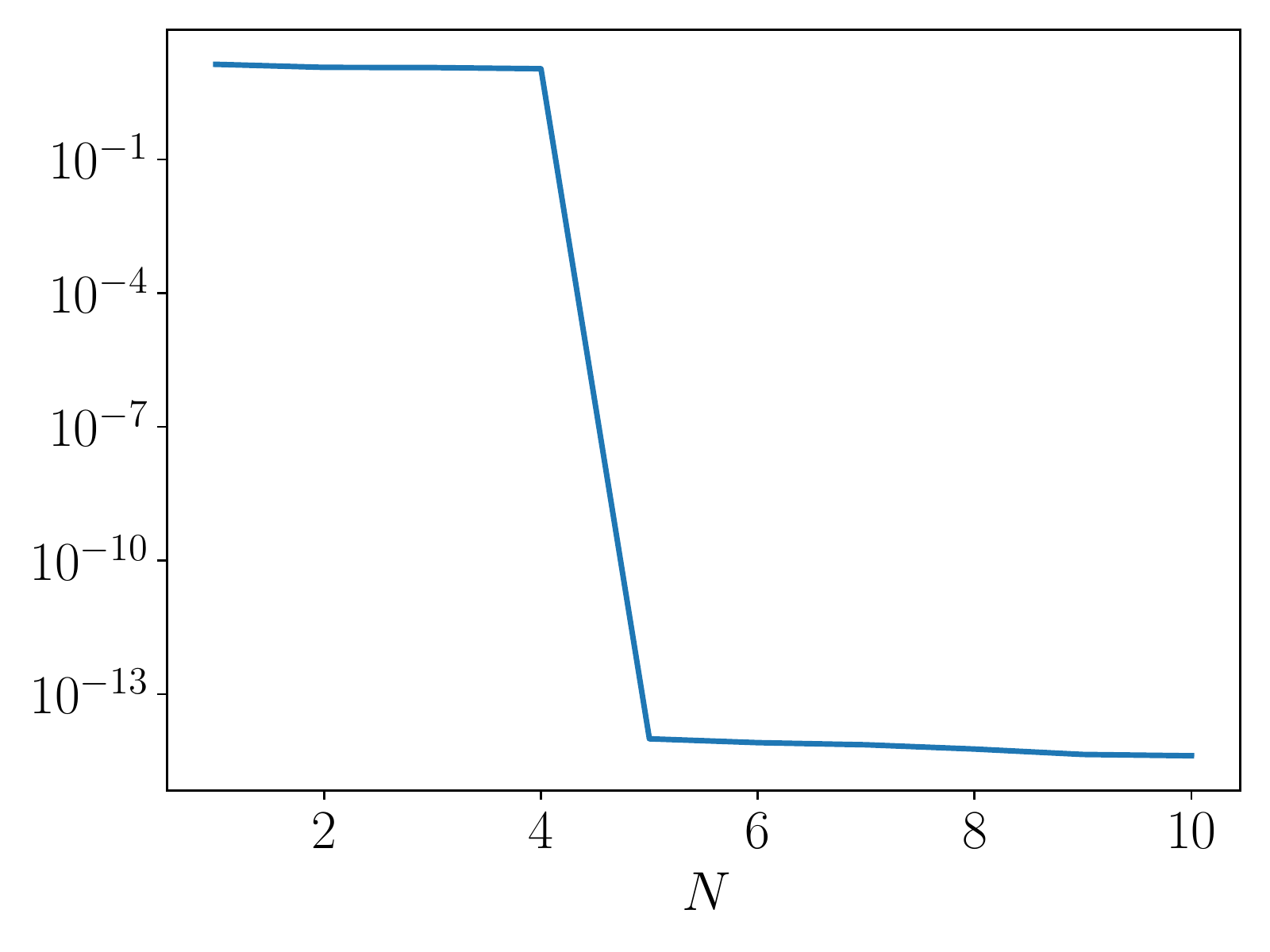}
         \caption{Transformation displacement}
         \label{fig:problem_2_eig_2}
     \end{subfigure}
    \caption{Eigenvalues of the correlation matrix for (a) weighted stress and (b) transformation displacement. Same as for the last example, the eigenvalues of weighted stress decay exponentially. For the transformation displacement only the first 4 modes are nonzero, meaning it can be represented with 4 basis functions.}
    \label{fig:problem_2_eigs}
\end{figure}

\paragraph{POD of Weighted Stress and Transformation Displacement}
In \cref{fig:problem_2_eigs}, the eigenvalues of the correlation matrix for both weighted stress and transformation displacement are depicted. An exponential decay can be observed for the weighted stress, while all but four eigenvalues are essentially zero for the transformation displacement.

\paragraph{Approximation Errors}
The average approximation errors on the 500 testing snapshots of the best approximation (projection of truth solution onto the reduced basis) with $N=50$ basis functions, PODGPR with $N=50$ basis functions and the four neural networks are reported in \cref{table:problem_2_errors}. From the results it can be observed that PODGPR nearly reaches the best approximation, showing that the first 50 coefficients are well approximated by the GPR models. Furthermore, it outperforms all neural network architectures by a factor of 2, reaching an average error of 0.131\% in effective stress.
\begin{table}[pth]
\centering
\caption{Mean errors for different methods. The trained surrogate models are tested on 500 testing snapshots. PODGPR outperforms all NN models by a factor 2.}
\begin{tabular}{c|c|c|c|c|c|c}
       & Best approximation & PODGPR & NN1 & NN2 & NN3 & NN4\\ \hline
$\bar{\mathcal{E}}_{\bm{P}}$ & $1.79\times 10^{-4}$ & $1.85\times 10^{-4}$ &   n.a.  & n.a. & n.a. & n.a. \\
$\bar{\mathcal{E}}_{\bar{\mathbf{P}}}$ & $1.09\times 10^{-3}$ &  $1.31\times 10^{-3}$ &  $3.3\times 10^{-3}$   &   $2.5\times 10^{-3}$  & $2.2\times 10^{-3}$ & $2.3\times 10^{-3}$
\end{tabular}
\label{table:problem_2_errors}
\end{table}

\subsection{Two-scale Cook's Membrane Problem}
\begin{figure}[ht]
     \centering
     \begin{subfigure}[b]{0.33\textwidth}
         \centering
         \includegraphics[width=\textwidth]{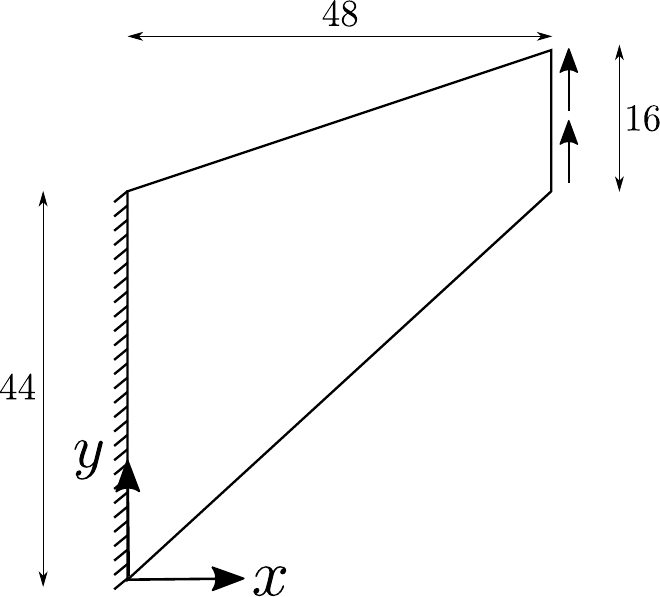}
         \caption{}
         \label{fig:problem_3_1}
     \end{subfigure}
     \begin{subfigure}[b]{0.22\textwidth}
         \centering
         \includegraphics[width=\textwidth]{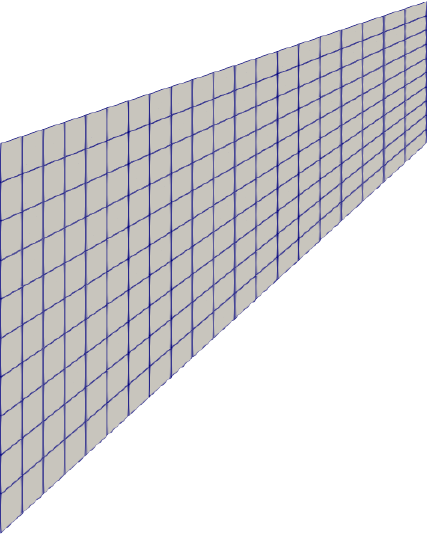}
         \caption{}
         \label{fig:problem_3_2}
     \end{subfigure}
     \qquad
     \begin{subfigure}[b]{0.35\textwidth}
         \centering
         \includegraphics[width=\textwidth]{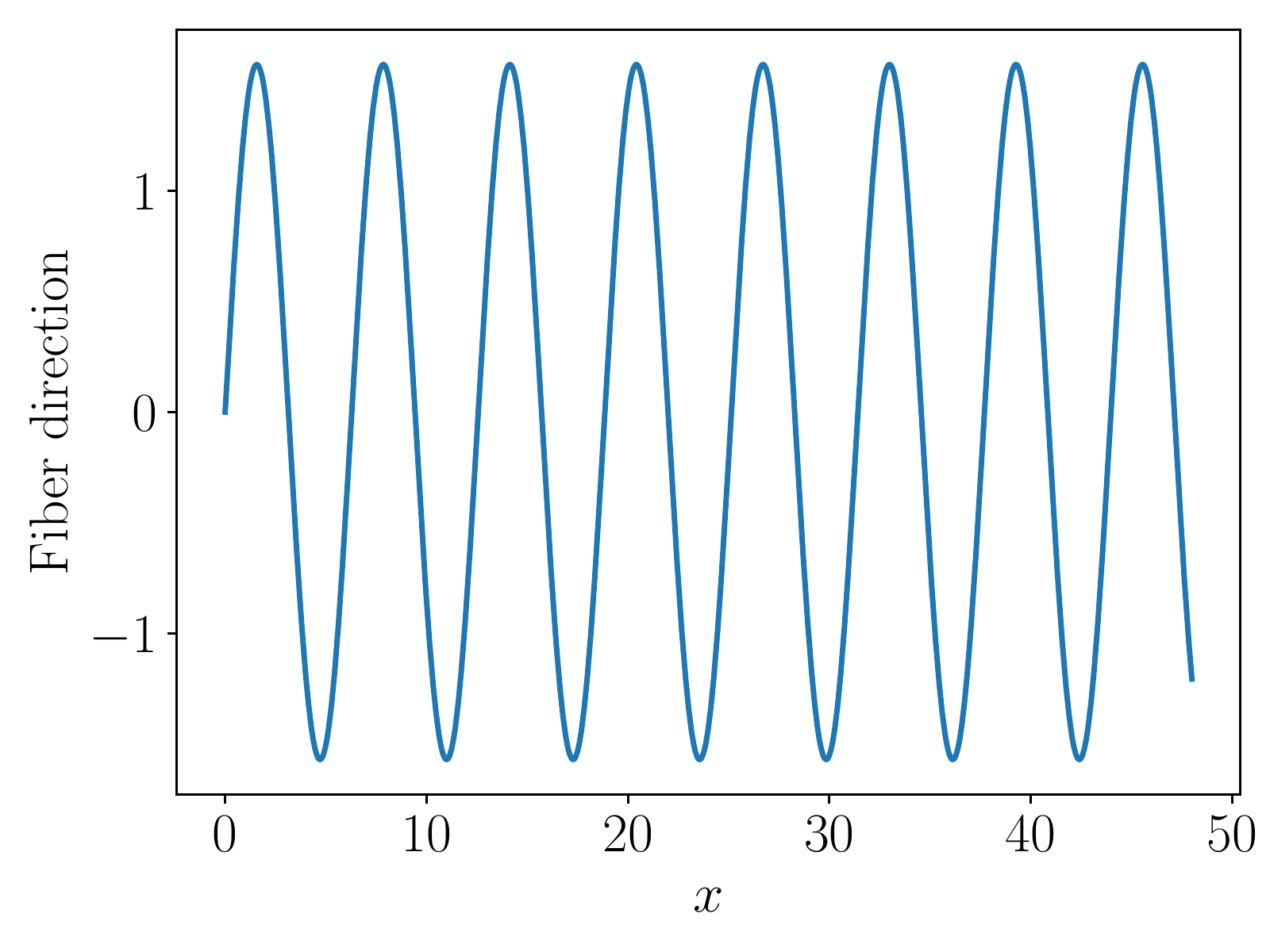}
         \caption{}
         \label{fig:problem_3_3}
     \end{subfigure}
    \caption{(a) The geometry of the Cook's membrane and (b) corresponding simulation mesh used. The mesh consists of 231 nodes and 200 4-node quadrilateral elements. (c) The fiber direction depends on the $x$-coordinate as $\theta(x)=\dfrac{\pi}{2}\sin{x}$.}
    \label{fig:cook_geo}
\end{figure}

\begin{figure}[ht]
     \centering
     \begin{subfigure}[c]{0.25\textwidth}
         \centering
         \includegraphics[width=\textwidth]{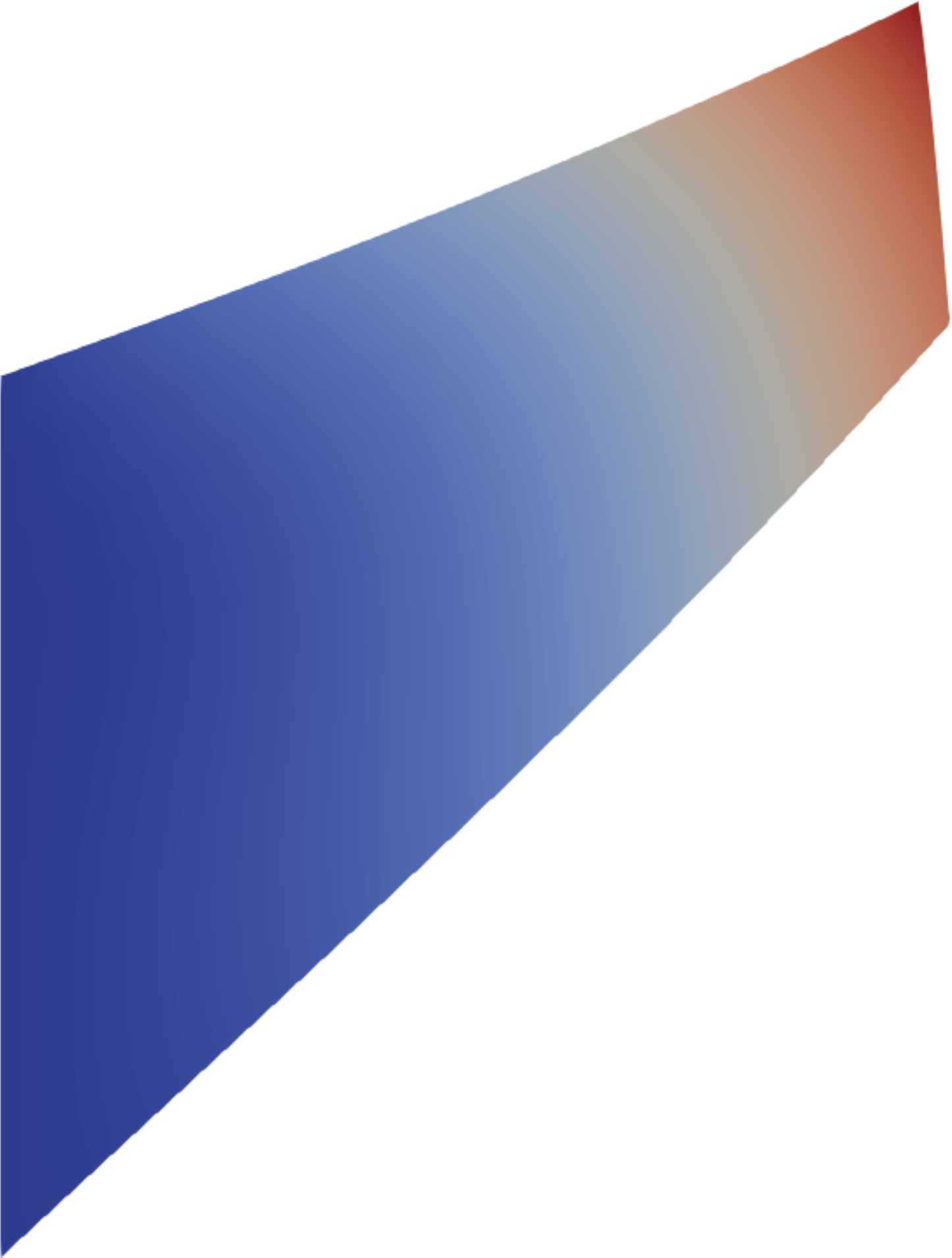}
         \caption{FE$^2$}
         \label{fig:cook_comparison1}
     \end{subfigure}
     \quad \begin{subfigure}[c]{0.25\textwidth}
         \centering
         \includegraphics[width=\textwidth]{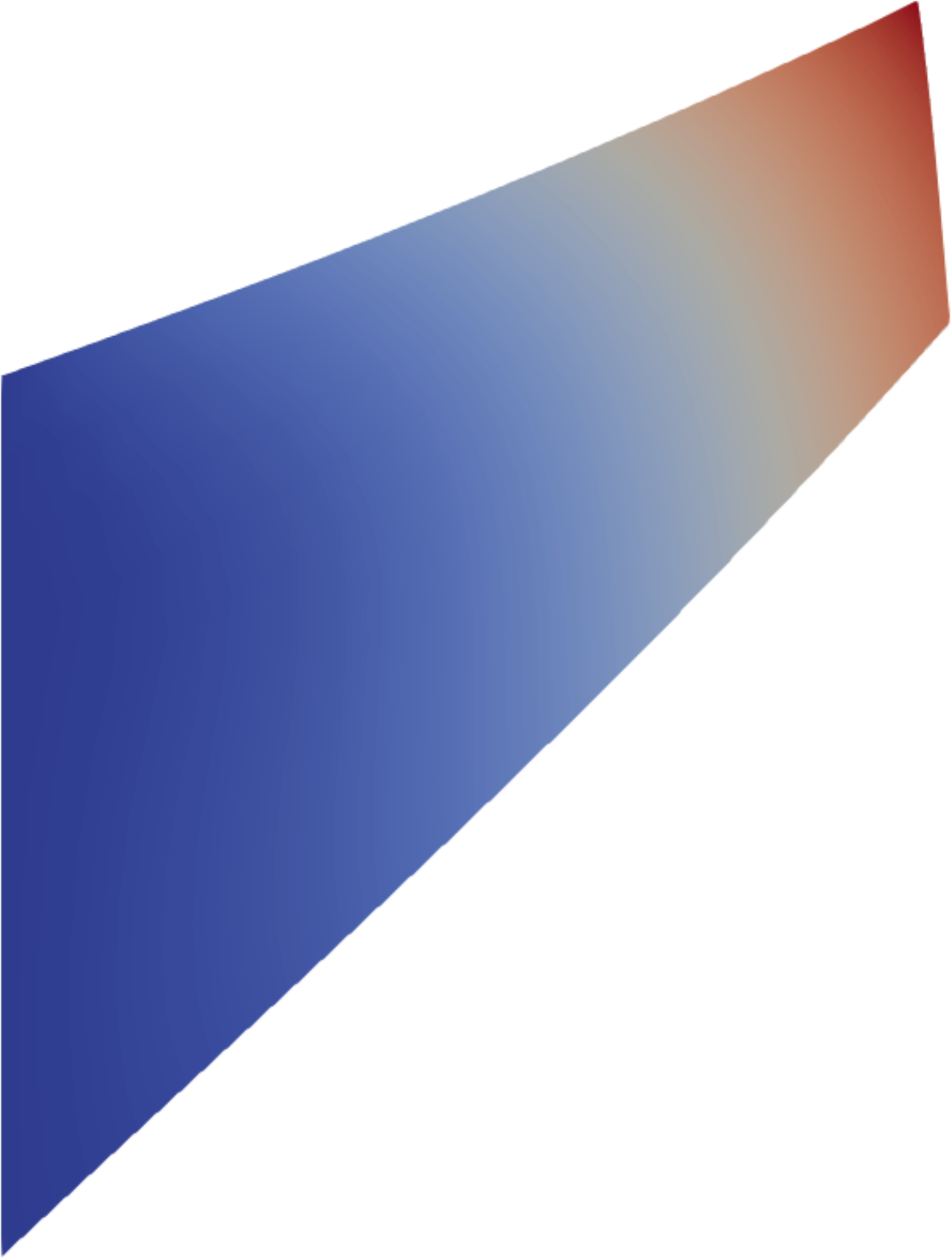}
         \caption{FE-PODGPR}
         \label{fig:cook_comparison2}
     \end{subfigure}
     \begin{subfigure}[c]{0.08\textwidth}
         \centering
         \includegraphics[width=\textwidth]{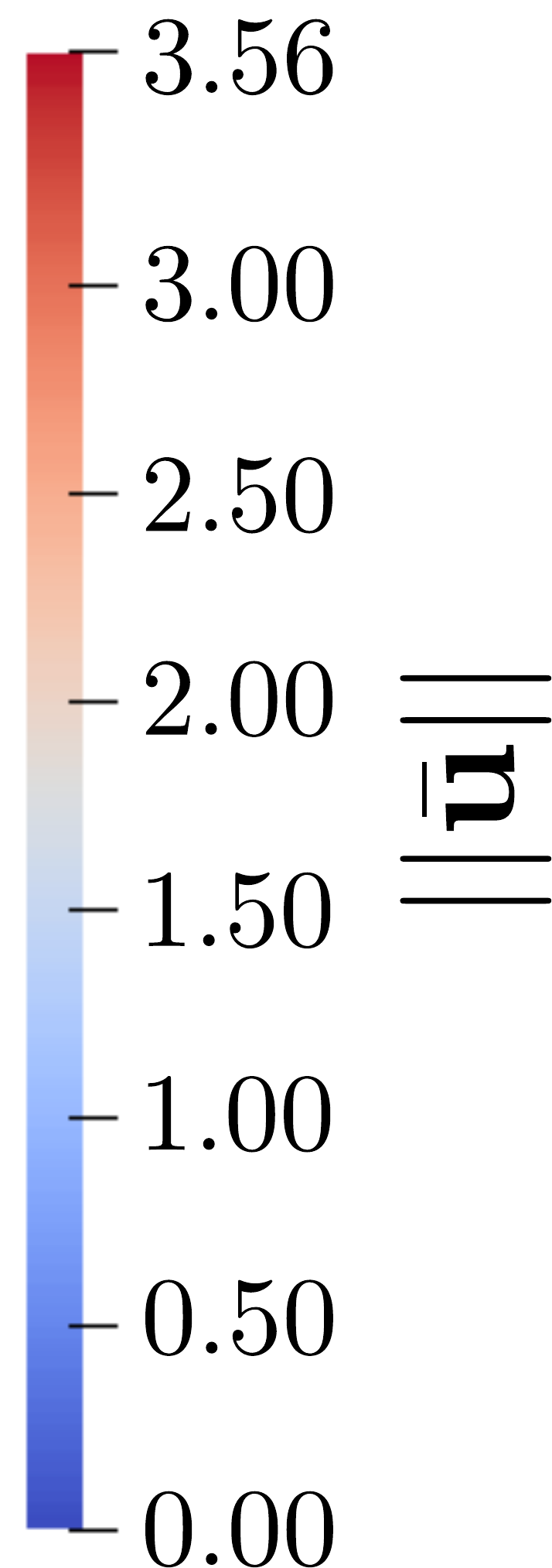}
         \label{fig:cook_comparison3}
     \end{subfigure}
     \hfill
     \begin{subfigure}[c]{0.26\textwidth}
         \centering
         \includegraphics[width=\textwidth]{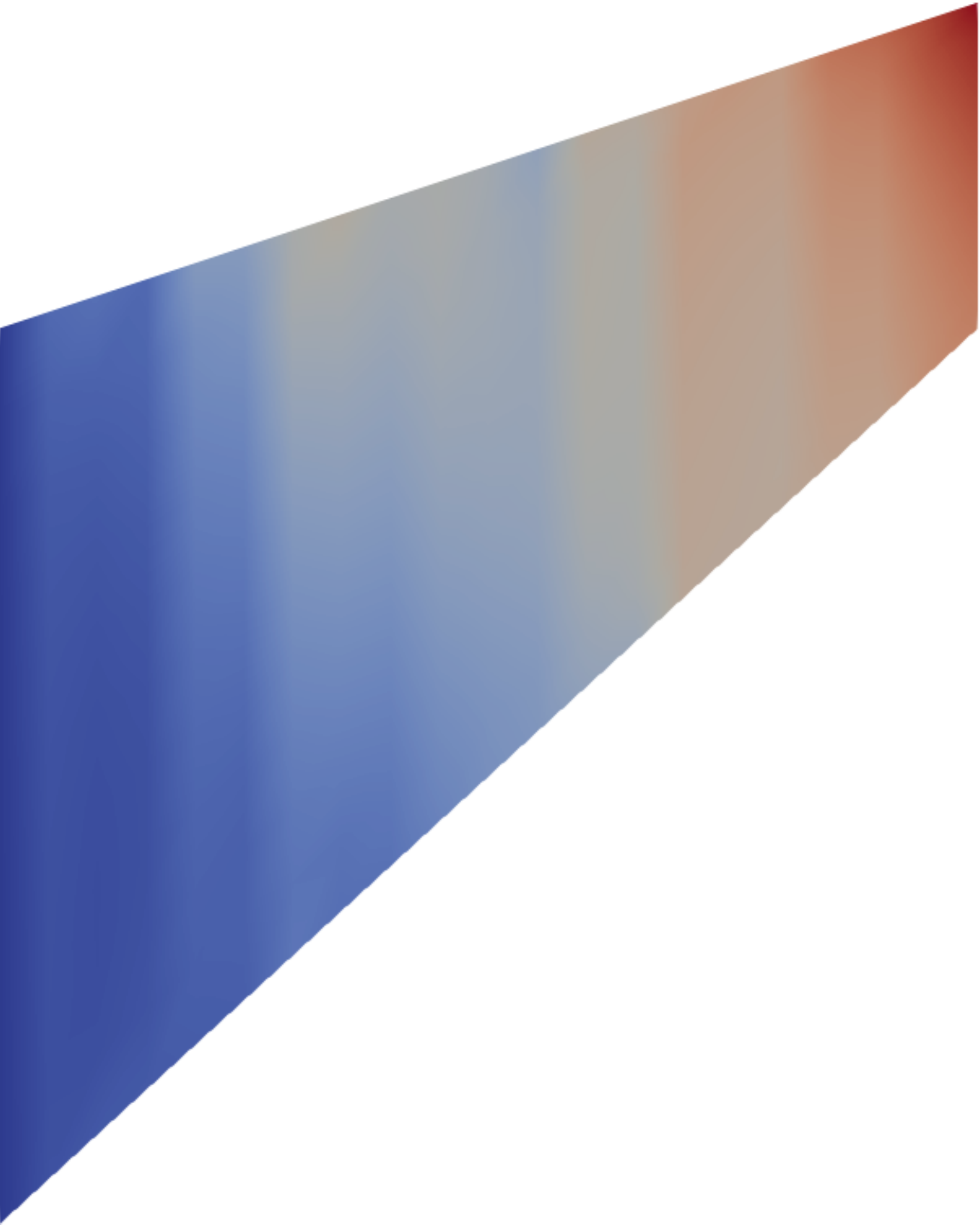}
         \caption{Absolute error}
         \label{fig:cook_comparison4}
     \end{subfigure}
     \begin{subfigure}[c]{0.08\textwidth}
         \centering
         \includegraphics[width=\textwidth]{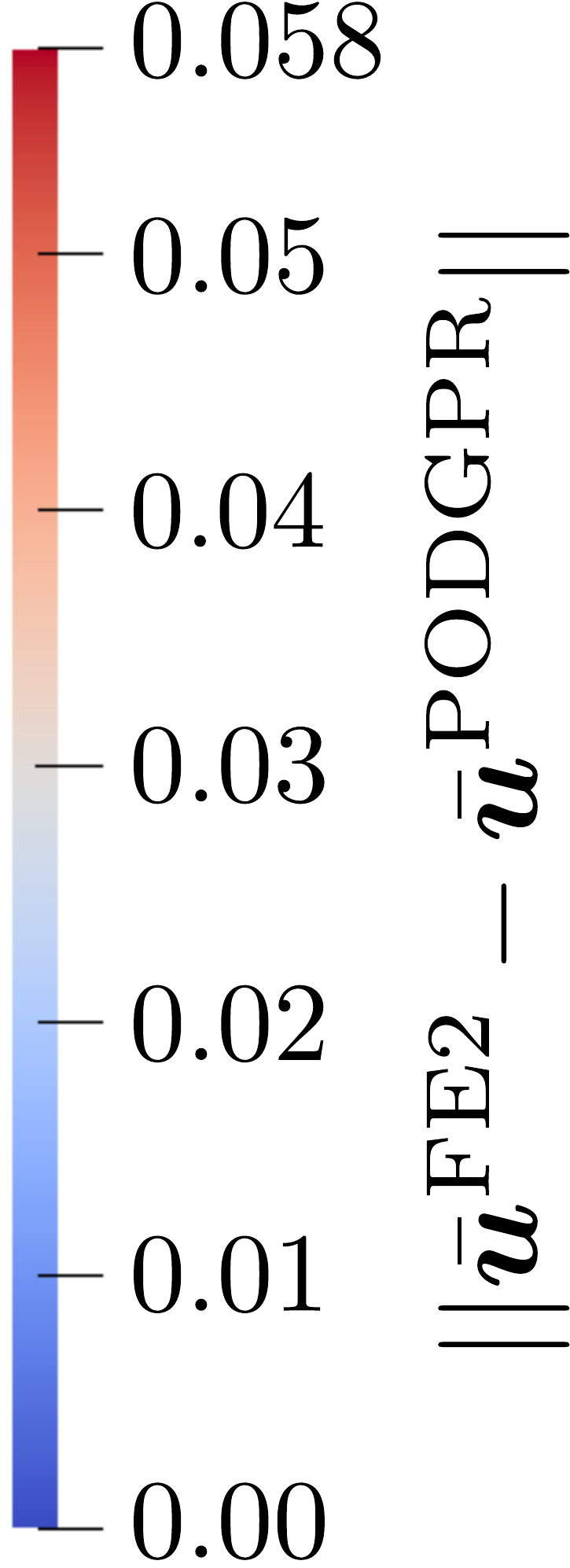}
         \label{fig:cook_comparison5}
     \end{subfigure}
    \caption{The norm of the displacement obtained for (a) FE$^2$ and (b) FE with PODGPR. The displacement fields are nearly identical. (c) The magnitude of the absolute error between both solutions. The error is around two orders of magnitude lower than the displacement values and increases from left to right. Vertical stripes are visible, corresponding to the rapidly varying fiber orientations, showing that some angles are better or worse approximated.}
    \label{fig:cook_comparison}
\end{figure}

While the last two examples dealt with the construction of the surrogate model for the microscale, in this example the surrogate model is employed in a full two-scale Cook's membrane problem. Here, the microstructure from \cref{subsec:problem_1} with an elliptical inclusion is considered. The geometry of the Cook's membrane and its mesh are shown in \cref{fig:cook_geo}. The mesh consists of 200 quadrilateral elements with 4 quadrature points, resulting in 800 microstructure evaluations required for a single Newton iteration. The microstructural parameters $a$ and $b$ are assumed to be constant with $a=0.35$ and $b=0.1$, corresponding to an ellipse, while the angle $\theta$ is a function in the $x$-coordinate with $\theta(x)=(\pi\sin{x})/2$, in order to test the performance of the surrogate model under rapidly varying fiber directions (see \cref{fig:problem_3_3} for an illustration of the function). A similar test problem was considered in \cite{Reddy2019AElasticity}. The left side of the membrane is fixed, while a vertical traction of $0.05$ is applied on the right edge, which leads to overall deformations within the training range of the surrogate model ($|\bar{U}_{xx}-1|, |\bar{U}_{yy}-1|, |\bar{U}_{xy}|<0.15$).

A full FE$^2$ simulation and a FE simulation using PODGPR are run and the obtained displacement fields $\bar{\bm{u}}$ are compared in \cref{fig:cook_comparison}. The PODGPR surrogate model constructed with $N=20$ basis functions from $N_s=1000$ training snapshots is used. The obtained displacement fields shown in \cref{fig:cook_comparison1,fig:cook_comparison2} are almost identical. To further quantify the error, the magnitude of the difference between both displacement solutions is shown in \cref{fig:cook_comparison4}. Here we observe that the highest absolute error is at the top right corner and the error increases from left to right. Comparing the error and the actual value at the top right corner, the relative error corresponds to $0.058/3.56 = 1.6\%$. Furthermore, vertical stripes with similar errors can be seen, which correspond to the quickly varying fiber directions in the $x$-coordinate, meaning that some angles are better or worse approximated. The quality of approximation for different angles depends on the sampling of the training snapshots. Moreover, the compliance $f_c$, defined as $f_c \coloneqq \mathbf{f}_{\text{ext}}^T \bar{\mathbf{u}}$ where $\mathbf{f}_{\text{ext}}$ corresponds to the externally applied vertical traction, is computed for both methods, yielding $f_c^{\text{FE2}} = 2.227$ and $f_c^{\text{PODGPR}} = 2.202$, resulting in a relative error of 1.1\%. This is an important quantity, often used in optimization problems.

The execution times\footnote{All operations were executed using four subprocesses on an Intel Core i7-8750H.} for both cases are reported in \cref{table:cook_runtimes}. For the construction of PODGPR, 1000 snapshots are generated, involving each time an auxiliary problem and a full simulation to be solved, taking roughly 4 hours. With the data available, the construction of PODGPR with 20 basis functions takes around 10 minutes. After this offline computation, the online speed up is on the order of 1000 as compared to the full two-scale simulation.

\begin{table}[pth]
\centering
\caption{Run times. The offline stage for constructing PODGPR takes slightly over 4 hours. Then, the Cook's membrane problem can be solved within 0.5 minutes, while at the same time the full FE$^2$ simulation takes around 1080 minutes for a single forward solution.}
\begin{tabular}{c|c|c}
        & FE$^2$              & FE-PODGPR                                                                                                                                                      \\ \hline
Offline & n.a                 & \begin{tabular}[c]{@{}c@{}}Auxiliary Problem: $\approx\SI{10}{min}$\\ Snapshot Generation: $\approx\SI{240}{min}$\\ PODGPR: $\approx\SI{10}{min}$\end{tabular} \\ \hline
Online  & $\approx\SI{1080}{min}$ & $\approx\SI{0.5}{min}$                                                                                                                                         
\end{tabular}
\label{table:cook_runtimes}
\end{table}

%%%%%%%% CONCLUSION %%%%%%%%%%%
\section{Conclusions}
\label{sec:conclusion}
In order to successfully find optimal microstructures for complex engineering systems in a reasonable amount of time, it is necessary to find an accurate and fast-to-evaluate approximation of the microscopic simulation, which can handle a large design space of shapes and geometrical variations. In this work, we proposed a PDE-based transformation method for the treatment of geometrical parameters. By combining the transformation with the proper orthogonal decomposition and Gaussian process regression, we have developed a non-intrusive effective constitutive model that makes predictions which automatically fulfill the underlying microscopic governing equations for a wide variety of geometries. For the two considered microstructures, each described by several geometrical parameters, the surrogate model captures the local stress fields accurately with an average error less than 1\%. The effective model is embedded in a two-scale problem, where a high variation in parameters throughout the domain is considered, and accelerates the simulation by a factor of 1000 as compared to the full FE$^2$ simulation, while maintaining high accuracy. Sensitivities with respect to microstructural parameters are available, which can be utilized in two-scale shape optimization problems or solution of inverse problems, and the methodology can also be directly applied to 3D problems.

Although this method is powerful, several limitations exist. For geometrical parameters leading to very severe geometrical variations, the proposed transformation leads to highly irregular and distorted meshes. A possible remedy to resolve this issue is to correct the distorted elements by mesh refinement, or to use multiple parent geometries with local surrogate models. In the online stage, the closest cluster could be chosen to evaluate the surrogate model. Another challenge is the requirement of data. Even though the methodology proved to be more data efficient than feed-forward neural networks in our examples, still a rather large amount of training data is needed. Possible solutions are multi-fidelity methods or adaptive sampling schemes.

%%%%%%%% REST %%%%%%%%%%%
\section*{Data availability}
The data that support the findings of this study are available from the corresponding author upon request.
\section*{Acknowledgements}
This result is part of a project that has received funding from the European Research Council (ERC) under the European Union’s Horizon 2020 Research and Innovation Programme (Grant Agreement No. 818473).
\bibliography{references}

\begin{thebibliography}{10}
\expandafter\ifx\csname url\endcsname\relax
  \def\url#1{\texttt{#1}}\fi
\expandafter\ifx\csname urlprefix\endcsname\relax\def\urlprefix{URL }\fi
\expandafter\ifx\csname href\endcsname\relax
  \def\href#1#2{#2} \def\path#1{#1}\fi

\bibitem{Geers2010}
M.~G.~D. Geers, V.~G. Kouznetsova, W.~A.~M. Brekelmans,
  \href{https://linkinghub.elsevier.com/retrieve/pii/S0377042709005536}{{Multi-scale
  computational homogenization: Trends and challenges}}, Journal of
  Computational and Applied Mathematics 234~(7) (2010) 2175--2182.
\newblock \href {https://doi.org/10.1016/j.cam.2009.08.077}
  {\path{doi:10.1016/j.cam.2009.08.077}}.
\newline\urlprefix\url{https://linkinghub.elsevier.com/retrieve/pii/S0377042709005536}

\bibitem{Matous2017}
K.~Matou{\v{s}}, M.~G.~D. Geers, V.~G. Kouznetsova, A.~Gillman, {A review of
  predictive nonlinear theories for multiscale modeling of heterogeneous
  materials}, Journal of Computational Physics (2017).
\newblock \href {https://doi.org/10.1016/j.jcp.2016.10.070}
  {\path{doi:10.1016/j.jcp.2016.10.070}}.

\bibitem{Feyel1999}
F.~Feyel, {Multiscale FE2 elastoviscoplastic analysis of composite structures},
  Computational Materials Science 16~(1-4) (1999) 344--354.
\newblock \href {https://doi.org/10.1016/s0927-0256(99)00077-4}
  {\path{doi:10.1016/s0927-0256(99)00077-4}}.

\bibitem{Miehe2002}
C.~Miehe, A.~Koch, {Computational micro-to-macro transitions of discretized
  microstructures undergoing small strains}, Archive of Applied Mechanics
  72~(4-5) (2002) 300--317.
\newblock \href {https://doi.org/10.1007/s00419-002-0212-2}
  {\path{doi:10.1007/s00419-002-0212-2}}.

\bibitem{Kouznetsova2002}
V.~G. Kouznetsova,
  \href{http://www.narcis.nl/publication/RecordID/oai:library.tue.nl:560009}{{Computational
  homogenization for the multi-scale analysis of multi-phase materials}}, no.
  2002, 2002.
\newblock \href {https://doi.org/10.6100/IR560009}
  {\path{doi:10.6100/IR560009}}.
\newline\urlprefix\url{http://www.narcis.nl/publication/RecordID/oai:library.tue.nl:560009}

\bibitem{Moulinec1998}
H.~Moulinec, P.~Suquet, {A numerical method for computing the overall response
  of nonlinear composites with complex microstructure}, Computer Methods in
  Applied Mechanics and Engineering 157~(1-2) (1998) 69--94.
\newblock \href {https://doi.org/10.1016/S0045-7825(97)00218-1}
  {\path{doi:10.1016/S0045-7825(97)00218-1}}.

\bibitem{Mishra2016a}
N.~Mishra, J.~Vond{\v{r}}ejc, J.~Zeman, {A comparative study on low-memory
  iterative solvers for FFT-based homogenization of periodic media}, Journal of
  Computational Physics (2016).
\newblock \href {https://doi.org/10.1016/j.jcp.2016.05.041}
  {\path{doi:10.1016/j.jcp.2016.05.041}}.

\bibitem{Kirchdoerfer2016}
T.~Kirchdoerfer, M.~Ortiz,
  \href{http://dx.doi.org/10.1016/j.cma.2016.02.001}{{Data-driven computational
  mechanics}}, Computer Methods in Applied Mechanics and Engineering 304 (2016)
  81--101.
\newblock \href {https://doi.org/10.1016/j.cma.2016.02.001}
  {\path{doi:10.1016/j.cma.2016.02.001}}.
\newline\urlprefix\url{http://dx.doi.org/10.1016/j.cma.2016.02.001}

\bibitem{Kirchdoerfer2017}
T.~Kirchdoerfer, M.~Ortiz,
  \href{http://dx.doi.org/10.1016/j.cma.2017.07.039}{{Data Driven Computing
  with noisy material data sets}}, Computer Methods in Applied Mechanics and
  Engineering 326 (2017) 622--641.
\newblock \href {https://doi.org/10.1016/j.cma.2017.07.039}
  {\path{doi:10.1016/j.cma.2017.07.039}}.
\newline\urlprefix\url{http://dx.doi.org/10.1016/j.cma.2017.07.039}

\bibitem{Eggersmann2019}
R.~Eggersmann, T.~Kirchdoerfer, S.~Reese, L.~Stainier, M.~Ortiz,
  \href{https://doi.org/10.1016/j.cma.2019.02.016}{{Model-Free Data-Driven
  inelasticity}}, Computer Methods in Applied Mechanics and Engineering 350
  (2019) 81--99.
\newblock \href {https://doi.org/10.1016/j.cma.2019.02.016}
  {\path{doi:10.1016/j.cma.2019.02.016}}.
\newline\urlprefix\url{https://doi.org/10.1016/j.cma.2019.02.016}

\bibitem{Wu2020a}
L.~Wu, V.~D. Nguyen, N.~G. Kilingar, L.~Noels, {A recurrent neural
  network-accelerated multi-scale model for elasto-plastic heterogeneous
  materials subjected to random cyclic and non-proportional loading paths},
  Computer Methods in Applied Mechanics and Engineering 369 (2020) 113234.
\newblock \href {https://doi.org/10.1016/j.cma.2020.113234}
  {\path{doi:10.1016/j.cma.2020.113234}}.

\bibitem{Abueidda2021}
D.~W. Abueidda, S.~Koric, N.~A. Sobh, H.~Sehitoglu,
  \href{https://doi.org/10.1016/j.ijplas.2020.102852}{{Deep learning for
  plasticity and thermo-viscoplasticity}}, International Journal of Plasticity
  136~(August 2020) (2021) 102852.
\newblock \href {https://doi.org/10.1016/j.ijplas.2020.102852}
  {\path{doi:10.1016/j.ijplas.2020.102852}}.
\newline\urlprefix\url{https://doi.org/10.1016/j.ijplas.2020.102852}

\bibitem{Mozaffar2019}
M.~Mozaffar, R.~Bostanabad, W.~Chen, K.~Ehmann, J.~Cao, M.~A. Bessa, {Deep
  learning predicts path-dependent plasticity}, Proceedings of the National
  Academy of Sciences 116~(52) (2019) 26414--26420.
\newblock \href {https://doi.org/10.1073/pnas.1911815116}
  {\path{doi:10.1073/pnas.1911815116}}.

\bibitem{Le2015}
B.~A. Le, J.~Yvonnet, Q.~C. He, {Computational homogenization of nonlinear
  elastic materials using neural networks}, International Journal for Numerical
  Methods in Engineering 104~(12) (2015) 1061--1084.
\newblock \href {https://doi.org/10.1002/nme.4953}
  {\path{doi:10.1002/nme.4953}}.

\bibitem{Kumar2020Inverse-designedMetamaterials}
S.~Kumar, S.~Tan, L.~Zheng, D.~M. Kochmann,
  \href{http://dx.doi.org/10.1038/s41524-020-0341-6}{{Inverse-designed
  spinodoid metamaterials}}, npj Computational Materials 6~(1) (2020) 1--10.
\newblock \href {https://doi.org/10.1038/s41524-020-0341-6}
  {\path{doi:10.1038/s41524-020-0341-6}}.
\newline\urlprefix\url{http://dx.doi.org/10.1038/s41524-020-0341-6}

\bibitem{Fernandez2021a}
M.~Fern{\'{a}}ndez, M.~Jamshidian, T.~B{\"{o}}hlke, K.~Kersting, O.~Weeger,
  \href{https://doi.org/10.1007/s00466-020-01954-7}{{Anisotropic hyperelastic
  constitutive models for finite deformations combining material theory and
  data-driven approaches with application to cubic lattice metamaterials}},
  Computational Mechanics (2021).
\newblock \href {https://doi.org/10.1007/s00466-020-01954-7}
  {\path{doi:10.1007/s00466-020-01954-7}}.
\newline\urlprefix\url{https://doi.org/10.1007/s00466-020-01954-7}

\bibitem{Linka2021}
K.~Linka, M.~Hillg{\"{a}}rtner, K.~P. Abdolazizi, R.~C. Aydin, M.~Itskov, C.~J.
  Cyron, \href{https://doi.org/10.1016/j.jcp.2020.110010}{{Constitutive
  artificial neural networks: A fast and general approach to predictive
  data-driven constitutive modeling by deep learning}}, Journal of
  Computational Physics 429 (2021) 110010.
\newblock \href {https://doi.org/10.1016/j.jcp.2020.110010}
  {\path{doi:10.1016/j.jcp.2020.110010}}.
\newline\urlprefix\url{https://doi.org/10.1016/j.jcp.2020.110010}

\bibitem{Flaschel2020}
M.~Flaschel, S.~Kumar, L.~De~Lorenzis, \href{http://arxiv.org/abs/2010.13496
  https://linkinghub.elsevier.com/retrieve/pii/S0045782521001894}{{Unsupervised
  discovery of interpretable hyperelastic constitutive laws}}, Computer Methods
  in Applied Mechanics and Engineering 381~(2018) (2021) 113852.
\newblock \href {https://doi.org/10.1016/j.cma.2021.113852}
  {\path{doi:10.1016/j.cma.2021.113852}}.
\newline\urlprefix\url{http://arxiv.org/abs/2010.13496
  https://linkinghub.elsevier.com/retrieve/pii/S0045782521001894}

\bibitem{dvorak1992}
G.~J. Dvorak, {Transformation field analysis of inelastic composite materials},
  Proceedings of the Royal Society of London. Series A: Mathematical and
  Physical Sciences 437~(1900) (1992) 311--327.
\newblock \href {https://doi.org/10.1098/rspa.1992.0063}
  {\path{doi:10.1098/rspa.1992.0063}}.

\bibitem{Michel2003}
J.~C. Michel, P.~Suquet,
  \href{https://linkinghub.elsevier.com/retrieve/pii/S0020768303003469}{{Nonuniform
  transformation field analysis}}, International Journal of Solids and
  Structures 40~(25) (2003) 6937--6955.
\newblock \href {https://doi.org/10.1016/S0020-7683(03)00346-9}
  {\path{doi:10.1016/S0020-7683(03)00346-9}}.
\newline\urlprefix\url{https://linkinghub.elsevier.com/retrieve/pii/S0020768303003469}

\bibitem{Liu2016}
Z.~Liu, M.~A. Bessa, W.~K. Liu,
  \href{http://dx.doi.org/10.1016/j.cma.2016.04.004}{{Self-consistent
  clustering analysis: An efficient multi-scale scheme for inelastic
  heterogeneous materials}}, Computer Methods in Applied Mechanics and
  Engineering 306 (2016) 319--341.
\newblock \href {https://doi.org/10.1016/j.cma.2016.04.004}
  {\path{doi:10.1016/j.cma.2016.04.004}}.
\newline\urlprefix\url{http://dx.doi.org/10.1016/j.cma.2016.04.004}

\bibitem{Quarteroni2015}
A.~Quarteroni, A.~Manzoni, F.~Negri, {Reduced Basis Methods for Partial
  Differential Equations}, Vol.~92 of UNITEXT, Springer International
  Publishing, Cham, 2016.
\newblock \href {https://doi.org/10.1007/978-3-319-15431-2}
  {\path{doi:10.1007/978-3-319-15431-2}}.

\bibitem{Hesthaven2016}
J.~S. Hesthaven, G.~Rozza, B.~Stamm, {Certified Reduced Basis Methods for
  Parametrized Partial Differential Equations}, SpringerBriefs in Mathematics,
  Springer International Publishing, Cham, 2016.
\newblock \href {https://doi.org/10.1007/978-3-319-22470-1}
  {\path{doi:10.1007/978-3-319-22470-1}}.

\bibitem{Hernandez2014}
J.~A. Hern{\'{a}}ndez, J.~Oliver, A.~E. Huespe, M.~A. Caicedo, J.~C. Cante,
  {High-performance model reduction techniques in computational multiscale
  homogenization}, Computer Methods in Applied Mechanics and Engineering 276
  (2014) 149--189.
\newblock \href {https://doi.org/10.1016/j.cma.2014.03.011}
  {\path{doi:10.1016/j.cma.2014.03.011}}.

\bibitem{Soldner2017}
D.~Soldner, B.~Brands, R.~Zabihyan, P.~Steinmann, J.~Mergheim, {A numerical
  study of different projection-based model reduction techniques applied to
  computational homogenisation}, Computational Mechanics 60~(4) (2017)
  613--625.
\newblock \href {https://doi.org/10.1007/s00466-017-1428-x}
  {\path{doi:10.1007/s00466-017-1428-x}}.

\bibitem{Ryckelynck2009}
D.~Ryckelynck, {Hyper-reduction of mechanical models involving internal
  variables}, International Journal for Numerical Methods in Engineering 77~(1)
  (2009) 75--89.
\newblock \href {https://doi.org/10.1002/nme.2406}
  {\path{doi:10.1002/nme.2406}}.

\bibitem{Radermacher2013}
A.~Radermacher, S.~Reese, {A comparison of projection-based model reduction
  concepts in the context of nonlinear biomechanics}, Archive of Applied
  Mechanics 83~(8) (2013) 1193--1213.
\newblock \href {https://doi.org/10.1007/s00419-013-0742-9}
  {\path{doi:10.1007/s00419-013-0742-9}}.

\bibitem{Guo2018}
M.~Guo, J.~S. Hesthaven, {Reduced order modeling for nonlinear structural
  analysis using Gaussian process regression}, Computer Methods in Applied
  Mechanics and Engineering 341 (2018) 807--826.
\newblock \href {https://doi.org/10.1016/j.cma.2018.07.017}
  {\path{doi:10.1016/j.cma.2018.07.017}}.

\bibitem{Hesthaven2018}
J.~S. Hesthaven, S.~Ubbiali, {Non-intrusive reduced order modeling of nonlinear
  problems using neural networks}, Journal of Computational Physics 363 (2018)
  55--78.
\newblock \href {https://doi.org/10.1016/j.jcp.2018.02.037}
  {\path{doi:10.1016/j.jcp.2018.02.037}}.

\bibitem{Swischuk2019}
R.~Swischuk, L.~Mainini, B.~Peherstorfer, K.~Willcox, {Projection-based model
  reduction: Formulations for physics-based machine learning}, Computers {\&}
  Fluids 179 (2019) 704--717.
\newblock \href {https://doi.org/10.1016/j.compfluid.2018.07.021}
  {\path{doi:10.1016/j.compfluid.2018.07.021}}.

\bibitem{Guo2021b}
T.~Guo, O.~Roko{\v{s}}, K.~Veroy, \href{http://arxiv.org/abs/2104.04451
  https://linkinghub.elsevier.com/retrieve/pii/S0045782521002619
  http://dx.doi.org/10.1016/j.cma.2021.113924}{{Learning constitutive models
  from microstructural simulations via a non-intrusive reduced basis method}},
  Computer Methods in Applied Mechanics and Engineering 384 (2021) 113924.
\newblock \href {https://doi.org/10.1016/j.cma.2021.113924}
  {\path{doi:10.1016/j.cma.2021.113924}}.
\newline\urlprefix\url{http://arxiv.org/abs/2104.04451
  https://linkinghub.elsevier.com/retrieve/pii/S0045782521002619
  http://dx.doi.org/10.1016/j.cma.2021.113924}

\bibitem{Barrault2004}
M.~Barrault, Y.~Maday, N.~C. Nguyen, A.~T. Patera, {An ‘empirical
  interpolation’ method: application to efficient reduced-basis
  discretization of partial differential equations}, Comptes Rendus
  Mathematique 339~(9) (2004) 667--672.
\newblock \href {https://doi.org/10.1016/j.crma.2004.08.006}
  {\path{doi:10.1016/j.crma.2004.08.006}}.

\bibitem{Chaturantabut2010}
S.~Chaturantabut, D.~C. Sorensen, {Nonlinear model reduction via discrete
  empirical interpolation}, SIAM Journal on Scientific Computing 32~(5) (2010)
  2737--2764.
\newblock \href {https://doi.org/10.1137/090766498}
  {\path{doi:10.1137/090766498}}.

\bibitem{Rasmussen2004}
C.~E. Rasmussen, {Gaussian Processes in Machine Learning}, in: O.~Bousquet,
  U.~von Luxburg, G.~R{\"{a}}tsch (Eds.), Advanced Lectures on Machine
  Learning: ML Summer Schools 2003, Canberra, Australia, February 2 - 14, 2003,
  T{\"{u}}bingen, Germany, August 4 - 16, 2003, Revised Lectures, Springer
  Berlin Heidelberg, Berlin, Heidelberg, 2004, pp. 63--71.
\newblock \href {https://doi.org/10.1007/978-3-540-28650-9{\_}4}
  {\path{doi:10.1007/978-3-540-28650-9{\_}4}}.

\bibitem{Botsch2010}
M.~Botsch, L.~Kobbelt, M.~Pauly, P.~Alliez, B.~Levy,
  \href{https://www.taylorfrancis.com/books/9781439865316}{{Polygon Mesh
  Processing}}, Vol.~44, A K Peters/CRC Press, 2010.
\newblock \href {https://doi.org/10.1201/b10688} {\path{doi:10.1201/b10688}}.
\newline\urlprefix\url{https://www.taylorfrancis.com/books/9781439865316}

\bibitem{Sederberg1986Free-formModels}
T.~W. Sederberg, S.~R. Parry, {Free-form deformation of solid geometric
  models}, Proceedings of the 13th Annual Conference on Computer Graphics and
  Interactive Techniques, SIGGRAPH 1986 20~(4) (1986) 151--160.
\newblock \href {https://doi.org/10.1145/15922.15903}
  {\path{doi:10.1145/15922.15903}}.

\bibitem{buhmann2003radial}
M.~D. Buhmann, {Radial basis functions: theory and implementations}, Vol.~12,
  Cambridge university press, 2003.

\bibitem{Rozza2008}
G.~Rozza, D.~B. Huynh, A.~T. Patera, {Reduced basis approximation and a
  posteriori error estimation for affinely parametrized elliptic coercive
  partial differential equations: Application to transport and continuum
  mechanics}, Archives of Computational Methods in Engineering 15~(3) (2008)
  229--275.
\newblock \href {https://doi.org/10.1007/s11831-008-9019-9}
  {\path{doi:10.1007/s11831-008-9019-9}}.

\bibitem{Veroy2003}
K.~Veroy, \href{https://dspace.mit.edu/handle/1721.1/29583}{{Reduced-basis
  methods applied to problems in elasticity : analysis and
  applications}}~(2000) (2003).
\newline\urlprefix\url{https://dspace.mit.edu/handle/1721.1/29583}

\bibitem{Rozza2020}
G.~Rozza, M.~Hess, G.~Stabile, M.~Tezzele, F.~Ballarin,
  \href{https://www.degruyter.com/document/doi/10.1515/9783110671490-001/html}{{1
  Basic ideas and tools for projection-based model reduction of parametric
  partial differential equations}}, in: Snapshot-Based Methods and Algorithms,
  De Gruyter, 2020, pp. 1--47.
\newblock \href {https://doi.org/10.1515/9783110671490-001}
  {\path{doi:10.1515/9783110671490-001}}.
\newline\urlprefix\url{https://www.degruyter.com/document/doi/10.1515/9783110671490-001/html}

\bibitem{Negri2015EfficientInterpolation}
F.~Negri, A.~Manzoni, D.~Amsallem,
  \href{http://dx.doi.org/10.1016/j.jcp.2015.09.046}{{Efficient model reduction
  of parametrized systems by matrix discrete empirical interpolation}}, Journal
  of Computational Physics 303 (2015) 431--454.
\newblock \href {https://doi.org/10.1016/j.jcp.2015.09.046}
  {\path{doi:10.1016/j.jcp.2015.09.046}}.
\newline\urlprefix\url{http://dx.doi.org/10.1016/j.jcp.2015.09.046}

\bibitem{Manzoni2012ModelGeometries}
A.~Manzoni, A.~Quarteroni, G.~Rozza,
  \href{https://onlinelibrary.wiley.com/doi/10.1002/cnm.1465}{{Model reduction
  techniques for fast blood flow simulation in parametrized geometries}},
  International Journal for Numerical Methods in Biomedical Engineering
  28~(6-7) (2012) 604--625.
\newblock \href {https://doi.org/10.1002/cnm.1465}
  {\path{doi:10.1002/cnm.1465}}.
\newline\urlprefix\url{https://onlinelibrary.wiley.com/doi/10.1002/cnm.1465}

\bibitem{Washabaugh2016OnGeometries}
K.~Washabaugh, M.~J. Zahr, C.~Farhat, {On the use of discrete nonlinear
  reduced-order models for the prediction of steady-state flows past
  parametrically deformed complex geometries}, 54th AIAA Aerospace Sciences
  Meeting 0~(January) (2016) 1--16.
\newblock \href {https://doi.org/10.2514/6.2016-1814}
  {\path{doi:10.2514/6.2016-1814}}.

\bibitem{Demo2018ShapeDecomposition}
N.~Demo, M.~Tezzele, G.~Gustin, G.~Lavini, G.~Rozza, {Shape optimization by
  means of proper orthogonal decomposition and dynamic mode decomposition},
  Technology and Science for the Ships of the Future - Proceedings of NAV 2018:
  19th International Conference on Ship and Maritime Research (2018)
  212--219\href {https://doi.org/10.3233/978-1-61499-870-9-212}
  {\path{doi:10.3233/978-1-61499-870-9-212}}.

\bibitem{Stabile2020EfficientMethods}
G.~Stabile, M.~Zancanaro, G.~Rozza, {Efficient geometrical parametrization for
  finite-volume-based reduced order methods}, International Journal for
  Numerical Methods in Engineering 121~(12) (2020) 2655--2682.
\newblock \href {https://doi.org/10.1002/nme.6324}
  {\path{doi:10.1002/nme.6324}}.

\bibitem{Yao1989}
T.-M. Yao, K.~K. Choi,
  \href{https://onlinelibrary.wiley.com/doi/10.1002/nme.1620280209}{{3-D shape
  optimal design and automatic finite element regridding}}, International
  Journal for Numerical Methods in Engineering 28~(2) (1989) 369--384.
\newblock \href {https://doi.org/10.1002/nme.1620280209}
  {\path{doi:10.1002/nme.1620280209}}.
\newline\urlprefix\url{https://onlinelibrary.wiley.com/doi/10.1002/nme.1620280209}

\bibitem{Jasak2006AutomaticMethod}
H.~Jasak, Z.~Tukovi{\'{c}}, {Automatic mesh motion for the unstructured Finite
  Volume Method}, Transactions of Famena 30~(2) (2006) 1--20.

\bibitem{itskov2007tensor}
M.~Itskov, {Tensor algebra and tensor analysis for engineers}, Springer, 2007.

\bibitem{Kouznetsova2001}
V.~G. Kouznetsova, W.~A.~M. Brekelmans, F.~P.~T. Baaijens, {Approach to
  micro-macro modeling of heterogeneous materials}, Computational Mechanics
  27~(1) (2001) 37--48.
\newblock \href {https://doi.org/10.1007/s004660000212}
  {\path{doi:10.1007/s004660000212}}.

\bibitem{Saeb2016}
S.~Saeb, P.~Steinmann, A.~Javili, {Aspects of computational homogenization at
  finite deformations: A unifying review from Reuss' to Voigt's Bound}, Applied
  Mechanics Reviews 68~(5) (2016).
\newblock \href {https://doi.org/10.1115/1.4034024}
  {\path{doi:10.1115/1.4034024}}.

\bibitem{Temizer2008}
I.~Temizer, P.~Wriggers, {On the computation of the macroscopic tangent for
  multiscale volumetric homogenization problems}, Computer Methods in Applied
  Mechanics and Engineering 198~(3-4) (2008) 495--510.
\newblock \href {https://doi.org/10.1016/j.cma.2008.08.018}
  {\path{doi:10.1016/j.cma.2008.08.018}}.

\bibitem{Sarna2021a}
N.~Sarna, P.~Benner,
  \href{https://doi.org/10.1016/j.cma.2021.114168}{{Data-Driven model order
  reduction for problems with parameter-dependent jump-discontinuities}},
  Computer Methods in Applied Mechanics and Engineering 387 (2021) 114168.
\newblock \href {https://doi.org/10.1016/j.cma.2021.114168}
  {\path{doi:10.1016/j.cma.2021.114168}}.
\newline\urlprefix\url{https://doi.org/10.1016/j.cma.2021.114168}

\bibitem{Guenot2013}
M.~Gu{\'{e}}not, I.~Lepot, C.~Sainvitu, J.~Goblet, R.~F. Coelho, {Adaptive
  sampling strategies for non-intrusive POD-based surrogates}, Engineering
  Computations (Swansea, Wales) 30~(4) (2013) 521--547.
\newblock \href {https://doi.org/10.1108/02644401311329352}
  {\path{doi:10.1108/02644401311329352}}.

\bibitem{Kast2020}
M.~Kast, M.~Guo, J.~S. Hesthaven, {A non-intrusive multifidelity method for the
  reduced order modeling of nonlinear problems}, Computer Methods in Applied
  Mechanics and Engineering 364 (2020) 112947.
\newblock \href {https://doi.org/10.1016/j.cma.2020.112947}
  {\path{doi:10.1016/j.cma.2020.112947}}.

\bibitem{Berzins2020}
A.~B{\={e}}rzin{\v{s}}, J.~Helmig, F.~Key, S.~Elgeti, {Standardized
  Non-Intrusive Reduced Order Modeling Using Different Regression Models With
  Application to Complex Flow Problems}, arXiv (2020).

\bibitem{Yang2020}
M.~Yang, Z.~Xiao, {POD-based surrogate modeling of transitional flows using an
  adaptive sampling in Gaussian process}, International Journal of Heat and
  Fluid Flow 84~(December 2019) (2020) 108596.
\newblock \href {https://doi.org/10.1016/j.ijheatfluidflow.2020.108596}
  {\path{doi:10.1016/j.ijheatfluidflow.2020.108596}}.

\bibitem{gpy2014}
{GPy}, \href{https://github.com/SheffieldML/GPy}{{GPy: A Gaussian process
  framework in python}} (2012).
\newline\urlprefix\url{https://github.com/SheffieldML/GPy}

\bibitem{permann2020moose}
C.~J. Permann, D.~R. Gaston, D.~Andr{\v{s}}, R.~W. Carlsen, F.~Kong, A.~D.
  Lindsay, J.~M. Miller, J.~W. Peterson, A.~E. Slaughter, R.~H. Stogner, R.~C.
  Martineau, {MOOSE: Enabling massively parallel multiphysics simulation},
  SoftwareX 11 (2020) 100430.
\newblock \href {https://doi.org/10.1016/j.softx.2020.100430}
  {\path{doi:10.1016/j.softx.2020.100430}}.

\bibitem{Kingma2014}
D.~P. Kingma, M.~Welling, {Auto-encoding variational bayes}, 2nd International
  Conference on Learning Representations, ICLR 2014 - Conference Track
  Proceedings~(Ml) (2014) 1--14.

\bibitem{NEURIPS2019_9015}
A.~Paszke, S.~Gross, F.~Massa, A.~Lerer, J.~Bradbury, G.~Chanan, T.~Killeen,
  Z.~Lin, N.~Gimelshein, L.~Antiga, A.~Desmaison, A.~Kopf, E.~Yang, Z.~DeVito,
  M.~Raison, A.~Tejani, S.~Chilamkurthy, B.~Steiner, L.~Fang, J.~Bai,
  S.~Chintala,
  \href{http://papers.neurips.cc/paper/9015-pytorch-an-imperative-style-high-performance-deep-learning-library.pdf}{{PyTorch:
  An Imperative Style, High-Performance Deep Learning Library}}, in:
  H.~Wallach, H.~Larochelle, A.~Beygelzimer,
  F.~d{\textbackslash}textquotesingle Alch{\'{e}}-Buc, E.~Fox, R.~Garnett
  (Eds.), Advances in Neural Information Processing Systems 32, Curran
  Associates, Inc., 2019, pp. 8024--8035.
\newline\urlprefix\url{http://papers.neurips.cc/paper/9015-pytorch-an-imperative-style-high-performance-deep-learning-library.pdf}

\bibitem{sobol1967distribution}
I.~M. Sobol', {On the distribution of points in a cube and the approximate
  evaluation of integrals}, Zhurnal Vychislitel'noi Matematiki i
  Matematicheskoi Fiziki 7~(4) (1967) 784--802.

\bibitem{Bingol2019NURBS-Python:Python}
O.~R. Bingol, A.~Krishnamurthy, {NURBS-Python: An open-source object-oriented
  NURBS modeling framework in Python}, SoftwareX 9 (2019) 85--94.

\bibitem{Reddy2019AElasticity}
B.~D. Reddy, D.~van Huyssteen, {A virtual element method for transversely
  isotropic elasticity}, Computational Mechanics 64~(4) (2019) 971--988.
\newblock \href {https://doi.org/10.1007/s00466-019-01690-7}
  {\path{doi:10.1007/s00466-019-01690-7}}.

\end{thebibliography}
\appendix
%%%%%%%% APPENDIX %%%%%%%%%%%
\section{Proof of \Cref{eq:invariance}}
\label{sec:appendix_A}
With $\bm{B}_n$ the $n$-th basis function of the weighted stress, cf. \cref{eq:P_approx_new}, we want to prove that
\begin{align}
    \int_{\Omega^p} \bm{B}_n(\bm{X}^p) \bm{F}_{\bm{\mu}}^T d\bm{X}^p \stackrel{!}{=} \int_{\Omega^p} \bm{B}_n(\bm{X}^p) d\bm{X}^p \label{eq:part1}
\end{align}
for all $\bm{\mu}$ and all $n$. First of all, $\bm{F}_{\bm{\mu}}$ can be written as
\begin{align}
    \bm{F}_{\bm{\mu}} = \mathbf{I} + \frac{\partial \bm{d}_{\bm{\mu}}}{\partial \bm{X}^p}
\end{align}
with $\bm{d}_{\bm{\mu}}$ the transformation displacement, so the left-hand side of \cref{eq:part1} splits into
\begin{align}
    \int_{\Omega^p} \bm{B}_n(\bm{X}^p) \bm{F}_{\bm{\mu}}^T d\bm{X}^p = \int_{\Omega^p} \bm{B}_n(\bm{X}^p) d\bm{X}^p + \int_{\Omega^p} \bm{B}_n(\bm{X}^p) \left(\frac{\partial \bm{d}_{\bm{\mu}}}{\partial \bm{X}^p}\right)^T d\bm{X}^p.
\end{align}
To prove \cref{eq:part1}, we thus need to show that
\begin{align}
    \int_{\Omega^p} \bm{B}_n(\bm{X}^p) \left(\frac{\partial \bm{d}_{\bm{\mu}}}{\partial \bm{X}^p}\right)^T d\bm{X}^p \stackrel{!}{=} \bm{0}. \label{eq:part2}
\end{align}
Without loss of generality, assume only a single training snapshot, obtained for parameters $(\bar{\mathbf{U}}^*,\bm{\lambda}^*,\bm{\mu}^*)$ on a domain $\Omega^{\bm{\mu}^*}$, and a transformation map $\bm{\Phi}_{\bm{\mu}^*} : \Omega^p \rightarrow \Omega^{\bm{\mu}^*}, \bm{X}^{\bm{\mu}^*} = \bm{\Phi}_{\bm{\mu}^*}(\bm{X}^p)$. Thus, there is only a single basis function,
\begin{align}
    \bm{B}_1(\bm{X}^p) = \bm{P}(\bm{\Phi}_{\bm{\mu}^*}(\bm{X}^p);\bar{\mathbf{U}}^*,\bm{\lambda}^*,\bm{\mu}^*) \bm{F}_{\bm{\mu}^*}^{-T} \left|\det{\bm{F}_{\bm{\mu}^*}}\right|,
\end{align}
and \cref{eq:part2} becomes
\begin{align}
    \int_{\Omega^p} \bm{P}(\bm{\Phi}_{\bm{\mu}^*}(\bm{X}^p);\bar{\mathbf{U}}^*,\bm{\lambda}^*,\bm{\mu}^*) \bm{F}_{\bm{\mu}^*}^{-T} \left|\det{\bm{F}_{\bm{\mu}^*}}\right| \left(\frac{\partial \bm{d}_{\bm{\mu}}}{\partial \bm{X}^p}\right)^T d\bm{X}^p. \label{eq:part3}
\end{align}
We would now like to push the integral in \cref{eq:part3} forward onto the domain $\Omega^{\bm{\mu}^*}$. By introducing the inverse mapping,
\begin{align}
    \bm{\Phi}_{\bm{\mu}^*}^{-1} : \Omega^{\bm{\mu}^*} \rightarrow \Omega^p, \quad \bm{X}^p = \bm{\Phi}^{-1}_{\bm{\mu}^*}(\bm{X}^{\bm{\mu}^*}), \quad d\bm{X}^{\bm{\mu}^*} =  \left|\det{\bm{F}_{\bm{\mu}^*}}\right| d\bm{X}^p, \label{eq:inverse_mapping}
\end{align}
$\bm{F}_{\bm{\mu}^*}$ is transformed with
\begin{align}
    \bm{F}_{\bm{\mu}^*} &= \frac{\partial \bm{\Phi}_{\bm{\mu}^*}(\bm{X}^p)}{\partial \bm{X}^p} = \frac{\partial \bm{\Phi}_{\bm{\mu}^*}(\bm{\Phi}^{-1}_{\bm{\mu}^*}(\bm{X}^{\bm{\mu}^*}))}{\partial \bm{X}^{\bm{\mu}^*}} \left(\frac{\partial \bm{X}^p}{\partial \bm{X}^{\bm{\mu}^*}}\right)^{-1} = \left(\frac{\partial \bm{X}^p}{\partial \bm{X}^{\bm{\mu}^*}}\right)^{-1}, \\
    \shortintertext{from which it follows that}
    \bm{F}_{\bm{\mu}^*}^{-T} &= \left(\frac{\partial \bm{X}^p}{\partial \bm{X}^{\bm{\mu}^*}}\right)^{T}; \label{eq:DPhiT}
\end{align}
furthermore, the partial derivative $\dfrac{\partial \bm{d}_{\bm{\mu}}(\bm{X}^p)}{\partial \bm{X}^p}$ can be expressed as
\begin{align}
    \frac{\partial \bm{d}_{\bm{\mu}}(\bm{X}^p)}{\partial \bm{X}^p} &= \frac{\partial \bm{d}_{\bm{\mu}}(\bm{\Phi}^{-1}_{\bm{\mu}^*}(\bm{X}^{\bm{\mu}^*}))}{\partial \bm{X}^{\bm{\mu}^*}} \left(\frac{\partial \bm{X}^p}{\partial \bm{X}^{\bm{\mu}^*}}\right)^{-1}, \label{eq:dddXr}\\
    \shortintertext{or}
    \left(\frac{\partial \bm{d}_{\bm{\mu}}}{\partial \bm{X}^p}\right)^T &= \left(\frac{\partial \bm{X}^p}{\partial \bm{X}^{\bm{\mu}^*}}\right)^{-T} \left(\frac{\partial \bm{d}_{\bm{\mu}}(\bm{\Phi}^{-1}_{\bm{\mu}^*}(\bm{X}^{\bm{\mu}^*}))}{\partial \bm{X}^{\bm{\mu}^*}}\right)^T \label{eq:dddXrT}
\end{align}
by transposing both sides of \cref{eq:dddXr}. With \cref{eq:inverse_mapping,eq:dddXrT,eq:DPhiT}, \cref{eq:part3} becomes
\begin{align}
    &\int_{\Omega^{\bm{\mu}^*}} \bm{P}(\bm{X}^{\bm{\mu}^*};\bar{\mathbf{U}}^*,\bm{\lambda}^*,\bm{\mu}^*) \left(\frac{\partial \bm{X}^p}{\partial \bm{X}^{\bm{\mu}^*}}\right)^{T} \left(\frac{\partial \bm{X}^p}{\partial \bm{X}^{\bm{\mu}^*}}\right)^{-T} \left(\frac{\partial \bm{d}_{\bm{\mu}}(\bm{\Phi}_{\bm{\mu}^*}^{-1}(\bm{X}^{\bm{\mu}^*}))}{\partial \bm{X}^{\bm{\mu}^*}}\right)^T d\bm{X}^{\bm{\mu}^*} \\
    = & \int_{\Omega^{\bm{\mu}^*}} \bm{P}(\bm{X}^{\bm{\mu}^*};\bar{\mathbf{U}}^*,\bm{\lambda}^*,\bm{\mu}^*) \left(\frac{\partial \bm{d}_{\bm{\mu}}(\bm{\Phi}_{\bm{\mu}^*}^{-1}(\bm{X}^{\bm{\mu}^*}))}{\partial \bm{X}^{\bm{\mu}^*}}\right)^T d\bm{X}^{\bm{\mu}^*}. \label{eq:part4}
\end{align}
Utilizing the divergence theorem, \cref{eq:part4} can be rewritten as
\begin{align}
    \int_{\partial \Omega^{\bm{\mu}^*}} \bm{t} \otimes \bm{d}_{\bm{\mu}}(\bm{\Phi}_{\bm{\mu}^*}^{-1}(\bm{X}^{\bm{\mu}^*})) \ ds - \int_{\Omega^{\bm{\mu}^*}} \bm{d}_{\bm{\mu}}(\bm{\Phi}_{\bm{\mu}^*}^{-1}(\bm{X}^{\bm{\mu}^*})) \otimes \operatorname{Div}\bm{P}(\bm{X}^{\bm{\mu}^*};\bar{\mathbf{U}}^*,\bm{\lambda}^*,\bm{\mu}^*)d\bm{X}^{\bm{\mu}^*}, \label{eq:part5}
\end{align}
where $\bm{t}\coloneqq \bm{P}(\bm{X}^{\bm{\mu}^*};\bar{\mathbf{U}}^*,\bm{\lambda}^*,\bm{\mu}^*)\bm{n}$ is the traction vector with $\bm{n}$ the outer unit normal along the boundary $\partial \Omega^{\bm{\mu}^*}$ and $ds$ is an infinitesimal boundary element. Using the fact that the training snapshot fulfills the linear momentum balance $\operatorname{Div}\bm{P}(\bm{X}^{\bm{\mu}^*};\bar{\mathbf{U}}^*,\bm{\lambda}^*,\bm{\mu}^*) = \bm{0}$ on the domain $\Omega^{\bm{\mu}^*}$, the latter part of \cref{eq:part5} 
becomes $\bm{0}$, and therefore,
\begin{align}
    \int_{\Omega^{\bm{\mu}^*}} \bm{P}(\bm{X}^{\bm{\mu}^*};\bar{\mathbf{U}}^*,\bm{\lambda}^*,\bm{\mu}^*) \left(\frac{\partial \bm{d}_{\bm{\mu}}(\bm{\Phi}_{\bm{\mu}^*}^{-1}(\bm{X}^{\bm{\mu}^*}))}{\partial \bm{X}^{\bm{\mu}^*}}\right)^T d\bm{X}^{\bm{\mu}^*} = \int_{\partial \Omega^{\bm{\mu}^*}} \bm{t} \otimes \bm{d}_{\bm{\mu}}(\bm{\Phi}_{\bm{\mu}^*}^{-1}(\bm{X}^{\bm{\mu}^*})) \ ds. \label{eq:part6}
\end{align}
Due to the definition of the auxiliary problem in \cref{eq:auxiliary_problem}, $\bm{d}_{\bm{\mu}}(\bm{\Phi}_{\bm{\mu}^*}^{-1}(\bm{X}^{\bm{\mu}^*}))$ is $\bm{0}$ on the boundary $\partial \Omega^{\bm{\mu}^*}$ and therefore the boundary integral in \cref{eq:part6} always results in the zero tensor, meaning that the integral in \cref{eq:part3} vanishes for all $\bm{\mu}$.

Since each basis function $\bm{B}_n$ is a linear combination of converged weighted stress fields, from the linearity, the integral on the left hand side of \cref{eq:part2} vanishes as well, which is what we wanted to prove and \cref{eq:part2} holds.

\hfill $\blacksquare$
\end{document}